\documentclass[12pt]{amsart}
\usepackage{amssymb,multicol,latexsym}

\usepackage[mathscr]{eucal}
\usepackage{amscd}

\usepackage{hyperref}

\allowdisplaybreaks

\newcommand{\Inv}{\mathop{\mathrm{Inv}}\nolimits}
\newcommand{\Coinv}{\mathop{\mathrm{Coinv}}\nolimits}
\newcommand{\qconst}[1]{\mathsf{q}^{#1}}
\newcommand{\lconst}[1]{\mathsf{l}^{#1}}
\newcommand{\hconst}[1]{\mathsf{h}^{#1}}

\newcommand{\aA}{\mathfrak{a}}  
\newcommand{\MOD}[1]{\mathsf{MOD}_{#1}}  
\newcommand{\FA}[1]{\mathfrak{F}_{#1}}  

\newcommand{\fusionhelp}{\mathop\tensor}
\newcommand{\sfusion}{\mathop{
           \fusionhelp\limits^{\bullet}_{\scriptscriptstyle s\ell(2)}}}
\newcommand{\nfusion}{\mathop{
           \fusionhelp\limits^{\bullet}_{\scriptscriptstyle N=2}}}

\def\DX#1#2{\frac{\d^{#1}}{\d x_{#2}^{#1}}}
\def\F{\mathsf{F}}
\def\G{\mathsf{G}}
\def\x{\boldsymbol{x}}
\def\Skew#1{\oC\langle{#1}\rangle}

\newcommand{\V}{\mathsf{V}}
\newcommand{\cI}{\mathscr{I}}
\newcommand{\cJ}{\mathscr{J}}

\newcommand{\DG}[1]{\partial_{#1}}

\newcommand{\g}{\mathfrak{g}}

\def\Asymhelp{\mathop{\mathrm{Alt}}}
\newcommand{\Alt}[1]{\mathop{\Asymhelp\limits_{#1}}}

\def\Nn{\iota}

\newcommand{\cS}{\mathscr{S}}

\def\cP{\mathscr{P}}

\newcommand{\wM}{\mathsf{M}}

\newcommand{\W}{\mathsf{W}}

\def\F{\mathop{\mathsf{F}}\nolimits}
\def\Gr{\mathop{\mathsf{Gr}}\nolimits}

\def\chr{\mathop{\mathrm{char}}\nolimits}

\newcommand{\M}{\mathsf{M}}

\makeatletter
 \@addtoreset{equation}{section}
\def\@secnumfont{\bfseries}
\makeatother

\def\tilde{\widetilde}

\newcommand{\N}[1]{N\!=\!#1}
\newcommand{\SL}[1]{s\ell(#1)}
\newcommand{\tSL}[1]{\widehat{s\ell}(#1)}

\newcommand{\half}{\frac{1}{2}}

\newcommand{\oC}{\mathbb{C}}

\newcommand{\oP}{\mathbb{P}}

\newcommand{\oZ}{\mathbb{Z}}
\newcommand{\CP}{\mathbb{CP}}

\newtheorem{Thm}{Theorem}[section]
\newtheorem{Cor}[Thm]{Corollary}
\newtheorem{Lemma}[Thm]{Lemma}

\newtheorem{Prop}[Thm]{Proposition}
\newtheorem{Conj}[Thm]{Conjecture}

\theoremstyle{definition}

\newtheorem{Rem}[Thm]{Remark}

\newenvironment{prf}{%
  \noindent{\sc Proof}}%
{\noindent{\hfill\mbox{\rule{.5em}{.5em}}\,}\par\medskip}

\def\cE{\mathscr{E}}
\def\cF{\mathcal{F}}
\def\cG{\mathcal{G}}
\def\cH{\mathcal{H}}

\def\cK{\mathcal{K}}
\def\cL{\mathcal{L}}

\def\cO{\mathscr{O}}
\def\cQ{\mathcal{Q}}

\def\cX{\mathcal{X}}
\newcommand{\cY}{\mathcal{Y}}

\def\smm{\mathfrak}

\def\mM{{\smm{M}}}  
\def\mAs{{\smm{L}}} 
\def\mD{{\smm{D}}}  
\def\mL{{\smm{L}}}  
\def\mK{{\smm{K}}}  

\def\coinv{\smm{Z}} 

\newcommand{\charsl}[2]{{\chi_{
{{\phantom{h}\kern-3pt #2}}}^{\phantom{y}\kern-3pt #1}
}}  
\newcommand{\charn}[2]{{\omega_{
{{\phantom{h}\kern-3pt #2}}}^{\phantom{y}\kern-3pt #1}
}} 
\newcommand{\scharn}[2]{{\varpi_{
{{\phantom{h}\kern-3pt #2}}}^{\phantom{y}\kern-3pt #1}
}} 
\newcommand{\bcharsl}[2]{{\overline{\chi}_{
{{\phantom{h}\kern-3pt #2}}}^{\phantom{y}\kern-3pt #1}
}}
\newcommand{\bcharn}[2]{{\overline{\omega}_{
{{\phantom{h}\kern-3pt #2}}}^{\phantom{y}\kern-3pt #1}
}}

\def\ctop{\mathsf{c}}

\newcommand{\Tr}{\mathop{\mathrm{Tr}}^{\phantom{y}}\nolimits}

\def\bar{\overline}

\newcommand{\rln}{S} 

\def\tensor{\otimes}
\renewcommand{\d}{\partial}

\begin{document}
\raggedbottom
\hfuzz=1pt
\vfuzz=1.6pt

\addtolength{\baselineskip}{4pt}

\title[Semi-Infinite $\N2$ Modules]{\hfill
  {\lowercase{\tt hep-th/0004066}}\\[12pt]
  A Semi-Infinite Construction of Unitary $\N2$ Modules}

\author[Feigin]{B.~L.~Feigin} \address{Landau Institute for
  Theoretical Physics, Russian Academy of Sciences}

\author[Semikhatov, Tipunin]{A.~M.~Semikhatov} \address{Tamm Theory
  Division, Lebedev Physics Institute, Russian Academy of Sciences} 

\author[Semi-Infinite]{I.~Yu.~Tipunin}

\begin{abstract} We show that each unitary
  representation of the $\N2$ superVirasoro algebra can be realized in
  terms of ``collective excitations'' over a filled Dirac sea of
  fermionic operators satisfying a generalized exclusion principle.
  These are semi-infinite forms in the modes of one of the fermionic
  currents.  The constraints imposed on the fermionic operators have a
  counterpart in the form of a model one-dimensional lattice system,
  studying which allows us to prove the existence of a remarkable
  monomial basis in the semi-infinite space.  This leads to a
  Rogers--Ramanujan-like character formula.  We construct the $\N2$
  action on the semi-infinite space using a filtration by
  finite-dimensional subspaces (the structure of which is related to
  the supernomial coefficients); the main technical tool is provided
  by the dual functional realization.  As an application, we identify
  the coinvariants with the dual to a space of meromorphic functions
  on products of punctured Riemann surfaces with a prescribed
  behaviour on multiple diagonals.  For products of punctured $\CP^1$,
  such spaces are related to the unitary $\N2$ fusion algebra, for
  which we also give an independent derivation.
\end{abstract}

\maketitle

\flushcolumns
\setcounter{tocdepth}{2} 
\vspace*{-36pt}

\noindent
\parbox{.96\textwidth}{
  \begin{multicols}{2}
    {\footnotesize \tableofcontents}
  \end{multicols}
  }

\thispagestyle{empty}
\addtolength{\parskip}{2pt}

\section{Introduction} \label{sec:1} We construct a new,
\textit{semi-infinite}, realization of unitary representations of the
$\N2$ superconformal algebra, in which every state in the module is a
``collective excitation'' over a filled Dirac sea of fermionic
operators satisfying a nontrivial exclusion principle.  The $\N2$
superVirasoro algebra is spanned by Virasoro generators~$\cL_n$,
Heisenberg generators~$\cH_n$, and the modes of two fermionic currents
$\cQ_n$ and $\cG_n$, $n\in\oZ$.  The semi-infinite realization
takes~a~considerably different input: a given unitary~$\N2$ module is
spanned by the action of \textit{a single} fermionic current~$\cG_n$,
$n\in\oZ$, subject to the conditions $\rln^{p}_a=0$, where, for a
fixed positive integer~$p$,
\begin{equation}\label{1.1}
  \rln^{p}_a=
  \sum_{\substack{i_0<\dots<i_{p-2}\\ i_0+\dots+i_{p-2}=a}}
  \Bigl(\prod_{m<n}(i_m-i_n)\Bigr)
  \cG_{i_0}\dots\cG_{i_{p-2}},\qquad a\in\oZ.
\end{equation}

Informally, the semi-infinite construction can be rephrased by saying
that no other $\N2$ generators except $\cG_n$ are needed to span a
unitary $\N2$ representation.  The apparent ``mismatch in the number
of the degrees of freedom'' is resolved because the module is spanned
by \textit{semi-infinite forms} in $(\cG_n)_{n\in\oZ}$.  With the
``exclusion principle'' $\rln^{p}_a=0$ imposed in addition to the
standard Pauli principle for fermions, it is a nontrivial result (a
character identity) that there are precisely as many semi-infinite
forms as there are states in a unitary $\N2$ module.  Moreover, the
semi-infinite forms carry a representation of the $\N2$ algebra with
the central charge~$3(1-\frac{2}{p})$; although eventually related to
the structure of the right-hand side of~\eqref{1.1}, this algebra
action is highly non-obvious from the conditions $\rln^{p}_a=0$
imposed on the semi-infinite forms.

In physical terms, semi-infinite
realizations~\cite{ref:1}--\cite{ref:6} of representations of
infinite-dimensional algebras are collective effects of a
``quasiparticle'' type: the representation space is filled by
excitations over a Dirac sea of operators satisfying an exclusion
principle.  The resulting nonstandard realizations of representations
can be viewed from the utilitarian standpoint as a particular
quasiparticle basis in the space of states.
In general, it is in no way obvious from the construction that these
quasiparticle states exactly fill an irreducible representation of any
algebra.  But in the cases where this is so, calculating the character
in two ways (in accordance with the quasiparticle picture and in the
standard basis) gives a nontrivial identity of the type of generalized
Rogers--Ramanujan identities~\cite{ref:10},~\cite{ref:11}.  These
identities (often called the Rogers--Ramanujan--Gordon--Andrews
identities), originally motivated by combinatorial
correspondences,\footnote{We recall the classical statement
  (see~\cite{ref:10}): ``The partitions of an integer~$n$ for which
  the difference between any two parts is not smaller than two are
  equinumerous to the partitions of~$n$ into parts $\equiv1$ or
  $4\pmod5$.''  Relations between the Rogers--Ramanujan identities and
  conformal field theory characters go back
  to~\cite{ref:12},~\cite{ref:13}.}  were investigated by different
methods, in particular with regard to their relation to conformal
field theory~\cite{ref:14}--\cite{ref:21}.  Semi-infinite
constructions of representations thus give a representation-theory
interpretation of a number of nontrivial combinatorial
phenomena~\cite{ref:22},~\cite{ref:1}.  The corresponding character
formulas can be interpreted as a result of the simultaneous existence
and interaction of particles of several
types.\footnote{\label{foot:quasi}The standard logic is as
  follows~\cite{ref:5}.  Consider, for example, two types of bosons
  (of ``colors''~1 and~2); without the interaction between the bosons,
  the partition function is ${1}/{(q)^2_{\infty}}$, which equals
  $\sum_{n_1\geq0}\sum_{n_2\geq0}{q^{n_1+n_2}}/{(q)_{n_1}(q)_{n_2}}$
  by the Durfee formula (we use the standard notation in
  Eq.~\eqref{1.5}).  The ``semi-infinite'' character formulas have a
  similar form, however the exponent in the numerator acquires a
  quadratic form in $n_1,n_2,\dots$, which is interpreted as an
  \textit{interaction} between the different types of quasiparticles.}
Formulas of this type for the partition function on the torus are
related to the thermodynamic Bethe ansatz \cite{ref:23}--\cite{ref:28}
and were also investigated in~\cite{ref:22},
\cite{ref:29}--\cite{ref:31}.

In semi-infinite constructions, the representation space (rigorously
defined as an inductive limit) is generated by \textit{semi-infinite
  forms}, or the products
\begin{equation}
  V^{(s_1)}_{\alpha_1}V^{(s_2)}_{\alpha_2}\dots
  V^{(s_n)}_{\alpha_n}\dots\,.
\end{equation} 
The subscripts (``modes'') take an infinite number of values, for
example $\alpha_i\in\oZ$, and it can be assumed as a rule that
$\alpha_1\leq\alpha_2\leq\cdots$\,. The superscripts distinguish a
finite number ($s_1,s_2,\ldots\in\{1,2,\dots,M\}$) of ``types'' of
elements (all elements are the same type in the simplest case).  It is
assumed that starting with some number~$\iota$, the sequence
$(s_i,\alpha_i)_{i\geq\iota}$ is periodic, i.e., the shift
$\alpha_i\mapsto\alpha_i+\nu$ maps the sequence into itself for
some~$\nu$.  An essential point is that the semi-infinite forms are
then considered modulo some identifications, whose role amounts to the
possibility of expressing a semi-infinite form with a sequence
$(\alpha_i)$ that is ``too dense'' through a linear combination of
``thinned out'' semi-infinite forms.

In the known constructions of the semi-infinite type, the elements
$V^{(s)}$ are either some vertex operators for a given
infinite-dimensional algebra or some currents taking values in the
algebra.  Thus, in the semi-infinite construction
of~\cite{ref:1},~\cite{ref:32} for an affine algebra $\hat{\aA}$, the
elements~$V^{(s)}$ are a part of the currents with the values in the
nilpotent subalgebra of~$\aA$; a description of the same space in
terms of different operators is given in~\cite{ref:2}, where the
semi-infinite construction is intermediate (in a certain sense)
between the case where the operators $V^{(s)}$ are currents in the
algebra and the case where they are vertex operators acting between
different modules.  Semi-infinite constructions for
$\widehat{s\ell}(2)$ modules where $V^{(s)}$ are vertex operators
arise~\cite{ref:33},~\cite{ref:3} from the Haldane--Shastry spin
chains~\cite{ref:34},~\cite{ref:35} and the related Calogero model
with spin~\cite{ref:36}.  The corresponding quasiparticle basis is
obtained as a conformal limit (taken in the neighborhood of the
antiferromagnetic state) of the space of states in the
Haldane--Shastry model; this limit produces a direct sum of the two
integrable level-1 $\widehat{s\ell}(2)$ representations.  A
decomposition into irreducible representations of the Yangian then
leads to a new basis in level-1 representations that can be written
using spin-$1/2$ $\widehat{s\ell}(2)$ vertex operators, which in the
context of the Haldane--Shastry model are interpreted as the creation
operators for \textit{spinon} excitations, quasiparticles with the
spin~$1/2$ and with a half-integer statistics~\cite{ref:37},
\cite{ref:33}, \cite{ref:3},~\cite{ref:38}.
In the cases where $V^{(s)}_{\alpha}$ are fermionic operators (for
example, for the $\N2$ superconformal algebra), the semi-infinite
construction can be viewed as a generalization to the ``interacting''
fermions of the infinite-wedge representation, which is a classical
tool in investigating a number of problems in representation theory
and beyond.  
The ``interaction'' here is understood not quite literally but in the
sense that the semi-infinite forms are considered modulo
identifications, i.e., satisfy a set of relations (as a result, in
contrast to the infinite-wedge construction, semi-infinite
constructions give irreducible representations or finite direct sums
thereof).

Semi-infinite spaces can be investigated using \textit{filtrations} by
some subspaces that can be conveniently studied.  For example, one of
the filtrations involved in the semi-infinite construction
in~\cite{ref:1} for the level-$k$ vacuum representation of the
$\widehat{s\ell}(2)$ algebra consists of finite-dimensional spaces
$\wM^+[\iota]$, which as vector spaces are isomorphic to the
$(\iota+1)$-multiple product $\oC^{k+1}\otimes
\dots\otimes\oC^{k+1}$~\cite{ref:42}.  This is similar to what is
observed in the corner transfer matrix method, where the space of
states is a semi-infinite tensor product of finite-dimensional spaces
and its ``approximations'' have the form of the above tensor product.
We recall that the results obtained in integrable statistical
mechanics models indicate an intimate relation between the space of
states of the corresponding lattice model and a representation of some
infinite-dimensional algebra~\cite{ref:43}--\cite{ref:48}.  Although
the correspondence is not straightforward by far, it is very
interesting to investigate the relations of semi-infinite
constructions of representations to exactly integrable statistical
mechanics models (this requires a $q$-deformation of the algebra
action on the semi-infinite space).

For the $\N2$ super-Virasoro (superconformal) algebra, the possibility
of constructing the semi-infinite realization of its unitary modules
was pointed out in~\cite{ref:55}, and we now develop the recipe
sketched there.  We also note the interest in the $\N2$ supersymmetry
precisely originating in investigations of the generalized
Rogers--Ramanujan--Gordon--Andrews
identities~\cite{ref:14},~\cite{ref:16}.  As noted above, ``reducing
the number of generators'' in the semi-infinite $\N2$ realization
results in that only the~$\cG_n$ modes are needed for generating the
\textit{entire} module.  This counterintuitive statement is summarized
in Theorem~\ref{thm:1} below (its proof starts with
Theorem~\ref{thm:3} and is finished in Theorem~\ref{thm:7}).

Let $p\geq3$ be a fixed positive integer, and let $\G(p)$ be the
algebra generated by anticommuting elements $(\cG_n)_{n\in\oZ}$ and an
invertible operator $\mathsf{U}$ such that
$\mathsf{U}\cG_n\mathsf{U}^{-1}=\cG_{n+1}$, with~$\cG_n$ satisfying
the constraints $S^{p}_a=0$ \ $(a\in\oZ)$ with $S^{p}_a$ defined
in~\eqref{1.1}.\footnote{We consider only graded representations of
  $\G(p)$ of the form $W=\bigoplus_{i\geq0}W_i$ with $\cG_nW_i\subset
  W_{i-n}$.  Although the expression $S^{p}_a$ contains infinite sums,
  its action on such representations is well defined; following the
  standard abuse of terminology, we mean that the ``algebra generated
  by~$\cG_n$'' involves, in particular, infinite combinations with a
  well-defined action on graded spaces of the above form.}

\begin{Thm}\label{thm:1} Let $W$ be the representation of $\G(p)$
  induced from the trivial one-dimensional representation of the
  Grassmann algebra with the generators $(\cG_n)_{n\geq0}$ {\rm(}on
  the vacuum vector~$|0\rangle${\rm)}.  Let $r$ be a fixed integer
  such that~$1\leq r\leq p-1$ and let~$C_r$ be the $\G(p)$-submodule
  in~$W$ generated from the vector $\cG_{-r}\dots\cG_{-1}|0\rangle$
  and~$V_{r,p}$ the submodule generated from the set of vectors
  \begin{equation}
    \cG_{\alpha-p+1}\dots\cG_{\alpha-r-1}
    \cG_{\alpha-r+1}\dots\cG_{\alpha-1}|\alpha\rangle-
    |\alpha-p\rangle,\qquad\alpha\in\oZ,
  \end{equation}
  where $|\alpha\rangle=\mathsf{U}^{\alpha}|0\rangle$.  The quotient
  space $W/(V_{r,p}+C_r)$ is a representation of the $\N2$ algebra
  with the central charge
  \begin{equation}
    c=3\left(1-\frac{2}{p}\right)
    \label{1.2}
  \end{equation} 
  and, moreover, is isomorphic to a direct sum of unitary $\N2$
  representations,\footnote{We recall (see Sec.~\ref{sec:2.2}) that
    unitary $\N2$ representations $\mK_{r,p;\theta}$ with central
    charge~\eqref{1.2} are labeled by a pair of integers $(r,\theta)$
    such that $0\leq\theta\leq r-1$ and $1\leq r\leq p-1$. It can be
    assumed that $0\leq\theta\leq p-1$ with each representation then
    labeled twice and the summation in~\eqref{1.3} taken over the
    spectral flow orbit (see Sec.~\ref{sec:2.1}).}
  \begin{equation}
    \W(r,p)\equiv\frac{W}{V_{r,p}+C_r}\simeq\bigoplus_{\theta=0}^{p-1}
    \mK_{r,p;\theta}.
    \label{1.3}
  \end{equation}
\end{Thm}

The ``semi-infiniteness'' is here hidden in the relations
$|\alpha\rangle=\cG_{\alpha+1}\dots
\cG_{\alpha+p-r-1}\cG_{\alpha+p-r+1}\dots
\cG_{\alpha+p-1}\*|\alpha+p\rangle$ imposed via taking the quotient;
they can be applied recursively, leading to a representation
of~$\W(r,p)$ as the space of semi-infinite forms
in~$(\cG_n)_{n\in\oZ}$ satisfying a set of relations (of which the
most important ones follow from the conditions~$S^{p}_a=0$).  At the
level of characters, we have a combinatorial corollary of
Theorem~\ref{thm:1}: taking the characters of the representations
whose isomorphism is established in Sec.~\ref{sec:5.6}, we obtain
\begin{Cor}\label{Corollary} For $k\in\mathbb{N}$, \ $1\leq r\leq
  k+1$, and $0\leq\theta\leq k+1$, there is the identity
  \begin{multline}
    \sum\limits_{{N_1\geq\dots\geq N_{k}\in\oZ}}
    \frac{z^{\sum\limits_{m=1}^{k}N_m}q^{\frac{1}{2}
        \bigl(\,\sum\limits_{m=r}^{k}-\sum\limits_{m=1}^{r-1}\bigr)
        N_m-\theta\sum\limits_{m=1}^{k}N_m+\frac{3}{2}
        \sum\limits_{m=1}^{k}N_m^2+ \sum\limits_{1\leq m<n\leq
          k}N_mN_{n}}}
    {(q)_{N_1-N_2}(q)_{N_2-N_3}\dots(q)_{N_{k-1}-N_{k}}(q)_{\infty}}=
    \\
    = \prod_{\substack{n\geq1\\
        n\not\equiv-\theta,r-\theta\mod{(k+2)}}}
    \frac{1+zq^n}{1-q^n} \prod_{\substack{n\geq1\\
        n\not\equiv\theta,\theta-r\mod{(k+2)}}}
    \frac{1+z^{-1}q^n}{1-q^n}.
    \label{1.4}
  \end{multline}
\end{Cor}

This involves the standard notation
\begin{equation}
  (q)_n=(1-q)\dots(1-q^n).
  \label{1.5}
\end{equation} 
It is useful to note that the integer parameter~$k$ is $p-2$ in terms
of the parameter in Theorem~\ref{thm:1} (regarding the quasiparticle
interpretation of the left-hand side of~\eqref{1.4}, see
footnote~\ref{foot:quasi}).

Although isomorphism~\eqref{1.3} and therefore identity~\eqref{1.4}
are eventually derived from the structure of the right-hand side
of~\eqref{1.1}, the relation between~\eqref{1.1} and~\eqref{1.3} is
far from obvious.  In this paper, we describe in detail the methods
used in the construction and proofs.  The problems solved in what
follows can be put into a general perspective of semi-infinite
constructions:
\begin{itemize}
\item[--] choosing the elements whereby the space is generated and
  formulating a system of conditions (constraints) on these elements
  (for the $\N2$ algebra, these are the fermions~$\cG_n$ and
  conditions~\eqref{1.1} respectively);
  
\item[--] counting the number of states remaining in each grade after
  imposing the conditions, i.e., finding the character (the main
  complication here consists in taking the constraints into account,
  i.e., in working with a space that is \textit{not} freely
  generated);
  
\item[--] constructing an appropriate---\textit{monomial}---basis in
  the semi-infinite space, typically by a certain ``thinning out''
  procedure\footnote{This point is in fact more fundamental than may
    seem at first glance: once established, the existence of any
    monomial basis, in itself nonobvious, becomes a powerful tool in
    analyzing spaces that are not freely generated, similarly to
    semi-infinite ones.  The existence of a monomial basis depends on
    certain subtle properties of the imposed constraints
    and allows controlling the properties of mappings between
    different subspaces of the semi-infinite space.}  (we note that in
  the set of \textit{all} semi-infinite forms, there are linear
  dependences resulting from the imposed constraints);
  
\item[--] constructing the algebra action on the semi-infinite forms
  built from the elements satisfying the chosen constraints (an
  obvious complication occurs in the case where, as with our
  construction, the semi-infinite forms constructed from the modes of
  a ``small'' number of currents must carry a representation of the
  \textit{entire} algebra);
  
\item[--] decomposing the semi-infinite space into a (direct, if
  possible) sum of representations of the basic algebra;
  
\item[--] finding representations of some other algebraic structures
  on subspaces of the semi-infinite space.
\end{itemize}

Realizing this program (where we do not consider the last item)
requires using a combination of different means.  We now describe them
in more detail for the $\N2$ superconformal algebra, essentially
following the contents of this paper although in an order somewhat
different from the order of sections (these methods are also
interesting because the problems that we solve in constructing the
semi-infinite realization are typical of a number of semi-infinite
constructions beyond the present one (cf.\ the spinon basis
construction in~\cite{ref:3}).)

\begin{description}
\item[Inductive limit {\rm (Sec.~\ref{sec:3.1})}] The semi-infinite
  space~$\W_{r,p;\theta}$ that is eventually shown to be isomorphic to
  the unitary $\N2$ representation~$\mK_{r,p;\theta}$ is defined as
  the inductive limit of the spaces $\W_{r,p;\theta}(\iota)$
  generated, as~$\iota$ increases, from progressively more twisted
  highest-weight states, namely, from those states on which the
  progressively larger part of the generators~$\cG_n$ become creation
  operators as $\iota\to\infty$, with ``only'' the
  $(\cG_n)_{n\geq\iota p+\theta}$ generators remaining the
  annihilation operators.
  
\item[The dual space {\rm (Sec.~\ref{sec:4.1})}] A key role in
  studying the inductive limit is played by the dual space, which can
  be realized using ``polynomials in an infinite number of variables''
  (more precisely, polynomial differential forms)~\cite{ref:1}.  The
  dual to the quotient space in Eq.~\eqref{1.3} is a subspace of
  polynomials satisfying certain conditions on $(p-1)$-multiple
  diagonals $x_1=\dots=x_{p-1}$ and at zero.  Specifically, we
  investigate the space
  \begin{equation}
    \W_{r,p;0}(0)^*\subset\oC\oplus\oC[x]\,dx\oplus
    \oC\langle x_1,x_2\rangle\,dx_1dx_1\oplus
    \oC\langle x_1,x_2,x_3\rangle\,dx_1\,dx_2\,dx_3\oplus\cdots,
  \end{equation}
  where $\oC\langle x_1,\dots,x_n\rangle$ are antisymmetric
  polynomials in~$n$ variables and the space $\W_{r,p;0}(0)^*$ is
  singled out by the vanishing conditions on multiple diagonals and at
  zero.  The character of this space can be evaluated by introducing a
  filtration on each $\oC\langle x_1,\dots,x_n\rangle$ induced by the
  lexicographic order on partitions of~$n$.
  
\item[The characters {\rm (Sec.~\ref{sec:4.2})}] The sum over
  partitions gives a formula of the
  Rogers--Raman\-ujan--Gordon--Andrews type for the character
  of~$\W_{r,p;0}(0)$, and after applying $\N2$ algebra automorphisms,
  also for all the spaces $\W_{r,p;\theta}(\iota)$, \ $\iota\in\oZ$,
  involved in the inductive limit.  A remarkable property of these
  character formulas is that they admit the limit as $\iota\to\infty$.
  This limit is a candidate for the character of the semi-infinite
  space $\W_{r,p;\theta}=
  \varinjlim_{\iota\to\infty}\W_{r,p;\theta}(\iota)$, but proving that
  the limit is the character requires establishing that all the
  mappings $\W_{r,p;\theta}(\iota)\to \W_{r,p;\theta}(\iota+1)$ used
  in constructing the inductive limit are embeddings (and the limit of
  the characters is hence equal to the character of the inductive
  limit).
  
\item[The thin basis {\rm (Sec.~\ref{sec:3.2})}] The above statement
  regarding the embeddings follows from the existence of a remarkable
  monomial basis consisting of the states in which the modes~$\cG_n$
  are ``thinned out'' in accordance with the procedure described in
  Theorem~\ref{thm:2} below.  By the sequence of \textit{occupation
    numbers} associated with a semi-infinite form
  $\cG_{i_1}\dots\cG_{i_n}\dots$ \ ($i_1<\dots<i_n<\cdots$), we mean
  the sequence $\bigl(\gamma(n)\bigr)$ of elements labeled by integers
  $n\geq i_1$ that are equal to zero except for
  $\gamma(i_1)=\gamma(i_2)=\dots=1$.  The construction of the special
  monomial basis, called the \textit{thin basis}, is given in the
  following theorem (in Sec.~\ref{sec:3.2}, we prove an equivalent
  statement in Lemma~\ref{lemma:1}).

  \begin{Thm}\label{thm:2} There exists a basis in $\W_{r,p;\theta}$
    whose elements are semi-infinite forms~$\cG_{i_1}\cG_{i_2}\dots$
    satisfying the following conditions:

    \begin{enumerate}
    \item for any~$n$, $i_{n+p-2}-i_{n}\geq p$ {\rm(}hence, any
      segment $[n,n+p-1]\subset\oZ$ of the length~$p$ can contain at
      most $p-2$ nonzero occupation numbers{\rm)}.
      
    \item For $n\gg1$, the sequence of occupation numbers $\alpha(n)$
      is periodic with the period~$p$ and exactly $p-2$ occurrences
      of~$1$ per period and is
      \begin{equation}
        \overbrace{\underbrace{1,\dots,1}_{r-1},0,
          \underbrace{1,\dots,1}_{p-r-1},0}^p,\dots,
        \overbrace{\underbrace{1,\dots,1}_{r-1},0,
          \underbrace{1,\dots,1}_{p-r-1},0}^p,\dots\,.
      \end{equation}
    \end{enumerate}
  \end{Thm}
  
\item[Lattices, crosses, and recursions {\rm (Sec.~\ref{sec:4.3})}]
  Construction of the thin basis corresponds to a model statistical
  system on a semi-infinite one-dimensional lattice.  Each
  semi-infinite form $\cG_{i_1}\dots\cG_{i_n}\dots$ can be represented
  as a configuration of crosses on the lattice, for example
  \begin{equation}
    \times\quad\circ\quad\times\quad\times\quad\circ\quad
    \circ\quad\circ\quad 
    \times\quad\circ\quad\circ\quad\times\quad\cdots
    \label{1.6}
  \end{equation}
  (a site with a cross corresponds to the occupation number~1, others
  to~0).  The ``enumeration'' of thin basis elements then becomes the
  problem of finding the partition function of all configurations of
  crosses with \textit{any $p$ consecutive lattice sites carrying at
    most $p-2$ crosses}.  These partition functions can be analyzed
  using a version of the so-called Andrews--Schur method, which
  consists in establishing recursive relations and subsequently taking
  the limit as the ``finitization parameter'' goes to infinity.  This
  allows us to show that the semi-infinite forms described in
  Theorem~\ref{thm:2} indeed constitute a basis in the semi-infinite
  space.
  
\item[The algebra action on the semi-infinite space {\rm
    (Sec.~\ref{sec:5})}] It can be directly shown that the derived
  characters of the semi-infinite spaces $\W_{r,p;\theta}$ coincide
  with the characters of the corresponding unitary $\N2$
  representations $\mK_{r,p;\theta}$; from the existing mapping
  $\W_{r,p;\theta}\to\mK_{r,p;\theta}$, it is then easy to deduce that
  $\W_{r,p;\theta}$ is in fact isomorphic to $\mK_{r,p;\theta}$, which
  in turn implies the existence of the $\N2$ algebra action on
  $\W_{r,p;\theta}$.  However, the explicit form of this action then
  remains unknown.  We choose a more ``conceptual'' approach and
  directly construct the $\N2$ action on the semi-infinite space
  $\W_{r,p;\theta}$.  The definition of the semi-infinite space does
  not suggest that this space carries an action of the generators
  $\cQ_n$, $\cL_n$, and~$\cH_n$, with $n\in\oZ$ satisfying
  algebra~\eqref{2.1}; the existence of this action is extremely
  sensitive to the imposed constraints~$S^{p}_a=0$.  The methods for
  constructing the action developed here can also be useful in
  investigating other semi-infinite spaces where constraints of a
  different form can be compatible with the action of another algebra.
  These methods are as follows:
  
\item[A. The positive filtration {\rm (Sec.~\ref{sec:5.1})}] In
  constructing the $\N2$ algebra action on the semi-infinite space, we
  use a filtration of $\W_{r,p;\theta}$ by finite-dimensional spaces
  called the \textit{positive filtration}.  For affine algebras,
  similar subspaces are known as \textit{Demazure modules} and have
  been studied from different standpoints
  (see~\cite{ref:49}--\cite{ref:51} and the bibliography therein).
  
\item[B. Differential operators {\rm (Sec.~\ref{sec:5.2})}] On the
  above finite-dimensional subspaces, we construct the action of a
  part of the $\N2$ algebra generators via differential operators (in
  finitely many Grassmann variables~$\cG_n$, \ $0\leq n\leq N$).  More
  precisely, differential operators a priori act on a freely generated
  space, and we must therefore show that their action is compatible
  with taking the quotient with respect to the relations induced by
  the basic conditions~\eqref{1.1} on the subspace.
  
\item[C. The dual picture {\rm (Sec.~\ref{sec:5.3})}] To verify
  compatibility of the differential operator action with
  relations~\eqref{1.1}, we again use the functional realization.  The
  operators dual to the differential operators become the
  ``homology-type'' differentials on polynomials, which considerably
  simplifies the statements that must be proved.
  
\item[D. Gluing the action from pieces {\rm (Sec.~\ref{sec:5.4})}] The
  action of the $\N2$ generators on a vector of the semi-infinite
  space must not depend on the filtration term to which this vector is
  viewed to belong.  We prove this independence as well as the
  independence of the constructed action from any arbitrariness
  involved in the construction.  We actually construct the action of
  only a part of the $\N2$ generators that \textit{generate the entire
    algebra}, and we then prove that these generate precisely the
  $\N2$ superconformal algebra.
  
\item[E. Isomorphism with the unitary representation {\rm
    (Sec.~\ref{sec:5.6})}] As soon as it is establish\-ed that the
  semi-infinite space $\W_{r,p;\theta}$ is a module over the $\N2$
  algebra, it is easy to verify that the mapping
  $\W_{r,p;\theta}\to\mK_{r,p;\theta}$ into the unitary representation
  is an isomorphism of $\N2$ modules, which completes the
  semi-infinite construction.
\end{description}

The semi-infinite construction of unitary $\N2$ modules is closely
related to a similar construction of the unitary $\widehat{s\ell}(2)$
modules~\cite{ref:1}.  In particular, the method used in constructing
the $\N2$ algebra action on the semi-infinite space also allows us to
define the $\widehat{s\ell}(2)$ algebra action on the corresponding
semi-infinite space (Sec.~\ref{sec:5.5}).  This involves the Demazure
modules; the corresponding characters are related to the generalized
Pascal triangles~\cite{ref:14} and supernomial
coefficients~\cite{ref:52}, and in Sec.~\ref{sec:6.1}, we also
consider a combinatorial construction of bases in the $\N2$
``Demazure'' subspaces (cf.~\cite{ref:49}--\cite{ref:51}).  As another
aspect of the relation between $\widehat{s\ell}(2)$ and $\N2$
structures, we consider the correspondence between the modular
functors~\cite{ref:32} (Sec.~\ref{sec:2.3}) and use it in relating the
(unitary) $\widehat{s\ell}(2)$ and $\N2$ fusion rules.  The modular
functor can be described somewhat more explicitly in the semi-infinite
realization (see Sec.~\ref{sec:6.2}), because the representation is
generated by only a part of the currents, which allows identifying the
space dual to coinvariants with the space of meromorphic functions on
products of punctured Riemann surfaces, with the functions required to
possess a prescribed behavior on multiple diagonals (which is a
counterpart of the relations $S^{p}_a=0$) and at the punctures.  In
some cases, the dimensions of these functional spaces can be evaluated
directly; for the products of $\oC\oP^1$, on the other hand, these
dimensions follow from the fusion rules for the $\N2$ algebra.  The
unitary $\N2$ fusion rules have been obtained from the Verlinde
conjecture~\cite{ref:53}, but we give an independent derivation based
on acting on the $\widehat{s\ell}(2)$ fusion rules with the functor
that realizes the equivalence of
categories~\cite{ref:54},~\cite{ref:55},\footnote{This equivalence is
  ``modulo'' the spectral flows, and this is why the fusion algebras
  for the $\widehat{s\ell}(2)$ and $\N2$ theories are not identical;
  however, the equivalence statement is sufficiently powerful to
  \textit{derive} the $\N2$ fusion algebra from the
  $\widehat{s\ell}(2)$ fusion.  The correspondence between
  $\widehat{s\ell}(2)$ and $\N2$ fusion algebras can be considered in
  the framework of the correspondence between $\widehat{s\ell}(2)$ and
  $\N2$ modular functors in Sec.~\ref{sec:2.3}, which extends the
  equivalence of representation categories.} see Sec.~\ref{sec:6.3}.

\section{The $\N2$ algebra} \label{sec:2} In this section, we recall
the main facts pertaining to the $\N2$ superconformal algebra and
motivate the semi-infinite construction.  The reader may wish to go
directly to Sec.~\ref{sec:2.4} and use Secs.~\ref{sec:2.1},
\ref{sec:2.2}, and (partly)~\ref{sec:2.3} for reference.

\subsection{The $\N2$ algebra and the spectral flow} \label{sec:2.1}
The $\N2$ superconformal algebra is generated by the bosonic
operators~$\cL_n$ (Virasoro algebra generators) and~$\cH_n$
(Heisenberg algebra) and the fermionic operators~$\cG_n$ and~$\cQ_n$.
We consider the algebra in the basis where the nonvanishing
commutation relations are given by
\begin{alignat}{2}\label{2.1}
  {[}\cL_m,\cL_n]={}&(m-n)\cL_{m+n},&[\cH_m,\cH_n]={}&
  \tfrac{\ctop}{3}m\delta_{m+n,0},\notag\\
  {[}\cL_m,\cG_n]={}&(m-n)\cG_{m+n},&
  [\cH_m,\cG_n]={}&\cG_{m+n}, \notag\\
  {[}\cL_m,\cQ_n]={}&-n\cQ_{m+n},&
  [\cH_m,\cQ_n]={}&-\cQ_{m+n},\displaybreak[0]\\
  {[}\cL_m,\cH_n]={}&
  -n\cH_{m+n}+\tfrac{\ctop}{6}(m^2+m)\delta_{m+n,0},\kern-80pt
  \notag\\
  [\cG_m,\cQ_n]={}& 2\cL_{m+n}-2n\cH_{m+n}+
  \tfrac{\ctop}{3}(m^2+m)\delta_{m+n,0},\kern-80pt\notag
\end{alignat}
where $m,n\in\oZ$ and the bracket $[\cdot,\cdot]$ denotes the
supercommutator.  The central charge $c\neq3$ can be conveniently
parametrized as $c=3(1-{2}/{p})$ with $p\in\oC\setminus\{0\}$.

The algebra automorphisms include the group $\oZ$ of
automorphisms $\mathsf{U}_{\theta}$, \ $\theta\in\oZ$, called
the spectral flow~\cite{ref:56}. In the basis chosen in~\eqref{2.1},
the spectral flow acts as
\begin{equation}
  \mathsf{U}_{\theta}:\quad
  \begin{aligned}
    \cL_n&\mapsto\cL_n+\theta\cH_n+
    \frac{c}{6}(\theta^2+\theta)\delta_{n,0},&\qquad
    \cH_n&\mapsto\cH_n+\frac{c}{3}\theta\delta_{n,0},
    \\
    \cQ_n&\mapsto\cQ_{n-\theta},&\qquad \cG_n&\mapsto\cG_{n+\theta},
  \end{aligned}
  \quad\theta\in\oZ.
  \label{2.2}
\end{equation}
The action of the spectral flow on an $\N2$ module gives a
nonisomorphic representation in general.  The modules transformed by
the spectral flow are called the \textit{twisted} modules.  We use the
notation $\mD_{\bullet;\theta}=\mathsf{U}_{\theta}\mD_{\bullet}$, which is also applied to modules carrying other labels,
for example, $\mK_{r,p;\theta}=\mathsf{U}_{\theta}\mK_{r,p}$ for the
unitary representations considered in what follows.  The
parameter~$\theta$ is called the twist.  We omit zero twist from the
notation.

The character of an $\N2$ module $\mD$ is defined as
\begin{equation}
  \omega_{\mD}(z,q)=\Tr_{\mD}(z^{\cH_0}q^{\cL_0}),
  \label{2.3}
\end{equation} 
where taking the trace over $\mD$ involves a sesquilinear
form~\cite{ref:57} that can be found in~\cite{ref:55} in our current
notation.  Under the spectral flow action, characters transform as
\begin{equation}
  \omega_{\mathsf{U}_{\theta}\mD}(z,q)=
  z^{-\frac{c}{3}\theta}q^{\frac{c}{6}(\theta^2-\theta)}
  \omega_{\mD}(zq^{-\theta},q).
  \label{2.4}
\end{equation}

We define the \textit{twisted highest-weight vector}
$|h,p;\theta\rangle$ as a state satisfying the annihilation conditions
\begin{align}\label{2.5}
  \cQ_{-\theta+m}|h,p;\theta\rangle=
  \cG_{\theta+m}|h,p;\theta\rangle=
  \cL_{m+1}|h,p;\theta\rangle=\cH_{m+1}|h,p;\theta\rangle=0,\qquad
  m\in\mathbb{N}_0
\end{align}
and also the conditions
\begin{align}
  &\left(\cH_0+\frac{c}{3}\theta\right)|h,p;\theta\rangle=
  h|h,p;\theta\rangle,
  \\
  &\left(\cL_0+\theta\cH_0+\frac{c}{6}(\theta^2+\theta)\right)
  |h,p;\theta\rangle=0
\end{align}
(where the second one follows from the annihilation conditions).

Anticipating some of what follows, we note that the semi-infinite
construction also involves twisted states on which, however, only the
action of the~$\cG_n$ operators is defined such that the same
vanishing conditions as for the corresponding states in the unitary
module are satisfied, but the action of the remaining generators must
be reconstructed (such that all the relations that hold in the unitary
module are satisfied).

\subsection{Unitary representations of the $\N2$
  algebra\mdseries\selectfont\kern-4pt}\cite{ref:55}.
\label{sec:2.2} The unitary $\N2$ representations~\cite{ref:57} are
periodic under the spectral flow with the period~$p$; that is, the
spectral flow $\mathsf{U}_{p}$ produces an isomorphic representation,
\begin{equation}
  \mK_{r,p;\theta+p}\approx\mK_{r,p;\theta}.
  \label{2.6}
\end{equation} 
For unitary representations, the twist parameter~$\theta$ can
therefore be considered modulo~$p$.  Moreover, for a given~$p$, there
are only $p(p-1)/2$ nonisomorphic unitary
representations~$\mK_{r,p;\theta}$, which can be labeled by
$0\leq\theta\leq r-1$ and $1\leq r\leq p-1$, because there are the
\textit{isomorphisms of} $\N2$ modules
\begin{equation}
  \mK_{r,p;\theta+r}\approx\mK_{p-r,p;\theta},\qquad
  1\leq r\leq p-1,\qquad\theta\in\oZ_p.
  \label{2.7}
\end{equation} Representations with zero twist are denoted by 
$\mK_{r,p}\equiv\mK_{r,p;0}$ for brevity.

The characters $\omega_{r,p;\theta}^{\mK}\equiv\omega_{\mK_{r,p;\theta}}$ of the unitary $\N2$ representations $\mK_{r,p;\theta}$ in the notation in~\cite{ref:55} (see
also~\cite{ref:58}--\cite{ref:60}) are given by
\begin{multline}
  \omega_{r,p;\theta}^{\mK}(z,q)=
  z^{\frac{2\theta-r+1}{p}-1}q^{\frac{\theta r-\theta^2}{p}+\theta}
  \frac{\eta(q^p)^3}{\eta(q)^3}
  \frac{\vartheta_{1,0}(z,q)\vartheta_{1,1}(q^r,q^p)}
  {\vartheta_{1,0}(zq^{-\theta},q^p)
    \vartheta_{1,0}(zq^{r-\theta},q^p)},
  \\
  1\leq r\leq p-1,\qquad0\leq\theta\leq r-1,
  \label{2.8}
\end{multline} 
where we introduce the theta functions
\begin{align}
  \vartheta_{1,1}(z,q)&=q^{\frac{1}{8}}\sum_{m\in\oZ}(-1)^m
  q^{\frac12(m^2- m)}z^{-m}=
  q^{\frac18}\prod_{m\geq0}(1-z^{-1}q^m)
  \prod_{m\geq1}(1-zq^m)\prod_{m\geq1}(1-q^m),
  \\
  \vartheta_{1,0}(z,q)&=q^{\frac18}\sum_{m\in\oZ}q^{\frac12(m^2-m)}z^{-m}=
  q^{\frac18}\prod_{m\geq0}(1+z^{-1}q^m)\prod_{m\geq1}(1+zq^m)
  \prod_{m\geq1}(1-q^m).
\end{align}

These characters (where we can set~$\theta=0$ for simplicity, because
the behavior of characters under the spectral flow transformations is
determined by~\eqref{2.4}) can also be expanded with respect to the
theta functions $\vartheta_{1,0}(\cdot,q^{p(p-2)})$ as
\begin{multline}
  \omega_{r,p}^{\mK}(z,q)={}z^{\frac{1+2\theta-r}{p}-\theta}
  q^{\frac{1}{2}(1-\frac{2}{p})(\theta^2-\theta)-\frac{1-r}{p}\theta-
    \frac{p(p-2)}{8}}\times
  \\
  \times \sum_{a=0}^{p-3}z^{a}q^{\frac12(a^2-(2\theta+1)a)}
  \vartheta_{1,0}\bigl(z^{p-2}q^{pa-(r-1)-(p-2)\theta-
    \frac{(p-1)(p-2)}{2}},q^{p(p-2)}\bigr)C^a_{r,p}(q),
  \label{2.9}
\end{multline}
where the \textit{string functions} $C^a_{r,p}(q)$ depend only on~$q$.

\subsection{Equivalence of categories and related~issues}
\label{sec:2.3} A linear combination of the unitary $\N2$
characters belonging to the same spectral flow orbit gives the
characters $\chi_{r,k}^{\mAs}$ of the unitary $\widehat{s\ell}(2)_k$
representations~$\mAs_{r,k}$,
\begin{multline}
  \chi_{r,p-2}^{\mAs}(z,q)\vartheta_{1,0}(zy,q)
  =q^{\frac{r^2-1}{4p}-\frac{p}{4}+\frac{1}{8}}y^{\frac{r-1}{p}}
  z^{\frac{r-1}{2}}\times
  \\
  \times\sum_{a=0}^{p-1}
  \omega_{r,p}^{\mK}(yq^{-a},q)y^{-a}z^{-a}q^{\frac{a^2-a}{2}-
    \frac{r-1}{p}a}\vartheta_{1,0}(z^py^{2}q^{-p+r-2a},q^{2p}).
  \label{2.10}
\end{multline}
This follows from of the isomorphism of $\N2$
modules~\cite{ref:54},~\cite{ref:55}
\begin{equation}
  \mAs_{r,k}\otimes\Omega\approx
  \bigoplus_{\theta=0}^{k+1}\mK_{r,k+2;\theta}\otimes
  \Biggl(\,\bigoplus_{\nu\in\sqrt{2(k+2)}\oZ}
  \cH^+_{\sqrt{\frac{2}{k+2}}(\frac{r-1}{2}-\theta)-\nu}\Biggr),
  \label{2.11}
\end{equation} 
where $\Omega$ is the free-fermion module and $\cH^+_a$ are Fock
spaces (the sum over the lattice $\sqrt{2(k+2)}\,\oZ$ defines a
vertex operator algebra).  Equation~\eqref{2.11} allows us to
establish the functorial correspondence
\begin{equation}
  \mAs_{r,k;\bullet}\rightsquigarrow\mK_{r,k+2;\bullet}
  \label{2.12}
\end{equation} 
between the spectral flow orbits of each algebra.  This is an
equivalence between categories whose objects are spectral flow
orbits~\cite{ref:54}; in other words, this is the equivalence of
representation categories of the algebras obtained by adding the
spectral flow operator to the universal enveloping algebra (of the
$\widehat{s\ell}(2)$ and the $\N2$ algebras respectively).

Expanding on the equivalence of categories, we now describe the
correspondence between modular functors for the $\widehat{s\ell}(2)$
and $\N2$ theories.  This correspondence is interesting, in
particular, because the $\widehat{s\ell}(2)$ modular functor has the
well-known geometric interpretation in terms of the moduli spaces of
$SL_2$ vector bundles, whereas no a priori geometric interpretation is
known for the $\N2$ modular functor.

For a given Riemann surface~$\cE$, the modular functor
$\MOD{\widehat{s\ell}(2)}(k,\cE)$ of the level-$k$
$\widehat{s\ell}(2)$ Wess--Zumino--Witten theory and the modular
functor $\MOD{\N2}(p,\cE)$ of the theory based on the $\N2$ algebra
with central charge~\eqref{1.2} are related by
\begin{equation}
  \MOD{N=2}(p,\cE)=\Coinv_{H^1(\cE,\oZ_2)/V_+}^{\phantom{y}}
  \bigl(\Inv_{V_+}^{\phantom{y}}
  \bigl(\MOD{\widehat{s\ell}(2)}(p-2,\cE)\otimes
  \MOD{\text{free}}(p,\cE)\bigr)\bigr),
  \label{2.13}
\end{equation} 
where $\MOD{\text{free}}(p,\cE)$ is the modular functor of the free
theory whose vertex operator algebra is associated with the lattice
$\sqrt{2p}\,\oZ$ (see~\eqref{2.11}).  The $\widehat{s\ell}(2)$ modular
functor $\MOD{\widehat{s\ell}(2)}(k,\cE)$ carries a projective action
$\cO(\cdot)$ of the group $H^1(\cE,\oZ_2)$ such that
$\cO(\alpha\beta)=\cO(\alpha)\cO(\beta)
(-1)^{k\langle\alpha,\beta\rangle}$ for the cycles~$\alpha$
and~$\beta$ with the intersection form $\langle\alpha,\beta\rangle$
or, in other words, there is the action of the Heisenberg
group~$\Gamma$ with the center~$\oZ_2$ such that $\Gamma/
\oZ_2=H^1(\cE,\oZ_2)$.  The modular functor of the free theory, which
is the space of level-$2p$ theta functions on the Jacobian of~$\cE$,
is an irreducible representation of the Heisenberg group obtained as
the central extension of $H^1(\cE,\oZ_{2p})\simeq\oZ_{2p}^{2g}$ with
the help of~$\oZ_{2p}$, where $g$ is the genus of~$\cE$.  It is
therefore a projective representation of the subgroup
$H^1(\cE,\oZ_2)\subset H^1(\cE,\oZ_{2p})$ (with the embedding induced
by $\oZ_2\to\oZ_{2p}$).  Therefore, the group~$\Gamma$ acts on the
tensor product $\MOD{\widehat{s\ell}(2)}(p-2,\cE)\otimes
\MOD{\text{free}}(p,\cE)$.  Moreover, the center then acts trivially,
and the tensor product becomes a representation of the
$H^1(\cE,\oZ_2)$ group.  In this group, which is isomorphic
to~$\oZ_{2}^{2g}$, one must choose the maximal
$\langle\cdot,\cdot\rangle$-isotropic subspace $V_+$ and take 
\textit{invariants} with respect to~$V_+$ in the tensor product.  The
invariants carry the action of the quotient $\oZ_{2}^{2g}/V_+$, and
one must then take \textit{coinvariants} with respect to this action.
This explains the notation used in~\eqref{2.13}.

\subsection{Motivation for the semi-infinite construction}
\label{sec:2.4} The semi-infinite construction of the unitary $\N2$
representations can be motivated by the following observations. In the
vacuum representation, we consider the decoupling condition for the
singular vector $\cQ_{1-p}\cQ_{2-p}\dots\cQ_{-1}|0,p;0\rangle$.  The
corresponding field
\begin{equation}
  \partial^{p-2}\cQ(z)\dots\partial\cQ(z)\cQ(z)
\end{equation} 
(with $\partial=\partial/\partial z$ and $\cQ(z)=
\sum_{n\in\oZ}\cQ_nz^{-n-1}$) has vanishing correlation functions with
all the fields in the corresponding ``minimal'' model
(cf.~\cite{ref:61}).  The identity
$\partial^{p-2}\cQ(z)\dots\partial\cQ(z)\cQ(z)=0$ therefore holds in
unitary representations as an ``operator'' equality in the sense that
it is satisfied when acting on any vector of any unitary
representation.  Similarly, the ``symmetric'' relation
$\partial^{p-2}\cG(z)\dots\partial\cG(z)\cG(z)=0$ also holds for
$\cG(z)=\sum_{n\in\oZ}\cG_n z^{-n-2}$.  Any of these relations can be
chosen as the starting point of the semi-infinite construction.
Remarkably, this suffices for reconstructing the entire
representation.  We choose the $\cG$-relation (which is a matter of
convention and/or application of $\N2$ algebra automorphisms).
Therefore, we define
\begin{equation}
  S^{p}(z)=\partial^{p-2}\cG(z)\dots\partial\cG(z)\cG(z),\qquad
  \cG(z)=\sum_{n\in\oZ}\cG_nz^{-n-2},
  \label{2.14}
\end{equation} 
and rewrite the condition $S^{p}(z)=0$ in terms of the modes $\cG_n$
as $S^{p}_a=0$ (see~\eqref{1.1}). For example, for $p=3$, the first
several relations satisfied on the vacuum vector are given by
\begin{multline}
  \cG_{-2}\cG_{-1}=0,\qquad\cG_{-3}\cG_{-1}=0,\qquad
  3\cG_{-4}\cG_{-1}+\cG_{-3}\cG_{-2}=0,
  \\
  4\cG_{-5}\cG_{-1}+2\cG_{-4}\cG_{-2}=0,\qquad
  5\cG_{-6}\cG_{-1}+3\cG_{-5}\cG_{-2}+
  \cG_{-4}\cG_{-3}=0,\qquad\dots\,.
  \label{2.15}
\end{multline}

\begin{figure}[tb]
  \begin{center}
 \unitlength=.8pt
    \begin{picture}(400,125)
      \put(0,10){
        \linethickness{.1pt}
        \multiput(0,-5)(20,0){6}{\line(0,1){115}}
        \multiput(-5,0)(0,20){6}{\line(1,0){115}}
        \thicklines
        \put(57.5,97){$\bullet$}
        \put(60,100){\vector(-1,-1){18}}
        \put(37.5,77){$\bullet$}
        \put(40,80){\vector(-1,-2){19}}
        \put(17.5,37){$\bullet$}
        \put(20,40){\vector(-1,-3){19}}
        }
      \put(40,-30){
        \put(248,144){$\bullet$}
        \put(348,77){$\bullet$}
        \put(148,57){$\bullet$}
        \linethickness{.9pt}
        \bezier{200}(150,60)(170,100)(190,120)
        \bezier{200}(190,120)(220,140)(250,148)
        \bezier{200}(250,148)(273,146)(312,124)
        \bezier{200}(312,124)(332.8,110)(351,80)
        \bezier{200}(351,80)(364,50)(377,10)
        }
    \end{picture}    
  \end{center}
  \caption[Extremal diagram of a unitary $N=2$ module]{\label{fig:1}
    Extremal diagram of a unitary $\N2$ module.}
\end{figure}
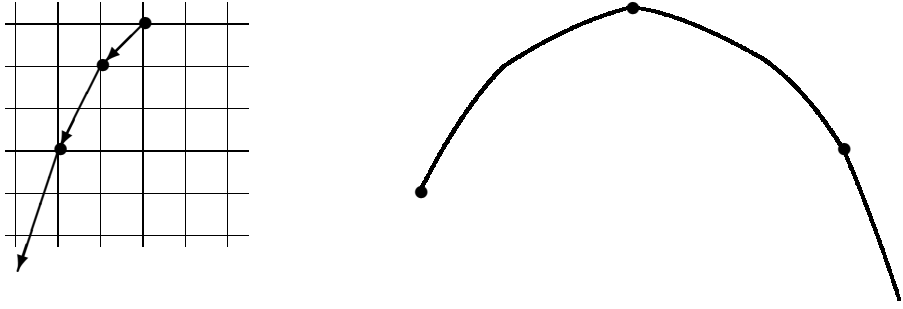

In addition, the highest-weight vector
$|r,p;\theta\rangle_{\mathrm{irr}}$ of a given unitary representation
$\mK_{r,p;\theta}$ satisfies another vanishing condition,
$\cG_{\theta-r}\dots\cG_{\theta-1}
|r,p;\theta\rangle_{\mathrm{irr}}=0$, which is the decoupling
condition for another singular vector.  Next, the state
$\cG_{\theta-p+1}\dots\cG_ {\theta-r-1}
\cG_{\theta-r+1}\dots\cG_{\theta-1}|r,p;\theta\rangle_{\mathrm{irr}}$
satisfies the twisted highest-weight conditions~\eqref{2.5} with the
twist $\vartheta=\theta-p$.  Acting on this state with the operators
$\cG_{-p+\theta-p+1}\dots\cG_{-p+\theta-r-1}
\cG_{-p+\theta-r+1}\dots\cG_{-p+\theta-1}$, we obtain a twisted
highest-weight state with the twist~$\vartheta=\theta-2p$, etc.
Acting on $|r,p;\theta\rangle_{\mathrm{irr}}$ with the~$\cQ_n$ modes,
we similarly obtain twisted highest-weight states with the twists
$\theta+p$, $\theta+2p$,~$\dots$\,.  We label these \textit{extremal
  states} as $|r,p;\theta|\iota\rangle$, where the number
$\iota\in\oZ$ determines the twist via $\vartheta=\theta+\iota p$.
Explicitly, the charge--level coordinates of
$|r,p;\theta|\iota\rangle$ are read off from
\begin{align}
  \cL_0|r,p;\theta|\iota\rangle=&\left(\theta\frac{r-1}{p}+
    \frac12\frac{p-2}{p}(\theta^2-\theta)-\iota\frac{p}{2}+
    \iota^2\frac{p(p-2)}{2}+\iota r+\iota(p-2)\theta\right)
  |r,p;\theta|\iota\rangle,
  \label{2.16}
  \\
  \cH_0|r,p;\theta|\iota\rangle=&\left(-\frac{r-1}{p}-
    \frac{p-2}{p}\theta-(p-2)\iota\right) |r,p;\theta|\iota\rangle.
  \label{2.17}
\end{align}
These states satisfy, in particular, the conditions
\begin{equation}
  \begin{aligned}
    \cG_n|r,p;\theta|\iota\rangle&=0\quad\text{for}\quad n\geq\iota
    p+\theta,
    \\
    \cG_{\iota p+\theta-r}\dots\cG_{\iota p+\theta-1}
    |r,p;\theta|\iota\rangle&=0,
    \\
    \cG_{\iota p+\theta-p+1}\dots\cG_{\iota p+\theta-r-1} \cG_{\iota
      p+\theta-r+1}\dots\cG_{\iota p+\theta-1}
    |r,p;\theta|\iota\rangle&=|r,p;\theta|\iota-1\rangle,
  \end{aligned}
  \label{2.18}
\end{equation} 
where $\iota\in\oZ$. In the charge--level coordinates on the plane, we
represent the states $\dots\cG_{-3}\cG_{-2}\cG_{-1}\*|r,p;0|0\rangle$
with arrows (see Fig.~\ref{fig:1}) or somewhat more schematically, by
replacing several consecutive arrows with sections of parabolas;
states~\eqref{2.18} are then the \textit{cusps} where the sections of
parabolas join (half of the cusps correspond to the mode $\cG_{\iota
  p+\theta-r}$ omitted in~\eqref{2.18} and the other half are the
states $|r,p;\theta|\iota\rangle$).

The idea behind the semi-infinite construction is to ``generate the
module from the state $|r,p;\infty\rangle$'' located infinitely far in
the bottom right in Fig.~\ref{fig:1}, using only the~$\cG_n$ modes;
this amounts to introducing semi-infinite forms in the fermionic
operators $(\cG_i)_{i\in\oZ}$.  We see in what follows that any
unitary representation~$\mK_{r,p;\theta}$ can be realized via this
semi-infinite construction.

\section{The semi-infinite space $\W_{r,p;\theta}$} \label{sec:3}
We fix an integer $p\geq3$ and also $r$ and $\theta$ such that $1\leq
r\leq p-1$ and $0\leq\theta\leq r-1$.  We let $\g(p)$ denote the
quotient of the Grassmann algebra of the operators~$\cG_n$, \ 
$n\in\oZ$, over the ideal~$\cS^{p}$ generated by
$(S^{p}_a)_{a\in\oZ}$, see Eq.~\eqref{1.1} (eventually, we
identify~$\cG_n$ with the corresponding $\N2$ generators).

\subsection{The inductive limit} \label{sec:3.1} The semi-infinite
space $\W_{r,p;\theta}$ is spanned by polynomials in $\cG_n\in\g(p)$,
\ $n\in\oZ$, acting on the states
\begin{equation}
  \bigl(|r,p;\theta|\iota\rangle_{\infty/2}\bigr)_{\iota\in\oZ},
  \label{3.1}
\end{equation} 
such that
\begin{align}
  \cG_n|r,p;\theta|\iota\rangle_{\infty/2}&=0,\qquad n\geq\iota
  p+\theta,
  \label{3.2}
  \\
  \cG_{\iota p+\theta-r}\dots\cG_{\iota p+\theta-1}
  |r,p;\theta|\iota\rangle_{\infty/2}&=0,
  \label{3.3}
  \\
  \cG_{\iota p+\theta-p+1}\dots\cG_{\iota p+\theta-r-1} \cG_{\iota
    p+\theta-r+1}\dots\cG_{\iota p+\theta-1}
  |r,p;\theta|\iota\rangle_{\infty/2}&=
  |r,p;\theta|\iota-1\rangle_{\infty/2}.
  \label{3.4}
\end{align}
Therefore, $\W_{r,p;\theta}$ is the linear span of states of the form
$\cG_{i_1}\cG_{i_{2}}\dots\cG_{i_m}\*
|r,p;\theta|\iota\rangle_{\infty/2}$ considered modulo the
ideal~$\cS^{p}$.  We call the elements of~$\W_{r,p;\theta}$ the
\textit{semi-infinite forms}; we also, somewhat loosely, use the term
``semi-infinite form'' for polynomials in~$\cG_n$ acting on
$|r,p;\theta|\iota\rangle_{\infty/2}$, rather than for their images in
the quotient.  We write~$\W_{r,p}$ instead of~$\W_{r,p;0}$. Obviously,
$\W_{r,p;\theta}$ is a module over $\g(p)$.

\begin{Rem}\label{rem:1} In less formal terms, the state
  $|r,p;\theta|\iota\rangle_{\infty/2}$ can be viewed as the
  semi-infinite product
  \begin{multline}
    \cG_{\iota p+\theta+1}\dots\cG_{\iota p+\theta+p-r-1} \cG_{\iota
      p+\theta+p-r+1}\dots\cG_{\iota p+\theta+p-1}\times
    \\
    \times \cG_{\iota p+\theta+p+1}\dots\cG_{\iota
      p+\theta+2p-r-1} \cG_{\iota p+\theta+2p-r+1}\dots\cG_{\iota
      p+\theta+2p-1}\dots,
    \label{3.5}
  \end{multline}
  where the associated sequence of occupation numbers is periodic with
  the period~$p$.  The elements of $\W_{r,p;\theta}$ are then those
  semi-infinite forms of $(\cG_i)_{i\in\oZ}$ (modulo~$\cS^{p}$)
  for which the sequence of occupation numbers $\alpha(i)$ becomes
  periodic with the period~$p$ for $i\gg1$.
\end{Rem}

We let $\W_{r,p;\theta}(\iota)$ denote the space generated by
$(\cG_n)\in\g(p)$, $n\in\oZ$, from the state
$|r,p;\theta|\iota\rangle$ satisfying the same annihilation conditions
as those in~\eqref{3.2} and~\eqref{3.3}. There are the mappings
\begin{equation}
  \dots\rightarrow\W_{r,p;\theta}(-2)\rightarrow
  \W_{r,p;\theta}(-1)\rightarrow \W_{r,p;\theta}(0)\rightarrow
  \W_{r,p;\theta}(1)\rightarrow\cdots,
  \label{3.6}
\end{equation} 
induced by mapping the vacuum vectors as
\begin{equation}
  |r,p;\theta|\iota-1\rangle\mapsto\cG_{p\iota+\theta-p+2}\dots
  \cG_{p\iota+\theta-r-1}\cG_{p\iota+\theta-r+1}\dots
  \cG_{p\iota+\theta-1}|r,p;\theta|\iota\rangle,\qquad\iota\in\oZ.
  \label{3.7}
\end{equation} 
\textit{Mappings~\eqref{3.6} commute with the action of~$\cG_n$}.  The
mappings $\W_{r,p;\theta}(\iota)\rightarrow \W_{r,p;\theta}$ induced
by
$|r,p;\theta|\iota\rangle\mapsto|r,p;\theta|\iota\rangle_{\infty/2}$
make all the diagrams
\begin{equation}\label{3.8}
  \begin{array}{rcccl}
    \W_{r,p;\theta}(\iota)\!\!&&\longrightarrow&&
    \!\!\W_{r,p;\theta}(\iota+1)\!\!\\
    &\searrow&&\swarrow& \\
    &&\!\!\!\W_{r,p;\theta}\!\!\!&&
  \end{array}
\end{equation} 
commutative.  Therefore, the space \textit{$\W_{r,p;\theta}$ is the
  inductive limit of~\eqref{3.6}},
$\W_{r,p;\theta}=\varinjlim_{\iota\to\infty} \W_{r,p;\theta}(\iota)$.

In addition to the spectral flow action $\mathsf{U}_{p}$ given by
$\mathsf{U}_{p}\cG_n\mathsf{U}_{-p}=\cG_{n+p}$, we define
\begin{equation}
  \mathsf{U}_{\pm p}|r,p;\theta|\iota\rangle_{\infty/2}=
  |r,p;\theta|\iota\pm1\rangle_{\infty/2}.
  \label{3.9}
\end{equation} 
This makes the semi-infinite space into a module over $\G^p(p)$, the
semi-direct product of $\g(p)$ and $(\mathsf{U}_{pn})_{n\in\oZ}$.

The formal definition given above does not tell us anything about the
structure of the semi-infinite space, in particular, about whether
mappings~\eqref{3.6} are embeddings (and hence, whether the spaces
$\W_{r,p;\theta}(\iota)$ give any ``approximation'' to the limit).
For this to be the case, the vanishings occurring among the
polynomials in $\cG_n$ under mappings~\eqref{3.7} must precisely agree
with the vanishings due to taking the quotient over the ideal
generated by $(S^{p}_a)_{a\in\oZ}$.  The difficulty in showing
this is that the ideal is not described explicitly.  For example,
Eqs.~\eqref{2.15} imply additional relations
$\cG_{-5}\cG_{-3}\cG_{-2}=0$ and $\cG_{-6}\cG_{-3}\cG_{-2}=0$;
continuing the list in~\eqref{2.15}, we obtain more elements in the
ideal.  The situation becomes more complicated in the general
case~\eqref{1.1}.  Nevertheless, we show in what follows that the
mappings involved in the inductive limit \textit{are} embeddings.
This implies that the results established for the individual spaces
$\W_{r,p;\theta}(\iota)$ give an ``approximation'' to those for the
entire semi-infinite space~$\W_{r,p;\theta}$.

\begin{Thm}\label{thm:3} All the mappings in~\eqref{3.8} are
  embeddings, i.e., the space~$\W_{r,p;\theta}$ admits the filtration
  \begin{align}
    \cdots\subset\W_{r,p;\theta}(-2)\subset
    \W_{r,p;\theta}(-1)\subset\W_{r,p;\theta}(0)\subset
    \W_{r,p;\theta}(1)\subset\cdots\,.
    \label{3.10}
  \end{align}
\end{Thm}
This follows from the existence of a remarkable monomial basis in
$\W_{r,p;\theta}(\iota)$ that agrees with the mappings
$\W_{r,p;\theta}(\iota)\to \W_{r,p;\theta}(\iota+1)$, as we discuss
momentarily.

\subsection{The thin basis }\label{sec:3.2} 
In each space~$\W_{r,p;\theta}(\iota)$, we construct a monomial basis,
which we call the \textit{thin basis} because it consists of
semi-infinite forms where the modes~$\cG_n$ are ``thinned out'' as
explained in Theorem~\ref{thm:2}.  We reformulate the desired result
as follows (Theorem~\ref{thm:2} is reproduced by writing
$|r,p;\theta|\iota\rangle_{\infty/2}$ as the semi-infinite
product~\eqref{3.5}).

\begin{Lemma}\label{lemma:1} The set of states $\cG_{i_m}\dots\cG_{i_{2}}\cG_{i_1} |r,p;\theta|\iota\rangle_{\infty/2}$, for which
  $\iota p+\theta>i_1>i_{2}>\dots>i_m$, \ $i_{a}-i_{a+p-2}\geq p$ for
  any~$a$, and $\iota p+\theta-i_r>r$, constitute a basis in the space
  $\W_{r,p;\theta}(\iota)$.
\end{Lemma}

We refer to the thin basis elements as \textit{thin monomials}.
Theorem~\ref{thm:3} now follows from the observation that in terms of
thin monomials, the mappings
$\W_{r,p;\theta}(\iota)\to\W_{r,p;\theta}(\iota+1)$ are implemented~by
\begin{multline}
  \cG_{i_m}\dots\cG_{i_{2}}\cG_{i_1}
  |r,p;\theta|\iota\rangle_{\infty/2}\mapsto
  \cG_{i_m}\dots\cG_{i_{2}}\cG_{i_1}\times
  \\
  \times \cG_{\iota p+\theta+1}\dots\cG_{\iota
    p+\theta+p-r-1} \cG_{\iota p+\theta+p-r+1}\dots\cG_{\iota
    p+\theta+p-1} |r,p;\theta|\iota+1\rangle_{\infty/2}.\quad
  \label{3.11}
\end{multline} 
Indeed, once the indices $i_m,\ldots,i_1$ satisfy the conditions of
Lemma~\ref{lemma:1}, the indices $i_m,\dots,i_1$, \ $\iota
p+\theta+1$, \ $\dots$, \ $\iota p+\theta+p-r-1$, \ $\iota
p+\theta+p-r+1$, \ $\dots$, \ $\iota p+\theta+p-1$ also satisfy these
conditions.

\medskip

\begin{prf}{~of Lemma~\ref{lemma:1}} \ consists of two parts, the
  first of which is simple, but the second requires certain effort.
  We briefly describe the first part and then focus on the second.
  
  The first part of the proof amounts to asserting that any
  semi-infinite form can be rewritten as a linear combination of the
  states $\cG_{i_m}\cG_{i_{m-1}}\dots\cG_{i_1}\*
  |r,p;\theta|\iota\rangle_{\infty/2}$ satisfying the conditions of
  the Lemma. This follows from the fact that if any monomial
  in~$\cG_n$ contains a combination of modes violating the conditions
  of the lemma, then in accordance with the relations~$S^{p}_a=0$,
  this monomial can be expressed through a linear combination of other
  monomials \textit{each of which is lexicographically smaller than
    the original one}.  The new monomials, obviously, can also involve
  combinations of modes violating the conditions of the lemma, however
  the argument regarding the lexicographic ordering allows developing
  an iteration procedure.  It converges after a finite number of steps
  (and thus gives a linear combination of monomials satisfying the
  conditions of the Lemma) because
  \begin{enumerate}
  \item the space~$\overline{\W}_{r,p;\theta}(\iota)$ is bigraded via
    $\cG_{i_m}\cG_{i_{m-1}}\dots\cG_{i_1}\*
    |r,p;\theta|\iota\rangle_{\infty/2}\mapsto (m,i_1+i_2+\dots+i_m)$;
    
  \item each graded component is a finite-dimensional space;
    
  \item the procedure of expressing a monomial through a linear
    combination of others using the relations~$S^{p}_a=0$ preserves
    the bigrading.
  \end{enumerate}
  
  In the second part of the proof, it remains to show that the thin
  monomials are linearly independent.  This follows by comparing the
  characters.  The calculation of characters is rather involved, and
  we give it in Sec.~\ref{sec:4}.  The proof is completed by the
  statement of Lemma~\ref{lemma:3} in~Sec.~\ref{sec:4.3}.
\end{prf}

\subsubsection*{Examples} We give several examples of the
transformation to the thin basis.

\subsubsection*{A. $p=3$, \ $r=1$ }Directly eliminating
dense combinations of modes leads to
$\cG_{-15}\cG_{-14}\cG_{-12}\cG_{-13}\cG_{-4}\*
\cG_{-3}|1,3;0|0\rangle=3393\cG_{-21}|1,3;0|{-5}\rangle+
4185\cG_{-20}\cG_{-15}\*|1,3;0|{-4}\rangle+
5697{\cG_{-19}\cG_{-16}|1,3;0|{-4}\rangle}+
1755\*\cG_{-19}\cG_{-15}\cG_{-12}|1,3;0|{-3}\rangle+
315\cG_{-18}\cG_{-15}\cG_{-12}\cG_{-9}|1,3;0|{-2}\rangle$.

\subsubsection*{B. $p=4$, \ $r=1$ }\ As a simple example, we
have $\cG_{-5}\cG_{-4}\cG_{-2}|1,4;0|0\rangle=
-2\cG_{-6}|1,4;0|{-1}\rangle$. Similarly,
$\cG_{-5}\cG_{-4}\cG_{-3}|1,4;0|0\rangle=
-10\cG_{-7}\*|1,4;0|{-1}\rangle-
8{\cG_{-6}\*\cG_{-4}\*\cG_{-2}\*|1,4;0|0\rangle}$.

\subsubsection*{C. $p=5$, \ $r=1$ }Of the following two
examples, the second is an extension of the first:
$\cG_{-2}\cG_{-1}\cG_{0}\cG_{1}\*|1,5;-1|1\rangle=
-35{\cG_{-5}|1,5;-1|0\rangle}-
45{\cG_{-4}\cG_{-1}\cG_{1}\cG_{2}|1,5;-1|1\rangle}-
20{\cG_{-3}\cG_{-2}\cG_{1}\cG_{2}|1,5;-1|1\rangle}-
15\cG_{-3}\cG_{-1}\*\cG_{0}\cG_{2}|1,5;-1|1\rangle$ and
$\cG_{-5}\cG_{-4}\cG_{-2}\cG_{-1}\cG_{0}\*\cG_{1}\* |1,5;-1|1\rangle=
-45\cG_{-7}\*\cG_{-4}\*\cG_{-3}\*|1,5;-1|0\rangle-
({245}/{3})\cG_{-6}\cG_{-5}\*\cG_{-3}\*|1,5;-1|0\rangle-
75\cG_{-6}\*\cG_{-4}\*\cG_{-3}\*\cG_{-1}\*\cG_{1}\*
\cG_{2}\*|1,5;-1|1\rangle$.

\section{Functional realization and characters} \label{sec:4}
We use the realization of the (graded-)dual space to the semi-infinite
space in terms of polynomial differential forms.  This
\textit{functional realization} is a powerful tool in studying
properties of semi-infinite spaces.

\subsection{The functional realization of $\W_{r,p}(0)^*$}
\label{sec:4.1} We first note that for any space 
generated from a vacuum $|\,\rangle$ by fermionic generators
$\cG_{\leq-1}$, the graded-dual space can be identified with
differential forms in some variables $x_1,x_2,\dots$ as follows.  In
each graded component with a given charge (the number
of~$\cG_{\bullet}$) $n\geq1$, we arrange all the states into a
generating function $\cG(x_1)\dots\cG(x_n)|\,\rangle$, where
\begin{equation}
  \cG(x)=\sum_{m\leq-1}\cG_m x^{m-1}.
\end{equation} 
The states $\cG_{i_1}\dots\cG_{i_n}|\,\rangle$ are reproduced by
taking the integrals
\begin{equation}
  \oint dx_1\,x_1^{i_1}\ldots\oint dx_n\,x_n^{i_n}
  \cG(x_1)\dots\cG(x_n)|\,\rangle.
\end{equation} 
Any functional $\langle\ell|$ on the charge-$n$ subspace is determined
by all of its values,\footnote{Whether $\cG(x)$ is viewed as a
  1-differential or, for example, a 2-differential, is a matter of
  convention; however, this convention must agree with the choice of
  the functional spaces involved in the functional realization.  With
  the choice made in the text, all functional spaces are polynomials
  $\oC[x_1,\dots,x_n]$, rather than $(x_1\dots
  x_n)^{\nu}\oC[x_1,\dots,x_n]$ with some positive or
  negative~$\nu$.}
\begin{equation}
  \langle\ell|\cG(x_1)\dots\cG(x_n)|\,\rangle=
  f_{\ell}(x_1,\dots,x_n)\,dx_1\dots dx_n,
  \label{4.1}
\end{equation} 
where $f_{\ell}$ is an antisymmetric polynomial in $x_1,\dots,x_n$
(and the product of the differentials is symmetric).  Therefore, the
space dual to $W(0)$, the space \textit{freely} generated by the modes
$\cG_{n\leq-1}$, can be identified with polynomial differential forms
in $x_1,x_2,\dots$,
\begin{equation}
  W(0)^*=\oC\oplus\oC[x]\,dx\oplus\oC\langle x_1,x_2\rangle\,
  dx_1\,dx_1\oplus\oC\langle x_1,x_2,x_3\rangle\,dx_1\,dx_2\,dx_3
  \oplus\cdots,
  \label{4.2}
\end{equation} 
where $\oC\langle x_1,\dots,x_n\rangle$ are antisymmetric
polynomials.

For quotient spaces, the functional realization of the corresponding
dual space is given by a \textit{subspace} in the space of
polynomials.  Let $\W(0)_n$ denote the charge~$n$ subspace and
$\W(0)^*_n$ its dual, i.e., the subspace of polynomials in~$n$
variables in~\eqref{4.2}.  In the dual language, taking the quotient
with respect to the ideal generated by~$S^{p}_a$ (see~\eqref{1.1})
corresponds to restricting to those antisymmetric polynomials
$f(x_1,\dots,x_n)$ for which
\begin{equation}
  \frac{d^{p-2}}{\partial x_{i_1}^{p-2}}
  \frac{\partial^{p-3}}{\partial x_{i_2}^{p-3}}\dots
  \frac{\partial}{\partial x_{i_{p-2}}}f(x_1,x_2,\dots)
  \biggr|_{x_{i_1}=x_{i_2}=\dots=x_{i_{p-2}}=x_{i_{p-1}}}=0.
  \label{4.3}
\end{equation} 
Another condition is read off from Eq.~\eqref{3.3} (for
$\theta=\iota=0$) as
\begin{equation}
  \frac{\partial^{r-1}}{\partial x_{i_1}^{r-1}}
  \frac{\partial^{r-2}}{\partial x_{i_2}^{r-2}}\dots
  \frac{\partial}{\partial x_{i_{r-1}}}f(x_1,x_2,\dots)
  \biggr|_{x_{i_1}=x_{i_2}=\dots= x_{i_{r-1}}=x_{i_{r}}=0}=0.
  \label{4.4}
\end{equation} 
Let $\W_{r,p}(0)^*$ denote the space of antisymmetric polynomials $f$
satisfying these conditions.

Because any antisymmetric polynomial can be represented as
\begin{equation}
  f(x_1,x_2,\dots)=\Delta(x_1,x_2,\dots)\phi(x_1,x_2,\dots),\qquad
  \Delta(x_1,x_2,\dots)=\prod_{1\leq i<j}(x_i-x_j),
  \label{4.5}
\end{equation}
with a \textit{symmetric} polynomial~$\phi(x_1,x_2,\dots)$,
conditions~\eqref{4.4} and~\eqref{4.3} become the following
restrictions on symmetric polynomials:

\begin{enumerate}
  \def\theenumi{\textbf{P}$\boldsymbol{\arabic{enumi}}$}
\item \label{P-cond-r}
  $\phi(\underbrace{0,\dots,0\!}_{r},x_{r+1},\dots,x_n)=0$;
  
\item \label{P-cond-p}
  $\phi(\underbrace{x,x,\dots,x\!}_{p-1},x_{p},x_{p+1},\dots,x_n)=0$.
\end{enumerate}

\subsubsection*{Example} For $p=3$, condition~\ref{P-cond-p} can be
easily solved as
\begin{equation}
  f(x_1,\dots,x_n)\,dx_1\dots dx_n=
  \prod_{i<j}(x_i-x_j)^3\varphi(x_1,\dots,x_n)\,dx_1\dots dx_n,
  \label{4.6}
\end{equation} 
where $\varphi$ is a symmetric polynomial.  For $r=1$,
condition~\ref{P-cond-r} is immediately
solved by $\varphi(x_1,\dots,x_n)=x_1\dots
x_n\bar\varphi(x_1,x_2,\dots,x_n)$, whence it follows that the space
$\W_{1,3}(0)_n^*$ consists of the differential forms
\begin{align}
  \prod_{i<j}(x_i-x_j)^3x_1\dots x_n
  \bar\varphi(x_1,x_2,\dots,x_n)\,dx_1\dots dx_n,
  \label{4.7}
\end{align} 
where $\bar\varphi$ are arbitrary symmetric polynomials.  Such an
explicit description, however, is not available for~$p\geq4$.

\subsection{Characters of $\W_{r,p;\theta}(\iota)$}\label{sec:4.2}
The functional realization allows us to calculate the characters of
$\W_{r,p;\theta}(\iota)$.  The idea of~\cite{ref:1} consists in
finding, for each $n\geq1$, the character of the space
$\W_{r,p}(0)^*_n=\W_{r,p}(0)^*\cap \oC\langle
x_1,\dots,x_n\rangle$, which coincides with the character of
$\W_{r,p}(0)_n$ (the subspace in $\W_{r,p}(0)$ generated by~$n$
modes~$\cG_i$); then
\begin{equation}
  \chr\W_{r,p}(0)(z,q)=\sum_{n\geq0}z^n\chr\W_{r,p}(0)_n(q).
\end{equation} 
The dependence on both~$\iota$ and~$\theta$ can be reconstructed in
accordance with the spectral flow.

\subsubsection{Example: $p=3$ }\label{sec:4.2.1}The functional model
of $\W_{1,3}(0)_n^*$ is given by~\eqref{4.7}, and the space of
symmetric polynomials in $n$ variables is algebraically generated by
the elementary symmetric polynomials; the contribution of these
polynomials to the character of~$\W_{1,3}(0)_n^*$ is therefore
$1/(q)_n$.  Next, the product $\prod_{i<j}(x_i-x_j)^3$ contributes
$q^{3n(n-1)/2}$ to the character, the differentials $dx_1\dots dx_n$
contribute~$q^n$, and another~$q^n$ comes from~$x_1\dots x_n$.
Therefore,
\begin{equation}
  \chr\W_{1,3}(0)=\chr\W_{1,3}(0)^*=
  \sum_{n\geq0}\frac{z^nq^{\frac{3n^2+n}{2}}}{(q)_n}.
  \label{4.8}
\end{equation} 
Applying the spectral flow transform in accordance with~\eqref{2.4},
we find
\begin{equation}
  \begin{split}
    \chr\W_{1,3}(\iota)(z,q)&= z^{-\iota}q^{\frac{\iota}{2}(3\iota-1)}
    \chr\W_{1,3}(0)(zq^{-3\iota},q)=
    \\
    &=\sum_{n\geq0}\frac{z^{n-\iota}q^{\frac{3(n-\iota)^2+n-\iota}{2}}}{(q)_n}=
    \sum_{n\geq-\iota}\frac{z^{n}q^{\frac{3n^2+n}{2}}}{(q)_{n+\iota}}.
  \end{split}
  \label{4.9}
\end{equation} 
This expression admits the limit
\begin{equation}
  \lim_{\iota\to\infty}\chr\W_{1,3}(\iota)(z,q)=
  \sum_{n=-\infty}^{+\infty}\frac{z^{n}q^{\frac{3n^2+n}{2}}}
  {\prod_{m\geq1}(1-q^m)}=q^{-\frac{1}{3}}
  \frac{\vartheta_{1,0}(zq^{-1},{q^3})}{\eta(q)},
  \label{4.10}
\end{equation}
which coincides with~\eqref{2.8}.  Much more work is needed to show
this remarkable coincidence in the general case.

\subsubsection{The general case: $p\geq4$} \label{sec:4.2.2}
\begin{Lemma}\label{lemma:2} For $p\geq3$ and $1\leq r\leq p-1$, the
  character of the space $\W_{r,p}(0)$ is given by
  \begin{equation}
    \chr\W_{r,p}(0)(z,q)=\sum_{n\geq0}
    \sum_{\substack{N_1\geq\dots\geq N_{p-2}\geq0 \\ N_1+\dots+N_{p-2}=n}}
    \frac{z^{n-\frac{r-1}{p}}q^{\sum_{m=r}^{p-2}N_m}q^{\frac12 n(n-1)}
      q^{\sum_{m=1}^{p-2}N_m^2}}{(q)_{N_1-N_2}(q)_{N_2-N_3}\dots
      (q)_{N_{p-3}-N_{p-2}}(q)_{N_{p-2}}}.
    \label{4.11}
  \end{equation}
\end{Lemma}

\begin{prf}. In the functional realization of $\W_{r,p}(0)^*_n$, we
  consider all the partitions of the set $\{x_1,\dots,x_n\}$ and
  introduce the lexicographic ordering on the partition (i.e., we set
  $(r_1,r_2,\dots,r_k)\prec(r'_1,r'_2,\dots,r'_{k'})$ if $r_i<r'_i$
  for the first pair $(r_i,r'_i)$ such that $r_i\neq r'_i$).
  
  We fix a partition of~$n$ written as $\{1,\dots,n\}=
  M_1\cup\dots\cup M_{\ell}$ and let $|M_{\alpha}|=r_{\alpha}$.  We
  say that a polynomial $\varphi(x_1,\dots,x_n)$ vanishes on this
  partition if it vanishes whenever all the variables~$x_{i_{\alpha}}$
  for $i_{\alpha}\in M_{\alpha}$ take the same value,
  $x_{i_{\alpha}}=a_{\alpha}$ for all $\alpha=1,\dots,\ell$. We write
  this as $\varphi(a_1;\dots;a_{\ell})=0$.
  
  The character of~$\W_{r,p}(0)^*$ can be written as
  \begin{equation}
    \chr\W_{r,p}(0)(z,q)=z^{-\frac{r-1}{p}}
    \sum_{n\geq0}z^nq^{\frac{n^2-n}{2}}Z^n_{r,p}(q),
    \label{4.12}
  \end{equation} 
  where the factor $q^{(n^2-n)/{2}}$ corresponds to
  $\Delta(x_1,\dots,x_n)$ in~\eqref{4.5} and $Z^n_{r,p}(q)$ is the
  partition function of symmetric polynomials in~$n$ variables that
  vanish on the partition $\hat p=(p-1,1,1,\dots,1)$.
  
  For a partition $P$, we consider the set of symmetric polynomials
  that vanish on every partition~$P'\succ P$.  The lexicographic order
  on partitions then induces a filtration on symmetric polynomials
  in~$n$ variables.  If $Z^{n,P}_{r,p}(q)$ is the partition function
  of the \textit{associated graded factor}~$\Gr_P$, we have
  \begin{equation}
    Z^n_{r,p}(q)=\sum_{\{P\mid P\prec\hat p\}}Z^{n,P}_{r,p}(q).
    \label{4.13}
  \end{equation} 
  To find $Z^{n,P}_{r,p}(q)$, we first consider the case where
  $r=p-1$; condition~\ref{P-cond-r} is then a consequence
  of~\ref{P-cond-p}.  For a partition~$P$ with the parts $M_1$, \dots,
  $ M_\ell$ (and with $|M_\alpha|=r_\alpha$), the graded
  factor~$\Gr_P$ is spanned by polynomials
  $\varphi(a)\equiv\varphi(a_1;\dots;a_\ell)$ satisfying the
  conditions
  \begin{description}
  \item[(mult)] $\varphi(a)=0$ if $a_\alpha=a_\beta$, with the
    multiplicity of the zero equal to $2\min(r_\alpha,r_\beta)$,
    
  \item[(sym)] $\varphi(\dots;a_\alpha;\dots;a_\beta;\dots)=
    \varphi(\dots;a_\beta;\dots;a_\alpha;\dots)$ whenever
    $r_\alpha=r_\beta$.
  \end{description} 
  
  Recalling the differentials from~\eqref{4.1}, we can therefore see
  that the partition function $Z^{n,P}_{p-1,p}(q)$ of the graded
  factor coincides with the partition function of the space spanned by
  \begin{equation}
    \prod_{1\leq\alpha\leq\beta\leq\ell}
    (a_{\alpha}-a_{\beta})^{2r_{\beta}}
    \cdot\bar\varphi(a)\cdot\prod_{\alpha=1}^{\ell}
    (da_{\alpha})^{r_{\alpha}},
    \label{4.14}
  \end{equation} 
  where a polynomial $\bar\varphi$ satisfies symmetry
  requirement~\textbf{(sym)}.  Let $\nu_m$ be the number of parts
  $M_\alpha$ such that $r_\alpha=m$.  The contribution of the chosen
  partition $P$ to the partition function is then
  \begin{equation}
    \frac{q^{\sum_{\alpha<\beta}2r_{\beta}+\sum_{\alpha}r_{\alpha}}}
    {\prod_{j\geq1}(q)_{\nu_j}}=
    \frac{q^{\sum_{\beta=1}^{\ell}(2\beta-1)r_{\beta}}}
    {\prod_{j\geq1}(q)_{\nu_j}}=
    \frac{q^{\sum_{m=1}^{p-2}N_m^2}}{(q)_{N_1-N_2}(q)_{N_2-N_3}\dots},
    \label{4.15}
  \end{equation} 
  where $N_m=\nu_m+\nu_{m+1}+\dots+\nu_{\ell}$ are the elements of the
  partition $P$ transposed.  Therefore,
  \begin{equation}
    Z^{n}_{p-1,p}(q)=
    \sum_{\substack{N_1\geq\ldots\geq N_{p-2}\geq0 \\ N_1+\ldots+N_{p-2}=n}}
    \frac{q^{\sum_{m=1}^{p-2}N_m^2}}
    {(q)_{N_1-N_2}(q)_{N_2-N_3}\dots(q)_{N_{p-2}}}.
    \label{4.16}
  \end{equation}
  
  If $r\neq p-1$, we must additionally account for
  condition~\ref{P-cond-r}, which amounts to taking symmetric
  polynomials~${\bar\varphi}(a)$ satisfying
  ${\bar\varphi}(a)\bigr|_{a_1=a_2=\dots=a_r=0}=0$.  For this, we
  recall the filtration
  $\cP^n_{1}\subset\dots\subset\cP^n_{r-1}\subset\cP^n_{r}$ on the
  space~$\cP^n_{r}$ of symmetric polynomials in~$n$ variables
  satisfying condition~\ref{P-cond-r},
  \begin{equation}
    \cP^n_{r}=\sigma_n\oC[\sigma_1,\dots,\sigma_n]+
    \sigma_{n-1}\oC[\sigma_1,\dots,\sigma_{n-1}]+\dots+
    \sigma_{n-r+1}\oC[\sigma_1,\dots,\sigma_{n-r+1}],
    \label{4.17}
  \end{equation} 
  where $\sigma_{1}=x_1+\dots+x_n$, \ $\dots$, \ $\sigma_{n}=x_1\dots
  x_n$ are the elementary symmetric polynomials in~$n$ variables.  In
  the corresponding graded factor, the explicit $\sigma_i$ factors
  result in additionally multiplying partition function~\eqref{4.15}
  by $q^{\sum_{m=r}^{p-2}N_m}$.  Thus,
  \begin{equation}
    Z^n_{r,p}(q)=
    \sum_{\substack{N_1\geq\dots\geq N_{p-2}\geq0 \\ N_1+\dots+N_{p-2}=n}}
    \frac{q^{\sum_{m=r}^{p-2}N_m}q^{\sum_{m=1}^{p-2}N_m^2}}
    {(q)_{N_1-N_2}(q)_{N_2-N_3}\dots(q)_{N_{p-3}-N_{p-2}}(q)_{N_{p-2}}}.
    \label{4.18}
  \end{equation} 
  Inserting~\eqref{4.18} in~\eqref{4.12}, we obtain~\eqref{4.11}.
\end{prf}

Using relation~\eqref{2.4} with $\theta=p\iota$, we find the
characters of the spectral-flow transformed spaces as
\begin{multline}
  \label{4.19}
  \chr\W_{r,p}(\Nn)(z,q)={}z^{-(p-2)\Nn}\,q^{\frac{p-2}{2}(p\Nn^2-\Nn)}
  \chr\W_{r,p}(0)(z\,q^{-p\Nn},q)\\
  ={}q^{\frac{p-2}{2}(p\Nn^2-\Nn)+(r-1)\Nn}\!\!
  \sum_{n\geq-(p-2)\Nn}\sum_{\substack{N_1\geq\dots\geq N_{p-2}\geq0
      \\
      N_1+\dots+N_{p-2}=n + (p-2)\Nn}}\!\!\!\!
  q^{-p\Nn[n+(p-2)\Nn]}\,z^{n-\frac{r-1}{p}}\,\times\\
  \kern.15\textwidth{}\times q^{\half\sum_{m=r}^{p-2}N_m -
    \half\sum_{m=1}^{r-1}N_m} \frac{\displaystyle q^{\half (n^2 +
      2n(p-2)\Nn + (p-2)^2\Nn^2)}\, q^{\sum_{m=1}^{p-2} N_m^2}}{
    \displaystyle (q)_{N_1-N_2}\,(q)_{N_2-N_3}\dots
    (q)_{N_{p-3}-N_{p-2}}\,(q)_{N_{p-2}}}\\[8pt]
  ={}z^{\frac{1-r}{p}} \sum_{n\geq-(p-2)\Nn}
  \sum_{\substack{N_1\geq\dots\geq N_{p-2}\geq-\Nn\\
      N_1+\dots+N_{p-2}=n}} \frac{z^n\, q^{\half n^2}\,
    q^{\sum_{m=1}^{p-2} N_m^2+\sum_{m=r}^{p-2} N_m +
      \half\sum_{m=r}^{p-2}N_m - \half\sum_{m=1}^{r-1}N_m}}
  {\displaystyle (q)_{N_1-N_2}\,(q)_{N_2-N_3}\dots
    (q)_{N_{p-3}-N_{p-2}}\,(q)_{N_{p-2}+\Nn}}.
\end{multline}

\subsection{Linear independence of thin monomials} \label{sec:4.3} We
now show that the thin basis defined in Sec.~\ref{sec:3.2} is indeed a
basis in $\W_{r,p}(0 )$ (and hence, by the spectral flow, also a basis
in $\W_{r,p;\theta}(\iota)$); in other words, we show that thin
monomials defined in Lemma~\ref{lemma:1} are linearly independent in
$\W_{r,p}(0)$.  We actually prove that the partition function of thin
monomials coincides with~\eqref{4.11}. For this, we represent each
semi-infinite form in terms of the occupation numbers and hence as
configurations of crosses on the semi-infinite one-dimensional lattice
(see~\eqref{1.6}).  We are interested in the partition function of the
configurations satisfying the basic condition that any $p$ consecutive
lattice sites carry at most~$p-2$ crosses.

We first consider the same restriction on configurations on a lattice
with a finite number of sites~$L$, with the sites labeled by $1\leq
i\leq L$.  Each configuration of crosses at the sites
$(i_1,\dots,i_m)$ contributes $z^m q^{i_1+\dots+i_m}$ to the partition
function.  Let $\omega_{L, r, p}(z, q)$ be the partition function of
all the configurations of crosses on $L$ sites satisfying the
conditions that
\begin{enumerate}\def\theenumi{\textbf{C}$\boldsymbol{\arabic{enumi}}$}
\item \label{C-cond-r} there are no more than $r-1$ crosses at the
  sites $1$, \dots, $r$;
  
\item \label{C-cond-p} on any $p$ consecutive sites, there are no more
  than $p-2$ crosses.
\end{enumerate}
For $r=p-1$, the first condition follows from the second.  The
configurations of crosses encoding the thin basis elements are then
recovered in the $L\to\infty$ limit.  Remarkably, we have the
following lemma. 

\begin{Lemma}\label{lemma:3} The partition function of the
  configurations of crosses satisfying conditions~\ref{C-cond-r}
  and~\ref{C-cond-p} on a semi-infinite lattice is equal to the
  character $\chr\W_{r,p}(0)(z,q)$ defined in Lemma~\ref{lemma:2},
  \begin{equation}
    z^{-\frac{r-1}{p}}\lim_{L\to\infty}\omega_{L,r,p}(z,q)=
    \chr\W_{r,p}(0)(z,q).
    \label{4.20}
  \end{equation}
\end{Lemma}

\begin{prf}. The idea of the proof is as follows.  Expanding both
  sides of~\eqref{4.20} in powers of~$z$ gives two systems of
  functions, each of which is completely determined by a set of
  recursive relations and the appropriate ``initial values.''  These
  two sets of recursive relations and the initial values are
  identical, which implies~\eqref{4.20}.  We now proceed with the
  details.
  
  The partition function of the configurations of crosses satisfies
  the recursive relations
  \begin{equation}
    \omega_{L,r,p}(z,q)=\sum_{j=0}^{r-1}z^jq^{\frac{j(j+1)}{2}}
    \omega_{L-j-1,p-j-1,p}(zq^{j+1},q).
    \label{4.21}
  \end{equation} 
  Indeed, condition~\ref{C-cond-r} selects those configurations that
  are the disjoint union $\bigcup_{j=0}^{r-1}$ of configurations with
  \textit{exactly} $j$ occupied sites (and the site $j+1$ free).  For
  each $j$, cutting off these occupied sites and the adjacent free
  site leaves configurations of crosses on $L-j-1$ sites; these
  configurations satisfy a ``boundary condition'' that keeps track of
  the crosses at the sites $1,\dots,j$ on the original lattice:
  by~\ref{C-cond-p}, there can be no more than $p-j-2$ crosses in the
  beginning of the sublattice.  The overall factor $z^j q^{j(j+1)/2}$
  in~\eqref{4.21} is precisely the contribution of the crosses
  at~$1,\dots,j$, and the ``spectral flow'' transformation $z\mapsto z
  q^{j+1}$ in the argument accounts for relabeling the sites on the
  sublattice.  This shows~\eqref{4.21}.
  
  The initial conditions for the recursive are set on lattices with
  $L\leq p-2$, where condition~\ref{C-cond-p} does not apply, and
  therefore,
  \begin{equation}
    \omega_{L,r,p}(z,q)=\chi_{L-1,p}(zq,q)+
    \sum_{\substack{j=1\\L-j-1\geq0}}^{r-1}
    z^jq^{\frac{j(j+1)}{2}}\chi_{L-j-1,p}(zq^{j+1},q),
    \label{4.22}
  \end{equation} 
  where $L\leq p-2$ and
  \begin{equation}
    \chi_{\ell,p}(z,q)=\sum_{i=0}^{\ell}z^i
    q^{\frac{i(i+1)}{2}}\bmatrix{\ell}\\{i}\endbmatrix_{q}=(-qz)_{\ell}
    \qquad\text{for}\qquad0<\ell\leq p-2
    \label{4.23}
  \end{equation} 
  (and $\chi_{0,p}(z,q)=1$).  With these initial conditions, recursive
  relations~\eqref{4.21} completely determine~$\omega_{L,r,p}(z,q)$.
  
  In the limit as $L\to\infty$, we expand the partition function in
  powers of $z$ as
  \begin{equation}
    \lim_{L\to\infty}\omega_{L,r,p}(z,q)=\sum_{n\geq0}z^n
    q^{\frac{n(n-1)}{2}}B^n_{r,p}(q).
    \label{4.24}
  \end{equation} 
  It then follows from~\eqref{4.21} that
  \begin{equation}
    B^n_{r,p}(q)=\sum_{j=0}^{r-1}q^{n}B^{n-j}_{p-j-1,p}(q)
    \label{4.25}
  \end{equation} 
  (for $n<r-1$, the summation on the right-hand side goes from~0
  to~$n$; it is convenient to set $B^n_{r,p}(q)=0$ for $n<0$, which
  allows Eq.~\eqref{4.25} to be used in all cases).  These can be
  rewritten as recursive relations expressing $B^n_{r,p}(q)$ through
  $B^m_{r,p}(q)$ with $m<n$.  The initial values for these recursive
  relations are given by $B^m_{r,p}$ for $m<p-1$, where
  condition~\ref{C-cond-p} does not apply.  We already saw
  in~\eqref{4.22} and~\eqref{4.23} how to evaluate $\omega_{L, r,
    p}(z, q)$ on a lattice where condition~\ref{C-cond-p} is not
  imposed, and it only remains to take the $L\to\infty$ limit
  of~\eqref{4.22}. Let $\mathsf{P}$ denote the projector on the space
  of polynomials in~$z$ of the order $<p-1$.  We then have
  \begin{multline}
    \sum_{m=0}^{p-2}z^mq^{\frac{m(m-1)}{2}}B^m_{r,p}(q)=
    \mathsf{P}\sum_{j=0}^{r-1}z^jq^{\frac{j(j+1)}{2}}(-zq^{j+2})_{\infty}=
    \\
    = \mathsf{P}\sum_{j=0}^{r-1}z^jq^{\frac{j(j+1)}{2}}
    \biggl(1+\sum_{n\geq1}\frac{(zq^{j+2})^{n}q^{\frac{n(n-1)}{2}}}
    {(1-q)\dots(1-q^n)}\biggr)=
    \\
    = \mathsf{P}\sum_{n\geq0}z^nq^{\frac{n(n-1)}{2}}
    \sum_{\substack{j=0 \\ j\leq n}}^{r-1}
    \frac{q^{2n-j}}{(1-q)\dots(1-q^{n-j})}.
    \label{4.26}
  \end{multline} 
  The inner sum in the right-hand side therefore defines the initial
  values $B^n_{r,p}(q)$ for $n\leq p-2$.
  
  Turning to the character of $\W_{r,p}(0)(z,q)$, we recall
  expansion~\eqref{4.12}, where, as we have seen,
  \begin{equation}
    Z_{r,p}^n(q)=
    \sum_{\substack{n_1,\dots,n_{p-2}\geq0 \\ N_1+\dots+N_{p-2}=n}}
    \frac{q^{N_1^2+\dots+N_{p-2}^2+N_{r}+\dots+N_{p-2}}}
    {(q)_{n_1}\dots(q)_{n_{p-2}}}
    \label{4.27}
  \end{equation} 
  with $N_i=n_i+\dots+n_{p-2}$; these expressions satisfy the
  relation~\cite{ref:10}
  \begin{equation}
    Z^n_{r,p}(q)=\sum_{j=0}^{r-1}q^{n}Z^{n-j}_{p-j-1,p}(q).
    \label{4.28}
  \end{equation} 
  Indeed, using the notation in~\cite{ref:10}, Eq.~\eqref{4.27}
  becomes $Z^n_{i,p}(q)=R^n_{p-1,i}(q)$, where the generating
  functions $\sum_{n\geq0}x^n R^n_{\kappa,i}(q)=
  R_{\kappa,i}(x;q)=J_{\kappa,i}(0;x;q)$ are known to satisfy the
  identities
  \begin{align}
    &R_{\kappa,i}(x;q)-R_{\kappa,i-1}(x;q)=
    (xq)^{i-1}R_{\kappa,\kappa-i+1}(xq;q)\qquad\text{for}\qquad 1\leq
    i\leq\kappa,
    \\
    &R_{\kappa,0}(x;q)=0.
  \end{align}
  Summing these relations (with $\kappa=p-1$) over $i=1,\dots,r-1$, we
  obtain~\eqref{4.28}.
  
  To complete the proof, it remains to find the initial values for
  this recursive, namely, $Z^m_{r,p}(q)$ for $m<\kappa$.  We
  have~\cite{ref:10}
  \begin{equation}
    R_{\kappa,i}(x;q)=\sum_{n\geq0}
    \frac{x^{\kappa n}q^{\kappa n+\kappa n^2+n-in}(1-x^iq^{2ni+i})
      (-1)^nq^{\frac{n(n-1)}{2}}}{(q)_n(xq^{n+1})_{\infty}}.
    \label{4.29}
  \end{equation} 
  The coefficients entering the expansion in powers of $x^j$ for
  $j\leq\kappa-1$ follow only from the term with $n=0$ in this sum,
  and therefore,
  \begin{multline}
    \sum_{n=0}^{\kappa-1}x^nR^n_{\kappa,i}(q)=
    \mathsf{P}\frac{1-x^iq^{i}}{(xq)_{\infty}}=
    \mathsf{P}\frac{1-x^iq^{i}}{(1-xq)}\cdot
    \frac{1}{(1-xq^2)(1-xq^3)\dots}=
    \\
    =\mathsf{P}\sum_{m=0}^{i-1}x^mq^m
    \biggl(1+\sum_{n\geq1}\frac{q^{2n}x^n}{(1-q)\dots(1-q^n)}\biggr)=
    \\
    = \mathsf{P}\sum_{n\geq0}x^n\sum_{\substack{j=0 \\ j\leq n}}^{i-1}
    \frac{q^{2n-j}}{(1-q)\dots(1-q^{n-j})},
    \label{4.30}
  \end{multline} 
  where the inner sum on the right-hand side determines
  $R^{n}_{\kappa,i}\equiv Z_{i,\kappa+1}^n$ for $n<\kappa$.  These are
  seen to be the same as $B^n_{i,\kappa}$ (obviously, there is no
  dependence on $\kappa=p-1$ because it arises only in the higher
  $B^n_{r,\kappa}$ and $Z^n_{r,\kappa}$ due to
  conditions~\ref{C-cond-p} and~\ref{P-cond-p}).
  
  We can therefore see that $Z^m_{r,p}(q)$ satisfy $Z^m_{r,p}(q)=
  B^m_{r,p}(q)$ for $m<p$; by the recursive relations, this implies
  that $Z^n_{r,p}(q)= B^n_{r,p}(q)$ for all $n$ and hence
  Eq.~\eqref{4.20}.
\end{prf}

The coincidence of the characters shows that there are no linear
relations among thin monomials.  This completes the proof of
Lemma~\ref{lemma:1} and hence of Theorem~\ref{thm:3}.

\subsection{ The character of the semi-infinite space}\label{sec:4.4}
Theorem~\ref{thm:3} has an important corollary.
\begin{Thm}\label{thm:4} The character of the semi-infinite space
  $\W_{r,p;\theta}$ is given by
  \begin{multline}
    \chr\W_{r,p;\theta}(z,q)=
    z^{\frac{1-r+2\theta}{p}-\theta}q^{(1-\frac{2}{p})
      \frac{\theta^2-\theta}{2}-\theta\frac{1-r}{p}} \times
    \\
    \times \sum_{{N_1\geq\dots\geq N_{p-2}\in\oZ}}
    \frac{z^{\sum_{m=1}^{p-2}N_m} }{(q)_{N_1-N_2}(q)_{N_2-N_3}\dots
      (q)_{N_{p-3}-N_{p-2}}(q)_{\infty}}\times
    \\
    \times
    q^{\frac{1}{2}\left(\,\sum_{m=r}^{p-2}-\sum_{m=1}^{r-1}\right)N_m-
      \theta\sum_{m=1}^{p-2}N_m+\frac{3}{2}\sum_{m=1}^{p-2}N_m^2+
      \sum_{1\leq m<m'\leq p-2}N_mN_{m'}}.
    \label{4.31}
  \end{multline}
\end{Thm}

\begin{prf}. It is a direct consequence of~Theorem~\ref{thm:3} that
  \begin{multline}
    \chr\W_{r,p;\theta}(z,q)=\lim_{\iota\rightarrow\infty}
    \chr\W_{r,p;\theta}(\iota)(z,q)=
    \\
    = \lim_{\iota\rightarrow\infty}
    z^{-(p-2)\iota}q^{\frac{p-2}{2}(p\iota^2-\iota)}
    \chr\W_{r,p;\theta}(0)(zq^{-p\iota},q),
    \label{4.32}
  \end{multline}
  where we applied the spectral flow transform formula~\eqref{2.4} in
  the last equality.  Next, a remarkable property of Eq.~\eqref{4.19}
  is that it has a well-defined limit as $\iota\to\infty$,
  \begin{equation}
    \begin{split}
      \lim_{\iota\to\infty}\chr\W_{r,p;\theta}(\iota)(z,q)={}&
      z^{\frac{1-r+2\theta}{p}-\theta}
      q^{(1-\frac{2}{p})\frac{\theta^2-\theta}{2}-\theta\frac{1-r}{p}}
      \times
      \\
      &\times \sum_{n=-\infty}^{+\infty}
      \sum_{\substack{N_1\geq\dots\geq N_{p-2}\in\oZ \\ 
          N_1+\dots+N_{p-2}=n}} \frac{z^nq^{\frac12 n^2-\theta
          n+\sum_{m=1}^{p-2}N_m^2+
          \frac12\sum_{m=r}^{p-2}N_m-\frac12\sum_{m=1}^{r-1}N_m}}
      {(q)_{N_1-N_2}(q)_{N_2-N_3}\dots(q)_{N_{p-3}-N_{p-2}}(q)_{\infty}}
    \end{split}
    \label{4.33}
  \end{equation} 
  (where the $\theta$ dependence is reconstructed in accordance with
  the spectral flow).
\end{prf}

\begin{Rem}\label{rem:2} The existence of the thin basis has the following
  implication for the construction of antisymmetric polynomials
  satisfying the basic conditions~\eqref{4.3}.  Let $\oC\langle
  x_1,\dots,x_n\rangle^{(p)}$ be polynomials in $\oC\langle
  x_1,\dots,x_n\rangle$ satisfying~\eqref{4.3}.  There is an
  associative multiplication on antisymmetric polynomials: for
  $f\in\oC\langle x_1,\dots,x_n\rangle$ and $g\in\oC\langle
  x_1,\dots,x_{n'}\rangle$, we define $f*g\in\oC\langle
  x_1,\dots,x_{n+n'}\rangle$ as
  \begin{equation}
    (f*g)(x_1,\dots,x_{n+n'})
    =\Alt{x_1,\dots,x_{n+n'}}
    \Biggl(f(x_1,\dots,x_n)g(x_{n+1},\dots,x_{n+n'})
    \prod_{\substack{1\leq i\leq n \\ n+1\leq i'\leq n+n'}}(x_i-x_{i'})\Biggr),
    \label{4.34}
  \end{equation} 
  where $\Alt{}$ means alternation.  For $n=2$ and $n'=1$, for
  example,
  \begin{multline}
    (f*g)(x_1,x_2,x_3)=
    f({x_1},{x_2})g({x_3})({x_1}-{x_3})({x_2}-{x_3})+ \\
    +f({x_2},{x_3})g({x_1})({x_2}-{x_1})({x_3}-{x_1})
    -f({x_1},{x_3})g({x_2})({x_1}-{x_2})({x_3}-{x_2}).
    \label{4.35}
  \end{multline} 
  Now, if $f_1\in\oC\langle x_1,\dots,x_{n_1}\rangle^{(p_1)}$ and
  $f_2\in\oC\langle x_1,\dots,x_{n_2}\rangle^{(p_2)}$, it is easy to
  see that the polynomial $f_1* f_2$ is in $\oC\langle
  x_1,\dots,x_{n_1+n_2}\rangle^{(p_1+p_2-2)}$, i.e.,
  \begin{equation}
    *:\oC\langle x_1,\dots,x_{n_1}\rangle^{(p_1)}\times
    \oC\langle x_1,\dots,x_{n_2}\rangle^{(p_2)}\to
    \oC\langle x_1,\dots,x_{n_1+n_2}\rangle^{(p_1+p_2-2)}.
    \label{4.36}
  \end{equation} 
  As a consequence of the thin-basis lemma, we have a stronger
  statement: the mapping
  \begin{equation}
    *:\bigoplus_{\substack{n_1,n_2\geq1 \\ n_1+n_2=n}}
    \oC\langle x_1,\dots,x_{n_1}\rangle^{(p_1)}\times
    \oC\langle x_1,\dots,x_{n_2}\rangle^{(p_2)}\to
    \oC\langle x_1,\dots,x_{n}\rangle^{(p_1+p_2-2)}
    \label{4.37}
  \end{equation} 
  is an \textit{epimorphism}.  This construction is useful because for
  $p=3$, the polynomials satisfying~\eqref{4.3} are known explicitly
  (see~\eqref{4.7}).
\end{Rem}

\subsection{Semi-infinite construction for the
  $\protect\widehat{s\ell}(2)$
  algebra\mdseries\selectfont\kern-4pt}\cite{ref:1}. \label{sec:4.5}
The level-$k$ $\widehat{s\ell}(2)$ algebra
\begin{equation}
  \begin{aligned}
    {[}h_i,e_n]&=e_{n+i},&\qquad [h_i,f_n]&=-f_{n+i},
    \\
    {[}e_i,f_j]&=2h_{i+j}+ki\delta_{i+j,0},&\qquad [h_i,h_j]&=\frac12
    ki\delta_{i+j,0}
  \end{aligned}
  \label{4.38}
\end{equation} 
admits a semi-infinite realization of any unitary
representation~$\mAs_{r,k}$ (where $1\leq r\leq k+1$).  The key
elements of the construction are the semi-infinite forms in
\textit{commuting} elements $(f_n)_{n\in\oZ}$ satisfying the
constraints following from the conditions
\begin{equation}
  f(z)^{k+1}=0\qquad\text{for}\qquad f(z)=\sum_{n\in\oZ}f_nz^{-n-1}.
  \label{4.39}
\end{equation} 
The semi-infinite space is generated by $(f_n)_{n\in\oZ}$ from the
vectors $|r,k|\iota\rangle_{\widehat{s\ell}(2)}$ such that
$f_{2\iota+i}|r,k|\iota\rangle_{\widehat{s\ell}(2)}=0$, \ $i\geq1$,
and
\begin{align}
  &(f_{2\iota})^{r}|r,k|\iota\rangle_{\widehat{s\ell}(2)}=0,
  \label{4.40}
  \\
  &(f_{2\iota-1})^{k-r+1}(f_{2\iota})^{r-1}
  |r,k|\iota\rangle_{\widehat{s\ell}(2)}=
  |r,k|\iota-1\rangle_{\widehat{s\ell}(2)}.
  \label{4.41}
\end{align}

In more formal terms, there is a theorem parallel to
Theorem~\ref{thm:1}.  Let $\F(k)$ be the algebra generated by the
elements $(f_n)_{n\in\oZ}$ modulo the relations following~\eqref{4.39}
and by an invertible operator~$\mathsf{U}$ such that $\mathsf{U}
f_n\mathsf{U}^{-1}=f_{n+1}$.

\begin{Thm}\label{thm:5} Let $M$ be the representation of the algebra
  $\F(k)$ induced from the trivial one-dimen\-sion\-al representation
  of the algebra of $(f_n)_{n\geq0}$ {\rm(}on the vacuum vector
  $|0\rangle${\rm)}.  Let~$C_r$ \ $(1\leq r\leq k+1)$ be the
  $\F(k)$-submodule generated from the vector $f_{-1}^r|0\rangle$, and
  $N_{r,k}$ the $\F(k)$-submodule generated from the set of vectors
  \begin{equation}
    f_{\alpha-2}^{k-r+1}f_{\alpha-1}^{r-1}|\alpha\rangle-
    |\alpha-2\rangle,\qquad\alpha\in\oZ,
  \end{equation} 
  where $|\alpha\rangle=\mathsf{U}^{\alpha}|0\rangle$.  The quotient
  space $M/(N_{r,k}+C_r)$ is a representation of the
  $\widehat{s\ell}(2)_k$ algebra and, moreover, is isomorphic to a
  direct sum of unitary $\widehat{s\ell}(2)_k$ representations,
  \begin{equation}
    \mM(r,k)\equiv\frac{M}{N_{r,k}+C_r}=
    \bigoplus_{\theta=0}^{1}\mAs_{r,k;\theta}=
    \mAs_{r,k}\oplus\mAs_{k+2-r,k}.
    \label{4.42}
  \end{equation}
\end{Thm}

In the semi-infinite $\widehat{s\ell}(2)_k$ space, there also exists a
monomial basis consisting of ``thin'' monomials; the above proof
applies with minimal modifications.  The construction of the $\N2$
algebra action on the semi-infinite space can be easily carried over
to the case of the $\widehat{s\ell}(2)$ algebra (see
Sec.~\ref{sec:5.5}).

\section{The $\N2$ algebra action on $\W_{r,p;\theta}$}
\label{sec:5}
A priori, the conditions imposed on the semi-infinite construction do
not suggest that the space is a representation of any algebra; for the
constraints~\eqref{1.1}, however, this representation can be found.

\begin{Thm}\label{thm:6} The semi-infinite space $\W_{r,p;\theta}$
  is a module over the $\N2$ algebra.
\end{Thm}

The problem with constructing the $\N2$ action on~$\W_{r,p;\theta}$ is
nontrivial because we must define the action of $\cQ_n$, $\cL_n$,
and~$\cH_n$ on the states $\cG_{i_1} \cG_{i_{2}}\dots\cG_{i_m}\*
|r,p;\theta|\iota\rangle_{\infty/2}$ constructed only from~$\cG_n$ and
show that this action can be pushed forward to the quotient with
respect to the ideal~$\cS^{p}$ generated by~$S^{p}_a$.  The action of
the $\N2$ algebra on $\W_{r,p;\theta}$ is defined in several steps.
The main tool here is the \textit{positive filtration} on~$\W_{
  r,p;\theta}$ by finite-dimensional subspaces
$\W^+_{r,p;\theta}[\iota]$ (similar to the Demazure modules,
see~\cite{ref:49}--\cite{ref:51}) that allows rewriting any element
of~$\W_{r,p;\theta}$ as a linear combination of semi-infinite forms
involving only nonnegative modes $\cG_{n\geq0}$.  The next problem
consists in taking the quotient, i.e., in verifying that the action is
independent of the chosen representative of a state written in terms
of nonnegative $\cG$-modes.  In Sec.~\ref{sec:5.2}, we define the
action of a part of the $\N2$ generators using differential operators
acting on finite-dimensional subspaces whose quotients are the
subspaces in the positive filtration.  To prove that these
differential operators can be pushed forward to the quotient, we take
the dual space and use the functional realization
(Sec.~\ref{sec:5.3}).  In Sec.~\ref{sec:5.4}, we then show that the
action of the entire $\N2$ algebra on the entire semi-infinite space
can be obtained by consistently gluing together the ``partial''
actions on the subspaces.  In particular, this gives the action
of~$\cQ_{n\geq0}$; together with~$\cG_{n\leq0}$, which act on the
semi-infinite space by definition, these generate the entire $\N2$
algebra, and it only remains to verify that their action on
$\W_{r,p;\theta}$ is precisely the $\N2$ algebra action.  We show that
the necessary relations are satisfied on any vector from the
semi-infinite space.

\subsection{The positive filtration}\label{sec:5.1} Let
$\W^+_{r,p;\theta}[\iota]\subset\W_{r,p;\theta}$ be the subspace
generated from the extremal state
$|r,p;\theta|\iota\rangle_{\infty/2}$ by $\cG_{\geq0}\in\g(p)$.
Relations~\eqref{3.4} with $\iota\geq1$ determine the sequence of
embeddings
\begin{equation}
  \W^+_{r,p;\theta}[0]\subset\dots\subset
  \W^+_{r,p;\theta}[\iota]\subset
  \W^+_{r,p;\theta}[\iota+1]\subset\cdots\,.
  \label{5.1}
\end{equation}

\begin{Lemma}\label{lemma:4} Sequence~\eqref{5.1} is a filtration on
  the space~$\W_{r,p;\theta}$.
\end{Lemma}

Therefore, each state in $\W_{r,p;\theta}$ can be represented as a
linear combination of monomials involving only nonnegative modes
$\cG_{n\geq0}$.

\begin{prf}. Abusing the terminology, we say a ``semi-infinite form
  $\cG_{i_m}\dots\cG_{i_1}|r,p;\theta|\iota\rangle_{\infty/2}$''
  meaning in fact its \textit{representative} in the freely generated
  space whose quotient is~$\W_{r,p;\theta}$.  The statement of the
  lemma is that each semi-infinite form has a representative expressed
  only through nonnegative modes~$\cG_n$, i.e., a representative of
  each state $\cG_{i_m}\dots\cG_{i_1}\*
  |r,p;\theta|\iota\rangle_{\infty/2}\in\W_{r,p;\theta}$ can be chosen
  from some $\W^+_{r,p;\theta}[\iota']$.  If $\iota<0$,
  Eq.~\eqref{3.4} allows us to rewrite the state as
  $\cG_{i_m}\dots\cG_{i_1}\cdot\cG_{\iota p+\theta+1}\dots \cG_{\iota
    p+\theta+p-r-1}\dots \cG_{\iota'p+\theta-r+1}\dots \cG_{\iota'
    p+\theta-1}|r,p;\theta|\iota'\rangle_{\infty/2}$ with
  $\iota'\geq0$.  We can therefore assume that $\iota\geq0$.  We then
  choose the negative mode $\cG_{i_a}$ that is nearest to
  $|r,p;\theta|\iota\rangle_{\infty/2}$ (i.e., $i_a<0$, but $i_b\geq0$
  for all $a>b\geq1$) and consider the state $\cG_{i_a}\dots\cG_{i_1}
  |r,p;\theta|\iota\rangle_{\infty/2}$.  Rewriting it (again
  using~\eqref{3.4}) as
  \begin{multline}
    \cG_{i_a}\dots\cG_{i_1}\cG_{\iota p+\theta+1}\dots \cG_{\iota
      p+\theta+p-r-1} \cG_{\iota
      p+\theta+p-r+1}\dots\cG_{(\iota+1)p+\theta-1}
    |r,p;\theta|\iota+1\rangle_{\infty/2}=\\
    = (-1)^{a-1}\cG_{i_{a-1}}\dots\cG_{i_1} \underbrace{
      \cG_{i_a}\cG_{\iota p+\theta+1}\dots\cG_{\iota p+\theta+p-r-1}
      \cG_{\iota p+\theta+p-r+1}\dots\cG_{(\iota+1)p+\theta-1} }
    |r,p;\theta|\iota+1\rangle_{\infty/2},
    \label{5.2}
  \end{multline}  
  we apply the basic identity~$S^{p}_A=0$ with
  $A=(p-2)(p\iota+\theta+p)-{p(p-1)}/{2}+r+i_a$ to the selected modes.
  As the result, we obtain a linear combination of states in each of
  which the rightmost negative mode~$\cG_{i_{a'}}$ is such that
  $i_{a}<i_{a'}$ (this is so because in the typical term
  $\mathrm{coeff}\cdot\cG_{j_0}\cG_{j_1}\dots\cG_{j_{p-r-1}}
  \cG_{j_{p-r+1}}\dots\cG_{j_{p-1}}$ resulting from applying condition
  $S^{p}_A=0$, we have $j_b<\iota p+\theta+b$ for $1\leq b\leq p-1$, \ 
  $b\neq r$, whereas the sum $\sum_{b\geq0}j_b$ is fixed; therefore,
  $j_0>i_a$). Repeating this procedure with each of the terms
  obtained, we obtain the product $\cG_{i_m}\dots\cG_{i_{a+1}}$
  applied to a linear combination of states from some
  $\W^+_{r,p;\theta}[\iota']$.  This procedure is then repeated
  for~$\cG_{i_{a+1}}$ etc.  We finally obtain the states that belong
  to a \textit{finite} sum $\sum_{\iota'}\W^+_{r,p;\theta}[\iota']$,
  and hence, to some space $\W^+_{r,p;\theta}[\iota'']$ for a
  sufficiently large~$\iota''$.
\end{prf}

\begin{figure}[tb]
  \begin{center}
    \unitlength=1.6pt
    \begin{picture}(200,88)\thicklines
      \put(0,0){
        \put(100,2){\vector(-1,0){8}}\put(88,6){$\cG_0$}
        \put(100,0){$\bullet$}\put(107,0){$|r,p;
          \theta|\iota\rangle_{\infty/2}$}
        \put(101,6){\vector(-1,4){8}}\put(100,20){$\cG_{\iota
            p+\theta-1}$}
        \put(90,40){$\bullet$}
        \put(91,46){\vector(-1,3){8}}\put(90,60){$\cG_{\iota p
            +\theta-2}$}
        \put(70,80){$\ddots$}
        }
    \end{picture}
  \end{center}
  \caption[The subspace
  $\protect\mathsf{W}^+_{r,p;\theta}{[}\iota{]}$.]{\label{fig:2} The
    subspace $\W^{(0)}_{r,p;\theta}[\iota]= \W^+_{r,p;\theta}[\iota]$
    generated by the modes $\cG_0,\dots,\cG_{\iota p+\theta-1}$.}
\end{figure}

\begin{Rem}\label{rem:3} The choice of the modes $\cG_{n\geq0}$ is a
  matter of convention; for any $\mu\geq0$, there exists a similar
  filtration in terms of the spaces generated by~$\cG_{n\geq\mu}$. We
  denote it as
  \begin{equation}
    \W^{(\mu)}_{r,p;\theta}[0]\subset\dots\subset
    \W^{(\mu)}_{r,p;\theta}[\iota]\subset
    \W^{(\mu)}_{r,p;\theta}[\iota+1]\subset\cdots.
    \label{5.3}
  \end{equation} 
  In this notation, the above positive filtration corresponds to
  $\mu=0$, $\W^{+}_{r,p;\theta}[\iota]\equiv
  \W^{(0)}_{r,p;\theta}[\iota]$ (see Fig.~\ref{fig:2}).  These
  filtrations are crucial for constructing the~$\N2$ action.
\end{Rem}

We give several examples of rewriting semi-infinite forms in terms of
positive modes.  We recall that the $|r,p;\theta|\iota\rangle$ states
are defined in~\eqref{3.1}--\eqref{3.4}. For $p=3$ and $r=1$, we
have~$|1,3;0|{-1\rangle}=({1}/{24}){\cG_{1}\cG_{2}
  \cG_{3}\cG_{4}|1,3;0|3\rangle}- ({3}/{40})\*\cG_{1}\*\cG_{2}\*
\cG_{3}\*\cG_{6}\*\cG_{8}\*|1,3;0|4\rangle$.  Complexity of the
expressions involving only positive modes grows very fast, as, for
example, we see from
\begin{multline}
    {\cG_{-4}|1,3;0|0\rangle}={}{-\tfrac{9}{125}}
    {\cG_{1}\cG_{2}\cG_{3}\cG_{4}\cG_{8}|1,3;0|4\rangle}-
    {\tfrac{818}{17325}}{\cG_{1}\cG_{2}\cG_{3}\cG_{5}\cG_{7}
      |1,3;0|4\rangle}+
    \\
    +{\tfrac{4}{2475}}{\cG_{1}\cG_{2}\cG_{4}\cG_{5}\cG_{6}
      |1,3;0|4\rangle}-{\tfrac{66}{2275}}{\cG_{1}\cG_{2}\cG_{3}
      \cG_{4}\cG_{8}\cG_{12}|1,3;0|5\rangle}
    +{\tfrac{12498}{175175}} {\cG_{1}\cG_{2}\cG_{3}\cG_{5}\cG_{9}
      \cG_{11}|1,3;0|5\rangle}+\\
    {}+
    {\tfrac{342}{2275}}{\cG_{1}\cG_{2}\cG_{3}\cG_{6}
      \cG_{8}\cG_{11}|1,3;0|5\rangle}
    +{\tfrac{24769}{175175}}
    {\cG_{1}\cG_{2}\cG_{3}\cG_{6}\cG_{9}\cG_{10}
      |1,3;0|5\rangle}+{\tfrac{21}{3575}}{\cG_{1}\cG_{2}\cG_{5}
      \cG_{6}\cG_{8}\cG_{9}|1,3;0|5\rangle}-
    \\
    -{\tfrac{19}{3185}}{\cG_{1}\cG_{3}\cG_{4}\cG_{6}\cG_{8}
      \cG_{9}|1,3;0|5\rangle}+{\tfrac{4808}{175175}}{\cG_{1}\cG_{3}
      \cG_{5}\cG_{6}\cG_{7}\cG_{9}|1,3;0|5\rangle}
    -{\tfrac{3}{3185}}{\cG_{2 }\cG_{3}\cG_{4}\cG_{5}\cG_{8}
      \cG_{9}|1,3;0|5\rangle}-\\
    {}-{\tfrac{6}{2275}}{\cG_{2}\cG_{3}\cG_{5}\cG_{6}\cG_{7}
      \cG_{8}|1,3;0|5\rangle}
    +{\tfrac{9}{280}}{\cG_{1}\cG_{2}\cG_{3}
      \cG_{5}\cG_{9}\cG_{12}\cG_{15}|1,3;0|6\rangle}- {\tfrac{171}
      {980}}{\cG_{1}\cG_{2}\cG_{3}\cG_{6}\cG_{9}
      \cG_{12}\cG_{14}|1,3;0|6\rangle}-\\
    {}-{\tfrac{171} {1960}}{\cG_{1}\cG_{3}\cG_{5}\cG_{6}\cG_{9}
      \cG_{11}\cG_{12}|1,3;0|6\rangle}+{\tfrac{9}{980}}{\cG_{2}
      \cG_{3}\cG_{5}\cG_{6}\cG_{9}\cG_{10}\cG_{12} |1,3;0|6\rangle}.
\end{multline} 
We thus obtain states belonging to $\W^+_{1,3;0}[\iota=6]$.

\subsection{Differential operators for generators on subspaces}
\label{sec:5.2} To prove Theorem~\ref{thm:6}, we first construct the
action of \textit{a part} of the $\N2$ generators on each space
$\W^{(\mu)}_{r,p;\theta}[\iota]$ involved in~\eqref{5.3} (see
Fig.~\ref{fig:2} for $\mu=0$).  In $\W^{(\mu)}_{r,p;\theta}[\iota]$,
we define the action of the operators $\cQ_{-\mu},\cQ_{\mu+1},\dots,
\cQ_{\iota p+\theta-1}$, \ $\cL_1,\cL_2,\dots$, \ $\cH_1,
\cH_2,\dots$, and $\cG_{\mu},\cG_{\mu+1},\dots, \cG_{\iota
  p+\theta-1}$ (the latter act tautologically), which eventually
becomes a part of the $\N2$ algebra action on the entire semi-infinite
space~$\W_{r,p;\theta}$.

As the first step, we ``standardize'' the spaces by applying the
spectral flow mapping each $\W^{(\mu)}_{r,p;\theta}[\iota]$ into the
space~$\V_{r,p}^{N}$ generated by $\cG_{-1},\dots,\cG_{-N}$.  Let
$\V(N)$ (where $N$ is a positive integer) be the subspace in $\W(0)$
generated by $\cG_{-1},\cG_{-2},\dots, \cG_{-N}$ from the
corresponding vacuum vector~$|\,\rangle$.  As in Sec.~\ref{sec:4.1},
$\W(0)$ denotes the freely generated space, and $\W_{r,p}(0)$ is the
quotient of~$\W(0)$ with respect to the
subspace~$\cI_{r,p}|\,\rangle$, where~$\cI_{r,p}$ is the ideal
generated by the elements
\begin{equation}
  \overline S^{\,p}_a=
  \sum_{\substack{n_0<\dots<n_{p-2}\leq-1 \\ n_0+\dots+n_{p-2}=a}}
  \biggl(\,\prod_{i<j}(n_i-n_j)\biggr)\cG_{n_0}\dots\cG_{n_{p-2}},
  \qquad
  a=-\frac{p(p-1)}{2},-\frac{p(p-1)}{2}-1,\dots,
  \label{5.4}
\end{equation} 
and by the element $S_r=\cG_{-r}\cG_{-r+1}\dots\cG_{-1}$.  We define
$\V_{r,p}^{N}$ as the subspace in $\W_{r,p}(0)$ generated by the
modes~$\cG_{-1}$, \dots, $\cG_{-N}$ from the vacuum vector
$|\,\rangle\equiv|r,p;0|0\rangle$.  Therefore,
$\V_{r,p}^{N}=\V(N)/\cI_{r,p}(N)$, where
$\cI_{r,p}(N)=\V(N)\cap\cI_{r,p}$ (and we write
$\cI_{r,p}\equiv\cI_{r,p}|\,\rangle$), or equivalently,
\begin{equation}
  \begin{CD}
    0@>>>\cI_{r,p}@>>>\W(0)@>>>\W_{r,p}(0)@>>>0
    \\
    @.@AAA@AAA@AAA@.
    \\
    0@>>>\cI_{r,p}(N)@>>>\V(N)@>>>\V_{r,p}^N@>>>0.
  \end{CD}
  \label{5.5}
\end{equation}

We let $\partial_n$ denote the operator ${\partial}/{\partial\cG_n}$
acting on the freely generated space $\W(0)$ and its subspaces.
Although the operators~$\partial_n$ certainly do not act on the
quotient space~$\V_{r,p}^{N}$, there are differential operators
constructed from~$\partial_n$ and~$\cG_m$ that do, and these are a
part of the $\N2$ generators.

\begin{Lemma}\label{lemma:5} The differential operators
  \begin{alignat}{2}\label{5.6}
    \cQ_\ell^{(r, p, N)} ={}& \qconst{(r,p,N)}\delta_{\ell, N} \DG{-N}
    + \sum_{\substack{n,m=-N\\ m+n+\ell\leq-1}}^{-1}(m-n)
    \cG_{\ell+n+m} \DG{n} \DG{m},\quad &&\ell=N, N+1,\dots,
    2N-1,\\[-6pt]
    \begin{split}
      \cL^{(r,p,N)}_{n} ={}& \sum_{m=-N}^{-n-1}(n - m)\cG_{n +
        m}\DG{m}
      +\lconst{(r,p,N)}\delta_{n,0},\\
      \cH^{(r,p,N)}_{n} ={}& \sum_{m=-N}^{-n-1}\cG_{n + m}\DG{m} +
      \hconst{(r,p,N)}\delta_{n,0},
    \end{split}
    && n\geq0,\label{5.7}
  \end{alignat}  
  where (for an arbitrary~$\mathsf{h}^{(r,p,N)}$)
  \begin{align}
    &\mathsf{l}^{(r,p,N)}=
    \frac{p-2}{p}N+N\mathsf{h}^{(r,p,N)}-N\frac{p-r-1}{p},
    \label{5.8}
    \\
    &\mathsf{q}^{(r,p,N)}=\frac{p-2}{p}(N^2+N)-2N\frac{p-1-r}{p}
    \label{5.9}
  \end{align}
  have a well-defined action on the space~$\V_{r,p}^{N}$.  Together
  with~$\cG_{\ell}$ for~$\ell\geq-N$, these operators satisfy the
  corresponding commutation relations in~\eqref{2.1}.
\end{Lemma}

The spaces $\W^+_{r,p;\theta}[\iota]$ involved in the positive
filtration are related to $\V_{r,p}^{N}$ with $N=\iota p+\theta$ by
spectral flow transformations, namely,
\begin{equation}
  \mathsf{U}_{-\iota p-\theta}\W^+_{r,p;\theta}[\iota]\simeq
  \V_{r,p}^{\iota p+\theta}.
  \label{5.10}
\end{equation} 
These isomorphisms allow us to carry the operator action over to the
above spaces $\W^+_{r,p;\theta}[\iota]$.

The main complication in proving Lemma~\ref{lemma:5} is rooted in the
fact that the ideal $\cI_{r,p}(N)\subset\V(N)$ is not described
explicitly (cf.\ the remarks before Theorem~\ref{thm:3}).  The idea of
the (quite lengthy) proof of Lemma~\ref{lemma:5} is expressed in a
more compact form in Sec.~\ref{sec:5.5} for the $\widehat{s\ell}(2)$
algebra.

\subsection{Duals to differential operators on functional spaces}\label{sec:5.3}
To prove Lemma~\ref{lemma:5}, we first note that
$\mathsf{q}^{(r,p,N)}$ is fixed by the requirement that $\cQ_N^{(r,p,
  N)}$ preserve the relations ${\overline S}^{\,p}_a=0$ (which can be
seen by directly applying the differential operators); the form of
$\mathsf{l}^{(r,p,N)}$ is chosen such that the commutator
$\bigl[\cQ_N^{(r,p,N)}, \cG_{-N}^{(r,p,N)}\bigr]$ is the same as in
the $\N2$ algebra.  However, $\mathsf{h}^{(r,p,N)}$ is still
arbitrary; moreover, the central term arising in the last commutator
can be absorbed in~$\cL_0^{(r,p,N)}$ and~$\cH_0^{(r,p,N)}$, and
therefore, the $\N2$ central charge is still undetermined.  Now, the
operators~$\cG_{\ell}^{(r,p,N)}$, $\cQ_{\ell}^{(r,p,N)}$,
$\cL_{\ell}^{(r,p,N)}$, and $\cH_{\ell}^{(r,p,N)}$ preserve the
subspace~$\V(N)$, and it must be verified that they also preserve the
ideal~$\cI_{r,p}$.  We replace this with the dual statement and apply
the functional realization in~Sec.~\ref{sec:4.1}.
Dualizing~\eqref{5.5}, we obtain
\begin{equation}
  \begin{CD}
    @.0@.@.@.
    \\
    @.@VVV@.@.@.
    \\
    @.\W_{r,p}(0)^*\cap\cJ(N)@.@.@.
    \\
    @.@VVV@.@.@.
    \\
    0@>>>\W_{r,p}(0)^*@>>>\W(0)^*@>>>\cI_{r,p}^*@>>>0
    \\
    @.@VVV@VVV@VVV@.
    \\
    0@>>>{\V_{r,p}^N}^*@>>>\V(N)^*@>>>\cI_{r,p}(N)^*@>>>0
    \\
    @.@VVV@.@.@.
    \\
    @.0@.@.@.
  \end{CD}
  \label{5.11}
\end{equation} 
The property of the operators to have a well-defined action on the
quotient then reformulates as the condition for their duals to
\textit{preserve} the subspace~$\W_{r,p}(0)^*$ of antisymmetric
polynomials satisfying conditions~\eqref{4.3} and~\eqref{4.4}
\textit{modulo} the subspace $\cJ(N)$ spanned by the monomials
$x_1^{i_1}\dots x_n^{i_n}$ with $i_j\geq N$ for at least one~$j$ (more
precisely, modulo the intersection of this subspace with the space of
skew-symmetric polynomials).

As before, we let $\oC\langle x_1,\dots,x_M\rangle^{(p)}$ denote
polynomials in $\oC\langle x_1,\dots,x_M\rangle$ satisfying
condition~\eqref{4.3}.  We now establish that for $f\in\oC\langle
x_1,\dots,x_{M}\rangle^{(p)}$,
\begin{equation}
  {\cK_n^{(r,p,N)}}^*f\in\oC\langle x_1,\dots,x_{M}\rangle^{(p)},\qquad
  \cK_n=\cH_n\quad\text{or}\quad\cL_n,
  \label{5.12}
\end{equation} 
while $\cQ_{\ell}$ preserve the conditions modulo $\cJ(N)$,
\begin{equation}
  {\cQ_{\ell}^{(r, p, N)}}^*f\in
  \oC\langle x_1,\dots,x_{M+1}\rangle^{(p)}+\cJ(N),
  \label{5.13}
\end{equation} 
and hence the expression
\begin{equation}
  \frac{\partial}{\partial x_{2}}\frac{\partial^{2}}{\partial x_{3}^{2}}\dots
  \frac{\partial^{p-2}}{\partial x_{p-1}^{p-2}}
  \bigl({\cQ_{\ell}^{(r,p,N)}}^*f\bigr)(x_1,\dots,x_{M+1})
  \bigr|_{x_1=x_2=\dots=x_{p-1}}
\end{equation} 
vanishes modulo $\cJ(N-p+2)$.  For this, we derive a suitable form of
the dual operators.

The operators dual to $\cG_i$ act between the $M$-variable subspaces
in $\W(0)^*$ as
\begin{align}
  &\oC\langle x_1,\dots,x_{M-1}\rangle\,dx_1\dots dx_{M-1}
  \overset{\cG_i^*}\to{\longleftarrow} \oC\langle
  x_1,\dots,x_M\rangle\,dx_1\dots dx_M,
  \label{5.14}
  \\
  &\bigl({\cG_{-\ell}^{(r,p,N)}}^*f\bigr)(x_1,\dots,x_{M-1})=
  \frac{1}{(\ell-1)!}\frac{\partial^{\ell-1}}{\partial x_M^{\ell-1}}
  f(x_1,\dots,x_M)\biggr|_{x_M=0}.
  \label{5.15}
\end{align}
The duals to $\partial_{-\ell}={\partial}/{\partial\cG_{-\ell}}$ are
the pairwise anticommuting differentials
\begin{equation}
  \oC\langle x_1,\dots,x_M\rangle\,dx_1\dots dx_M
  \overset{\partial_{-\ell}^*}\to{\longrightarrow}
  \oC\langle x_1,\dots,x_{M+1}\rangle\,dx_1\dots dx_{M+1},
  \label{5.16}
\end{equation} 
acting as
\begin{equation}
  (\partial_{-\ell}^*f)(x_1,\dots,x_{M+1})=
  \sum_{j=1}^{M+1}(-1)^{M+1-j}f\bigl([\x]_j^{(M+1)}\bigr)x_j^{\ell-1},
  \label{5.17}
\end{equation} 
where we use the notation
\begin{align}
  &[\x]^{(M)}=(x_1,\dots,x_M),\qquad
  [\x]^{(M)}_i=(x_1,\dots,x_{i-1},x_{i+1},\dots,x_M),
  \\
  &[\x]^{(M)}_{ij}=(x_1,\dots,x_{i-1},x_{i+1},\dots,
  x_{j-1},x_{j+1},\dots,x_M),\qquad1\leq i<j\leq M.
\end{align}
For $f\in\Skew{x_1,\dots,x_M}$, we then have, for example,
\begin{equation}\label{dd-act}
  (\d^*_{-m}\d^*_{-n} f)(x_1,\dots,x_{M+2})={}
  \sum_{1\leq i<j\leq M+2}(-1)^{i+j+1}f([\x]^{(M+2)}_{ij})
  (x_i^{n-1} x_j^{m-1} - x_i^{m-1} x_j^{n-1}).    
\end{equation}

The dual action of $\cH_n$ and $\cL_n$ is given by
\begin{align}\label{5.18}
  \cH_n^*f &=\bigl(p_n+\mathsf{h}^{(r,p,N)}\delta_{n,0}\bigr)\cdot f,
  \\
  (\cL_n^*f)(x_1,\dots,x_M) &= \biggl((2n+1)p_n(x_1,\dots,x_M)+
  \label{5.19}\\ &\qquad{}+
  \mathsf{l}^{(r,p,N)}\delta_{n,0}+
  \sum_{j=1}^Mx_j^{n+1}\frac{\partial}{\partial x_{j}}\biggr)
  f(x_1,\dots,x_M),\notag
\end{align}
where $g\cdot h$ denotes the ``pointwise'' multiplication (of a
symmetric and an antisymmetric polynomial) and we introduce the
symmetric polynomials
\begin{equation}
  p_r(x_1,\dots,x_{M})=x_1^r+\dots+x_{M}^r.
  \label{5.20}
\end{equation} 
It is obvious that the operators $\cH_n$ with $n\geq0$ map
$\W_{r,p}(0)^*$ into itself.  For~$\cL_n$, this statement is verified
as follows.  We split the summation~\eqref{5.19} as
\begin{equation}
  \sum_{j=1}^M\frac{\partial}{\partial x_{j}}f(x_1,\dots,x_M)=
  \biggl(\,\sum_{j=1}^{p-1}+\sum_{j=p}^M\,\biggr)
  \frac{\partial}{\partial x_{j}}f(x_1,\dots,x_M).
  \label{5.21}
\end{equation} 
Each term in the second sum in the right-hand side of~\eqref{5.21}
preserves the conditions
\begin{equation}
  \frac{\partial}{\partial x_{2}}\frac{\partial^{2}}{\partial x_{3}^{2}}\dots
  \frac{\partial^{p-2}}{\partial x_{p-1}^{p-2}}f
  \biggr|_{x_1=x_2=\dots=x_{p-1}}=0.
\end{equation} 
To the first sum, we apply the operator
\begin{equation}
  \frac{\partial}{\partial x_{2}}\frac{\partial^{2}}{\partial x_{3}^{2}}\dots
  \frac{\partial^{p-2}}{\partial x_{p-1}^{p-2}},
\end{equation} 
commute it with
\begin{equation}
  \sum_{j=1}^{p-1}x_j^{n+1}\frac{\partial}{\partial x_{j}}
\end{equation} 
and restrict the expression obtained to the $(p-1)$-diagonal
$x_1=x_2=\dots=x_{p-1}=x$, after which the sum
\begin{equation}
  \sum_{j=1}^{p-1}x_j^{n+1}\frac{\partial}{\partial x_{j}}=
  x^{n+1}\sum_{j=1}^{p-1}\frac{\partial}{\partial x_{j}}
\end{equation} 
becomes the derivative along the diagonal, and hence,
\begin{equation}
  \sum_{j=1}^{p-1}\frac{\partial}{\partial x_{j}}
  \frac{\partial}{\partial x_{2}}\frac{\partial^{2}}{\partial x_{3}^{2}}
  \dots\frac{\partial^{p-2}}{\partial x_{p-1}^{p-2}}f
  \biggr|_{x_1=x_2=\dots=x_{p-1}}=0.
\end{equation} 
Property~\eqref{5.12} is thus established for~$\cL_n$.

Some more work is required with the ${\cQ_\ell^{(r, p, N)}}^*$
operators.  They act in the same direction as the differentials
in~\eqref{5.16}; using~\eqref{5.15} and~\eqref{dd-act}, we find
\begin{multline}\label{Q-star}
  ({\cQ_\ell^{(r, p, N)}}^*f)(x_1,\dots,x_{M+1}) =
  \qconst{(r,p,N)}\delta_{\ell, N}
  \sum_{j=1}^{M+1}(-1)^{M+1-j}f([\x]_j^{(M+1)})x_j^{\ell-1}
  +{}\\[-2pt]
  {}+ \sum_{1\leq i<j\leq M+1}\!\!(-1)^{i+j}\!\!
  \sum_{\substack{m,n=1\\ m+n-\ell\geq1}}^N \!\!\Bigl(x_j\DX{}{j} -
  x_i\DX{}{i}\Bigr) \frac{x_i^{n-1} x_j^{m-1} + x_i^{m-1}
    x_j^{n-1}}{(n+m-\ell-1)!}D^{n+m-\ell-1}_M f([\x]^{(M+1)}_{ij},0),
\end{multline}
where $D_i f(\xi_1,\dots \xi_M)=\frac{d}{d t}
f(\xi_1,\dots,\xi_{i-1},\xi_i+t,\xi_{i+1},\dots,\xi_M)\bigr|_{t=0}$.
Since we are only interested in terms modulo $\cJ(N)$, we can bring
the last equation to a much more tractable form by adding suitable
terms from~$\cJ(N)$.  For $\ell\geq N$ as in the condition of the
lemma, the Taylor series can be completed by terms from~$\cJ(N)$, and
${\cQ_\ell^{(r, p, N)}}^*$ can therefore be redefined modulo~$\cJ(N)$
to (omitting the ${}^{(M+1)}$ superscript for brevity)
\begin{multline*}
  ({\cQ_\ell^{(r, p, N)}}^*f)(x_1,\dots,x_{M+1})=
  \qconst{(r,p,N)}\delta_{\ell, N}
  \sum_{j=1}^{M+1}(-1)^{M+1-j}f([\x]_j)\,x_j^{\ell-1}
  + \\
  {}+\sum_{1\leq i<j\leq M+1}(-1)^{i+j} \Bigl(x_j\DX{}{j} -
  x_i\DX{}{i}\Bigr)\left (F_\ell(x_i, x_j) \left( f([\x]_{ij},x_i) +
      f([\x]_{ij},x_j) \right) \right),
\end{multline*}
where
\begin{equation}\label{F-ell}
  F_\ell(x,y)=\sum\limits_{n=0}^{\ell-1}x^n y^{\ell-1-n}=
  \frac{x^{\ell}-y^{\ell}}{x-y}.
\end{equation}
Note that for antisymmetric polynomials, $f([\x]_{ij}^{(M+1)},x_i)\!
=\!(-1)^{M-i} f([\x]_j^{(M+1)})$ and
$f([\x]_{ij}^{(M+1)},x_j)\!=\!(-1)^{M+1-j}f([\x]_i^{(M+1)})$ for
$i<j$.  Further, when acting with the derivatives on the sum of two
$f$ polynomials, we obtain
\begin{multline*}
  F_\ell(x_i,x_j)\Bigl(x_j\DX{}{j} - x_i\DX{}{i}\Bigr)\bigl(
  f([\x]_{ij},x_i) + f([\x]_{ij},x_j)\bigr) ={}\\
  = (x_j^\ell-x_i^\ell) \sum_{n\geq1}\frac{(-1)^{M+i}}{n!}\,
  (x_j-x_i)^{n-1}\frac{\d^n i[f]}{\d x_i^n} ([\x]_j)\in\cJ(N),
\end{multline*}
where $i[f](y_1,\dots,y_M)=y_iD_if(y_1,\dots,y_M)$.  Thus,
${\cQ_\ell^{(r, p, N)}}^*f$ can be redefined modulo~$\cJ(N)$ as
\begin{equation}\label{Q-rewritten}
  ({\cQ_\ell^{(r, p, N)}}^*f)(x_1,\dots,x_{M+1})
  =\sum_{j=1}^{M+1} (-1)^{M+j}f([\x]_j)
  \left(
    \sum_{i=1}^{M+1} A_\ell(x_j, x_i) -
    \qconst{(r,p,N)}\delta_{\ell, N}\,x_j^{\ell-1}
  \right)
\end{equation}
with
\begin{equation}  
  \begin{split}      
    A_\ell(x_j, x_i)={}& \sum_{m=1}^{\ell}(\ell+1-2m)x_j^{\ell-m}
    x_i^{m-1}= \frac{(\ell-1)(x_j^{\ell+1} - x_i^{\ell+1}) -
      (\ell+1)(x_j^{\ell}x_i - x_i^{\ell}x_j)}{(x_j - x_i)^2}\\
    ={}&(x_j - x_i)B_\ell(x_j, x_i),\qquad B_\ell(x_j, x_i) =
    \sum_{m=1}^{\ell-1} m(\ell-m)x_j^{\ell-m-1}x_i^{m-1}.
  \end{split}
\end{equation}

This can be expressed in a more ``invariant'' form as follows.  We
define the differential $d_m$ to be $\d^*_{-m-1}$ up to a sign,
\begin{equation}
  (d_m f)(x_1,\dots,x_{M+1})=\sum_{j=1}^{M+1}(-1)^{j+1}
  x_j^m f([x]_j),\qquad m\geq 0.
\end{equation}  
Up to a similar conventional sign factor $(-1)^{M+1}$ (which is
inessential for the vanishing property we want to show), we finally
rewrite~\eqref{Q-rewritten} as
\begin{equation}\label{Qstar-final}
  {\cQ_\ell^{(r, p, N)}}^*f=
  \sum_{m=1}^\ell(\ell+1-2m)p_{m-1}\cdot d_{\ell-m}f -
  \qconst{(r,p,N)}\delta_{\ell, N}d_{\ell-1}f.
\end{equation}  


Equation~\eqref{5.13} can now be shown by induction on the
number of variables $M$.  Assuming that~\eqref{5.13} holds for
all $f\in\Skew{x_1,\dots,x_{M-1}}^{(p)}$, we take a polynomial
$f\in\Skew{x_1,\dots,x_{M}}^{(p)}$ and consider
\begin{multline}\label{already-know}
  \cG_m^*{\cQ_\ell^{(r, p, N)}}^*f= -{\cQ_\ell^{(r, p,
      N)}}^*(\cG_m^*f) + \cL_{m+\ell}^*f-2\ell\cH_{m+\ell}^*f
  +\tfrac{p-2}{p}(m^2+m)\delta_{m+\ell,0}f,\\*
  m=-1,\dots,-N,
\end{multline}
where the right-hand side follows from the explicit expressions (most
easily, in the form of differential operators).
In~\eqref{already-know}, we already know that $\cL_{m+\ell}^*$ and
$\cH_{m+\ell}^*$ map $\Skew{x_1,\dots,x_{M}}^{(p)}$ into itself.
Moreover, we see from~\eqref{5.14} that $\cG_m^*f$ is a polynomial
in $M-1$ variables and, moreover, \textit{it belongs
  to}~$\Skew{x_1,\dots,x_{M-1}}^{(p)}$.  By the induction hypothesis,
therefore, ${\cQ_\ell^{(r, p,
    N)}}^*(\cG_m^*f)\in\Skew{x_1,\dots,x_{M}}^{(p)} + \cJ(N)$; thus,
the same is true for $\cG_m^*{\cQ_\ell^{(r, p, N)}}^*f$.  This in turn
tells us much about ${\cQ_\ell^{(r, p, N)}}^*f$: the first $N-1$ terms
in its Taylor expansion in $x_{M+1}$ are
in~$\cJ(N)+\Skew{x_1,\dots,x_{M}}^{(p)}$; the remaining terms do not
interest us, however, because they certainly are in~$\cJ(N)$.  It only
remains to prove the induction base, namely that ${\cQ_\ell^{(r, p,
    N)}}^*f\in\Skew{x_1,\dots,x_{p}}^{(p)}+\cJ(N)$ for
$f\in\Skew{x_1,\dots,x_{p-1}}^{(p)}$, which follows from the explicit
form~\eqref{Qstar-final}.

Finally, condition~\eqref{4.4} on polynomials is easiest to consider
in terms of the symmetric polynomial corresponding to a given
antisymmetric one (see~\eqref{4.5}); this condition states the
vanishing of the polynomial at zero, which can be reformulated in
terms of filtration~\eqref{4.17}.  It is then easy to directly verify
that the operators preserve these vanishing conditions.

\subsection{The algebra action on $\W_{r,p;\theta}$ by gluing the
  pieces together} \label{sec:5.4} To complete the proof of
Theorem~\ref{thm:6}, we construct the action of the $\N2$ generators
on the entire semi-infinite space $\W_{r,p;\theta}$ by gluing together
the actions constructed on each of the spaces involved in
filtration~\eqref{5.3}.  This is done in several steps.

Because the dependence on the spectral flow parameter~$\theta$ does
not affect the general structure of the results, we temporarily set
$\theta=0$ to simplify the formulas.  We therefore consider the
semi-infinite spaces $\W_{r,p}\equiv\W_{r,p;0}$, with the dependence
on~$\theta$ to be reconstructed in accordance with the spectral flow.
We write $|r,p|\iota\rangle_{\infty/2}$ for
$|r,p;0|\iota\rangle_{\infty/2}$.

\subsubsection{}\label{sec:5.4.1}For a given semi-infinite form
$|x\rangle\in\W_{r,p}$ with the charge--level bigrade $(h,\ell)$, we
choose~$\iota$ and $\mu<\iota p-1$ such that all the states in this
bigrade lie in $\W^{(\mu)}_{r,p}[\iota]$.  Wishing to define the state
$\cO_n|x\rangle$ for $\cO=\cL$, \ $\cH$, or~$\cQ$, we also require
that the space $\W^{(\mu)}_{r,p}[\iota]$ contain all the states in the
corresponding bigrade (given by $(h,\ell-a)$ for $\cO=\cL$ or~$\cH$
and $(h-1,\ell-a)$ for~$\cQ$).  This can be always achieved by
increasing~$\iota$ or~$\mu$.

Similarly to~\eqref{5.10}, the spectral flow transform can be used to
map the state $|r,p|\iota\rangle_{\infty/2}$ into
$|r,p|0\rangle_{\infty/2}$ and all the subspaces
$\W^{(\mu)}_{r,p}[\iota]$ into the corresponding spaces $\V_{r,p}^{N}$
in~Sec.~\ref{sec:5.2},
\begin{equation}
  \mathsf{U}_{-\iota p}\W^{(\mu)}_{r,p}[\iota]\simeq
  \V_{r,p}^{\iota p-\mu}.
  \label{5.25}
\end{equation} 
In $\V_{r,p}^{\iota p-\mu}$, we apply the operator $\mathsf{U}_{-\iota
  p}\cO_n\mathsf{U}_{\iota p}$ to $\mathsf{U}_{-\iota p}|x\rangle$
using the differential operators in Lemma~\ref{lemma:5}, and then use
the spectral flow to transform the result back into
$\W^{(\mu)}_{r,p}[\iota]$.  We thus set
\begin{equation}
  \cO_n|x\rangle=\mathsf{U}_{\iota p}\bigl(\bigl(\,\mathsf{U}_{-\iota p}
  \cO_n\mathsf{U}_{\iota p}\,\bigr)^{(r,p,\iota p-\mu)}
  \mathsf{U}_{-\iota p}|x\rangle\bigr),
  \label{5.26}
\end{equation} 
where
\begin{equation}
  \cO_n=\cL_{n},\quad n\geq0;\qquad
  \cO_n=\cH_{n},\quad n\geq0;\qquad
  \cO_n=\cQ_{n},\quad n\geq-\mu,
\end{equation} 
and the superscript in~\eqref{5.26} means that the corresponding
differential operator from Lemma~\ref{lemma:5} is applied.

The operators in Eqs.~\eqref{5.6} and~\eqref{5.7} depend on $N=\iota
p-\mu$ and involve an additional parameter $\mathsf{h}^{(r,p,N)}$.
Another free parameter arises when the spectral flow is applied: while
mapping \textit{semi-infinite forms} involves only the relations
$\mathsf{U}_{\theta}\cG_n\mathsf{U}_{-\theta}=\cG_{n+\theta}$ and
$\mathsf{U}_{-\iota p}|r,p|\iota\rangle_{\infty/2}=
|r,p|0\rangle_{\infty/2}$, which are a part of the definition of the
semi-infinite space, it is understood that the operator
$\mathsf{U}_{-\iota p}\cO_n\mathsf{U}_{\iota p}$ is evaluated using
the $\N2$ spectral flow formula~\eqref{2.2}. This gives rise to the
parameter~$c$, which is \textit{also free at this stage}.

Remarkably, all the free parameters are uniquely fixed by the
consistency requirements.  Moreover, it can then be proved that the
action defined above is independent of the chosen~$\iota$ and~$\mu$.
This is shown in Secs.~\ref{sec:5.4.2} and~\ref{sec:5.4.3}; in
Secs.~\ref{sec:5.4.4} and~\ref{sec:5.4.5}, we then establish that the
action of the generators constructed defines precisely the $\N2$
algebra.

\subsubsection{} \label{sec:5.4.2} Using Eqs.~\eqref{2.2}
and~\eqref{3.9}, we compare the eigenvalues of~$\cH_0$ on the states
$|r,p|\iota\rangle_{\infty/2}$ and $|r,p|\iota-1\rangle_{\infty/2}$.
In view of Eqs.~\eqref{3.4} (see also Fig.~\ref{fig:2}), these
eigenvalues differ by $p-2$.  On the other hand,
\begin{equation}
  \mathsf{U}_{p}\cH_0|r,p|\iota-1\rangle_{\infty/2}=
  \mathsf{U}_{p}\cH_0\mathsf{U}_{-p}|r,p|\iota\rangle_{\infty/2}=
  \left(\cH_0+\frac{c}{3}p\right)|r,p|\iota\rangle_{\infty/2},
  \label{5.27}
\end{equation} 
which gives central charge~\eqref{1.2} expressed through the
parameter~$p$ in the basic relation~\eqref{1.1}.

A similar argument applied to $\cL_0$ gives
\begin{equation}
  \mathsf{U}_{p}\cL_0|r,p|\iota-1\rangle_{\infty/2}=
  \mathsf{U}_{p}\cL_0\mathsf{U}_{p}^{-1}|r,p|\iota\rangle_{\infty/2}
  =
  \left(\cL_0+p\cH_0+\frac{c}{6}(p^2+p)\right)
  |r,p|\iota\rangle_{\infty/2},
  \label{5.28}
\end{equation}
where we already know $c$.  On the other hand, (minus) the difference
between the eigenvalues of~$\cL_0$ on the states
$|r,p|\iota-1\rangle_{\infty/2}$ and $|r,p|\iota\rangle_{\infty/2}$ is
given by (see~\eqref{3.4} and Fig.~\ref{fig:2})
\begin{equation}
  \bigl([\iota p-1]+\dots+\bigl[(\iota-1)p+1\bigr]\bigr)-[\iota p-r]=
  \iota p(p-1)-\frac12p(p-1)-\iota p+r,
\end{equation} 
whence it follows that the parameter $\mathsf{h}^{(r,p,N)}$ in
Lemma~\ref{lemma:5} is given by
\begin{equation}
  \mathsf{h}^{(r,p,N)}=-\frac{r-1}{p}.
  \label{5.29}
\end{equation} 
With this $\mathsf{h}^{(r,p,N)}$ in~\eqref{5.8}, it follows that
$\mathsf{l}^{(r,p,N)}=0$.

Restoring the dependence on $\theta$, we now have the relation
\begin{equation}
  \cH_0|r,p;\theta|\iota\rangle_{\infty/2}=
  \left(-\frac{r-1}{p}-\frac{p-2}{p}\theta-(p-2)\iota\right)
  |r,p;\theta|\iota\rangle_{\infty/2}
  \label{5.30}
\end{equation} 
holding in $\W_{r,p;\theta}$.  This gives the same eigenvalues as in
the unitary module (given by Eq.~\eqref{2.17}), and the same is easily
seen to be true for the eigenvalues of~$\cL_0$ on
$|r,p;\theta|\iota\rangle_{\infty/2}$ (given by Eq.~\eqref{2.16}).

\subsubsection{}\label{sec:5.4.3}In addition to $\cL_0$ and $\cH_0$,
the action of all the operators~$\cL_{\geq0}$ and~$\cH_{\geq0}$ is
carried over from $\V_{r,p}^{\iota p-\mu}$ to $\W_{r,p}$.  Similarly,
in each $\W^{(\mu)}_{r,p}[\iota]$, we obtain the action of the
operators~$\cQ_{n}$ for $n\geq-\mu$. A priori, they depend on the
space $\V_{r,p}^{\iota p-\mu}$ in which the differential operators are
applied.  Now, as noted above, Eq.~\eqref{5.9} follows from the
condition that the operator $\cQ_N^{(r,p,N)}$ preserve the relations
${\overline S}^{\,p}_a=0$.  It is remarkable that for
this~$\mathsf{q}^{(r,p,N)}$ and with $\mathsf{h}^{(r,p,N)}$ chosen
in~\eqref{5.29}, we can rewrite Eq.~\eqref{5.6} as
\begin{equation}
  \cQ_{\ell}^{(r,p,N)}=2\sum_{n=-N}^{-1}
  \left(\cL_{n+\ell}^{(r,p,N)}-\ell\cH_{n+\ell}^{(r,p,N)}+
    \frac12\frac{p-2}{p}(\ell^2-\ell)\delta_{\ell+n,0}\right)\partial_n.
  \label{5.31}
\end{equation} 
Moreover, for $\mathsf{h}^{(r,p,N)}$ of form~\eqref{5.29} (and
with~$\mathsf{l}^{(r,p,N)}=0$), the right-hand side of~\eqref{5.31}
depends on~$N$ only through the summation limits.  This implies that
the differential operators in Lemma~\ref{lemma:5} commute with the
embeddings $\V_{r,p}^{N-1}\to\V_{r,p}^{N}$.  Therefore, for $n\geq0$,
the operators $\cL_n$, $\cH_n$, and~$\cQ_{-\mu+n}$ acting in
$\W_{r,p}$ as defined in~\eqref{5.26} are independent of~$\iota$
and~$\mu$.

\subsubsection{}\label{sec:5.4.4}We have seen that for any $\mu\in\oZ$, the operators $\cH_{\geq0}$, $\cL_{\geq0}$, $\cG_{\geq\mu}$,
and~$\cQ_{\geq-\mu}$ are well-defined on each state
in~$\W_{r,p;\theta}$; moreover, these operators satisfy the $\N2$
commutation relations.  Hence, there is a family of subalgebras of the
$\N2$ algebra acting on the semi-infinite space~$\W_{r,p;\theta}$. It
remains to define the action of~$\cH_{<0}$ and $\cL_{<0}$ and show
that all the $\N2$ commutation relations are satisfied for the
operators~$\cH_n$, $\cL_n$, $\cG_n$, and~$\cQ_n$ with $n\in\oZ$.

We still have not found the commutators $[\cQ_m,\cG_n]$ with
$m+n\leq-1$, which do not follow from the argument based on the
spectral flow.  We find them by ``solving the Jacobi identities.''  We
first show that the operators~$\cQ_n$ and~$\cG_n$, \ $n\in\oZ$,
satisfy the relations
\begin{align}
  &\bigl[\cG_m,[\cG_n,\cQ_{\ell}]\bigr]=2(m-n)\cG_{m+n+\ell},
  \label{5.32}
  \\
  &\bigl[\cQ_{\ell},[\cQ_m,\cG_n]\bigr]=2(\ell-m)\cQ_{\ell+m+n}.
  \label{5.33}
\end{align}

Starting with~\eqref{5.33}, we set
\begin{equation}
  \cX(\ell,m,n)=\bigl[\cQ_{\ell},[\cQ_m,\cG_n]\bigr]-
  2(\ell-m)\cQ_{\ell+m+n}
  \label{5.34}
\end{equation} 
and show that this expression acts by zero on any state
$|\alpha\rangle\in\W_{r,p;\theta}$.  It suffices to consider monomial
states~$|\alpha\rangle$.  For given $\ell$, $m$, and~$n$, there exists
a positive integer~$\mu$ such that $n\geq\mu$ and $\ell,m\geq-\mu$.
We choose any such~$\mu$ and use filtration~\eqref{5.3}.  Let
$|\alpha\rangle\in\W^{(\mu)}_{r,p;\theta}[\iota_0]$. This means that
the state is represented as $|\alpha\rangle=
\cG_{a_1}\dots\cG_{a_{\nu}}|r,p;\theta|\iota_0\rangle_{\infty/2}$,
where $a_j\geq \mu$.  We let~$a$ denote one of~$a_j$.

Because $a+\ell\geq0$, \ $a+m\geq0$, and $a+\ell+m+n\geq0$, the
commutation relations found so far allow us to evaluate
\begin{multline}
  \bigl[\cG_a,\cX(\ell,m,n)\bigr]={}
  2\bigl[\cL_{a+\ell}-\ell\cH_{a+\ell},[\cQ_m,\cG_n]\bigr]-
  2\bigl[\cQ_{\ell},[\cL_{a+m}-m\cH_{a+m},\cG_n]\bigr]-
  \\
  {}-4(\ell-m)\bigl(\cL_{a+\ell+m+n}-(\ell+m+n)\cH_{a+\ell+m+n}\bigr).
  \label{5.35}
\end{multline} 
Using the already established commutation relations again, we then
have
\begin{multline}
  \bigl[\cG_a,\cX(\ell,m,n)\bigr]= 2(\ell-m)[\cQ_{a+\ell+m},\cG_n]+
  2(a-n)[\cQ_m,\cG_{a+\ell+n}]+
  \\
  {}+2(a-n)[\cQ_{\ell},\cG_{a+m+n}]
  -4(\ell-m)\bigl(\cL_{a+\ell+m+n}-(\ell+m+n)\cH_{a+\ell+m+n}\bigr).
  \label{5.36}
\end{multline} 
We also obtain $a+\ell+m\geq-\mu$, \ $a+\ell+n\geq\mu$, and
$a+m+n\geq\mu$, and once again applying the already known commutation
relations, we see that the right-hand side of Eq.~\eqref{5.36}
vanishes.

The action of $\cX(\ell,m,n)$ on the state
$|\alpha\rangle=\cG_{a_1}\dots\cG_{a_{\nu}}
|r,p;\theta|\iota_0\rangle_{\infty/2}$ is thus given by
$(-1)^{\nu}\cG_{a_1}\dots\*\cG_{a_{\nu}}\*\cX(\ell,m,n)
|r,p;\theta|\iota_0\rangle_{\infty/2}$.  But it can be assumed
(possibly at the expense of increasing~$\iota_0$) that
$\cX(\ell,m,n)|r,p;\theta |\iota\rangle_{\infty/2}=0$, because
the operator $\cX(\ell, m,n)$ maps each state
$|r,p;\theta|\iota\rangle_{\infty/2}$ with $\iota\gg0$ into a bigrade
containing no states.  Therefore, $\cX(\ell,m,n)=0$
in~$\W_{r,p;\theta}$.

To prove relations~\eqref{5.32}, we denote $\cY(m,n,\ell)=\bigl[\cG_m,
[\cG_n,\cQ_{\ell}]\bigr]- 2(m-n)\cG_{m+n+\ell}$ and following
\eqref{5.35} and~\eqref{5.36} mutatis mutandis, directly verify that
the commutators $\bigl[\cG_a,\cY(m,n,\ell)\bigr]$ and
$\bigl[\cQ_a,\cY(m,n,\ell)\bigr]$ vanish for all~$a$ satisfying the
conditions
\begin{equation}
  a+\ell\geq0,\qquad a+n\geq0,\qquad a+m\geq0,\qquad a+\ell+m+n\geq0.
  \label{5.37}
\end{equation} 
Next, examining the gradings shows that
$\cY(m,n,\ell)|r,p;\theta|\iota'\rangle_{\infty/2}=0$ for
$\iota'\ll-1$.  However, this does not directly imply that
$\cY(m,n,\ell)$ vanishes on any state from the semi-infinite space
(compared to the case with~$\cX(m,n,\ell)$, the problem is with the
apparent asymmetry of the semi-infinite construction with respect to
the ~$\cG_n$ and~$\cQ_n$ generators).  It remains to be shown that
$\cY(m,n,\ell)\*|r,p;\theta|\iota\rangle_{\infty/2}=0$ for
$\iota\gg1$, because any state in the semi-infinite space can be
generated by the modes~$\cG_a$ from
$|r,p;\theta|\iota\rangle_{\infty/2}$ with a sufficiently
large~$\iota$.

It suffices to show that for any~$\iota\gg1$, the state
$|r,p;\theta|\iota\rangle_{\infty/2}$ can be obtained by acting with
the modes~$\cG_a$ and~$\cQ_a$ satisfying conditions~\eqref{5.37} on
the state $|r,p;\theta|\iota'\rangle_{\infty/2}$ with some~$\iota'$
for which $\ell+m+n\geq p\iota'+\theta$. Let~$M$ denote the minimal
integer~$a$ satisfying~\eqref{5.37}.  We fix~$\iota$ such that
$p\iota\geq M$.  We then have
\begin{equation}
  \cG_{\iota p+\theta+1}\dots\cG_{\iota p+\theta+p-r-1}\cQ_{M+\mu}
  \dots\cQ_{M+1}\cQ_{M}|r,p;\theta|\iota'\rangle_{\infty/2}=
  F|r,p;\theta|\iota\rangle_{\infty/2}
  \label{5.38}
\end{equation} 
for
\begin{align}
  &\iota'=-(p-1)\iota-\theta-M-p+r+1,
  \label{5.39}
  \\
  &\mu=(p-2)(p\iota+\theta+M+p-r-1)+p-r-2,
  \label{5.40}
\end{align}
which follows because both sides of~\eqref{5.38} belong to the same
eigenspace of~$\cH_0$ and~$\cL_0$ and this eigenspace is a
$1$-dimensional subspace in~$\W_{r,p;\theta}$.  If we show that
$F\neq0$ in Eq.~\eqref{5.38}, we can conclude that
$\cY(m,n,\ell)|r,p;\theta|\iota\rangle_{\infty/2}=0$ for $\iota\gg1$,
because it has already been shown that $\cY(m,n,\ell)$ commutes with
all ~$\cG_a$ and~$\cQ_a$ for $a\geq M$ and
$\cY(m,n,\ell)|r,p;\theta|\iota'\rangle_{\infty/2}=0$.

To verify that the coefficient $F$ is nonvanishing, it is easiest to
use the mapping $\W_{r,p;\theta}\to \mK_{r,p;\theta}$ to the unitary
$\N2$ module, under which each state~\eqref{3.1} goes into the
corresponding extremal state
$|r,p;\theta|\iota\rangle\in\mK_{r,p;\theta}$ (see
Sec.~\ref{sec:2.4}).  In~$\mK_{r,p;\theta}$, the image of~\eqref{5.38}
is satisfied with a nonvanishing~$F$.  Therefore, the coefficient~$F$
in~\eqref{5.38} cannot vanish, which implies that
$\bigl[\cG_m,[\cG_n,\cQ_{\ell}]\bigr]= 2(m-n)\cG_{m+n+\ell}$ in the
semi-infinite space.

\subsubsection{}\label{sec:5.4.5}Equations~\eqref{5.32}
and~\eqref{5.33} imply all the commutation relations~\eqref{2.1}.
Indeed, for $a<0$, we can define the operators
\begin{align}
  \cL_a^{m,n}&=\frac{1}{2(m-n)}\bigl(m[\cG_{n+a},\cQ_{-n}]-
  n[\cG_{m+a},\cQ_{-m}]\bigr),\qquad m\neq n,
  \label{5.41}
  \\
  \cH^{m,i,n}_a&=\frac{1}{2m}
  \bigl([\cG_{m+a},\cQ_{-m}]-2\cL^{i,n}_a\bigr),\qquad m\neq0.
  \label{5.42}
\end{align}
Let $A$ be the algebra generated by~$\cQ_n$ and~$\cG_n$ satisfying
relations~\eqref{5.32} and~\eqref{5.33}.  It follows from \eqref{5.32}
and \eqref{5.33} that for any fixed $m,i,n\in\oZ$, the operators
\begin{alignat*}2
  &\cL_a^{m,n},\quad\cH_a^{m,i,n},&\qquad&a<0,
  \\
  &\cL_j,\quad\cH_j,&\qquad&j\geq0,
  \\
  &\cQ_j,\quad\cG_j,&\qquad&j\in\oZ,
\end{alignat*}
satisfy commutation relations~\eqref{2.1}.  In particular, the
operators $\cL_a^{m,n}-\cL_a^{m',n'}$ and $\cH_a^{m,i,n}-
\cH_a^{m',i',n'}$ commute with~$\cQ_j$ and~$\cG_{\ell}$ and generate a
commutative ideal~$Z$ in~$A$; moreover,~$Z$ is in the center of~$A$,
and the quotient of~$A$ with respect to~$Z$ coincides with the $\N2$
algebra.  Thus, $A$ is a central extension of the~$\N2$ algebra of the
form
\begin{equation}
  [\cG_m,\cQ_n]=2\cL_{m+n}-2n\cH_{m+n}+
  \frac{c}{3}(m^2+m)\delta_{m+n,0}+f_{n,m},
  \label{5.43}
\end{equation} 
where $f_{n,m}\neq0$ only for $n<0$ and $m<0$.  In view of the Jacobi
identities, $f_{n,m}=0$, and therefore, $Z=0$ and the
operators~$\cL_a^m$ and~$\cH_a^{m,i,n}$ are independent of~$m$, $i$,
and~$n$.  \smallskip

This completes the proof of Theorem~\ref{thm:6}.  This sufficiently
strong result leads to the statement of Theorem~\ref{thm:1}.  Before
considering that theorem, however, we consider a similar construction
on the $\widehat{s\ell}(2)$ semi-infinite space.

\subsection{The $\widehat{s\ell}(2)$ action on the semi-infinite
  space}\label{sec:5.5} As for the $\N2$ algebra, we consider the
filtration of the $\widehat{s\ell}(2)$ semi-infinite space by
finite-dimensional subspaces
\begin{equation}
  \wM^+_{r,k}[0]\to\wM^+_{r,k}[1]\to\dots\to
  \wM^+_{r,k}[\iota]\to\cdots,
  \label{5.44}
\end{equation} 
where $\wM^+_{r,k}[\iota]$ is generated by $f_0,f_1,\dots,f_{2\iota}$
from~$|r,k|\iota\rangle_{\widehat{s\ell}(2)}$ (up to a spectral flow
transform, $\wM^+_{r,k}[\iota]$ are the Demazure
modules~\cite{ref:49}--\cite{ref:51}).  On each subspace
$\M^+_{r,k}[\iota]$, we can define a part of the $\widehat{s\ell}(2)$
generators as differential operators with respect to
$f_0,f_1,\dots,f_{2\iota}$; after the appropriate spectral flow
transform, we obtain differential operators with respect to
$f_{-1},f_{-2},\dots,f_{-2\iota+1}$.  It must be shown that they
preserve the ideal generated by the elements
\begin{equation}
  \sum_{\substack{-1\leq i_1,\dots,i_{k+1}\leq-2\iota+1 \\ i_1+\dots+i_{k+1}=a}}
  f_{i_1}\dots f_{i_{k+1}}=0,\qquad a=-k-1,-k-2,\dots\,.
  \label{5.45}
\end{equation} 

On each finite-dimensional space generated by
$f_{-1},f_{-2},\dots,f_{-N}$ from the vacuum vector
$|0\rangle_{\widehat{s\ell}(2)}$, the operators~$h_{\geq0}$
and~$e_{\geq N}$ are represented by the differential operators
\begin{align}
  &h_{\ell}=-\sum_{n=-N}^{-\ell-1}f_{\ell+n}\partial_n,\qquad\ell\geq0,
  \label{5.46}
  \\
  &e_{\ell}=-\sum_{\substack{n,m=-N \\ m+n+\ell\leq-1}}^{-1}
  f_{\ell+m+n}\partial_n\partial_m+kN\delta_{\ell,N}\partial_{-N},\qquad
  \ell\geq N,
  \label{5.47}
\end{align}
where $\partial_m={\partial}/{\partial f_{m}}$.

We now go over to the dual formulation, where the problem, similarly
to~Sec.~\ref{sec:5.3}, essentially reduces to finding the action of
the generators on the space of symmetric polynomials.  The dual
generators are given by
\begin{align}
  (f_{-\ell}^*\phi)(x_1,x_2,\dots,x_{M-1})&=
  \frac{1}{(\ell-1)!}\frac{\partial^{\ell-1}}{\partial x_{M}^{\ell-1}}
  \phi(x_1,x_2,\dots,x_{M})\biggr|_{x_M=0},
  \label{5.48}
  \\
  (h_{\ell}^*\phi)(x_1,x_2,\dots,x_{M})&=
  \sum_{i=1}^{M}x_i^{\ell}\phi(x_1,x_2,\dots,x_{M}),
  \label{5.49}
  \\
  (e_{\ell}^*\phi)(x_0,x_1,x_2,\dots,x_{M})&=-\sum_{0\leq i<j\leq M}
  F_{\ell}(x_i,x_j)\bigl(\phi\bigl([\x]_i\bigr)+
  \phi\bigl([\x]_j\bigr)\bigr)+ \label{5.50}
  \\*
  &\qquad\qquad\qquad{} +kN\delta_{\ell,N}\sum_{0\leq i\leq
    M}x_i^{N-1}\phi\bigl([\x]_i\bigr) \notag
\end{align}
with $F_{\ell}$ defined in~\eqref{F-ell}.

As for the $\N2$ algebra, the formula for $e_{\ell}^*$ was obtained by
adding some terms from the ideal $\cJ(N)$ generated by the monomials
$x_1^{i_1}\dots x_n^{i_n}$ in which $i_j\geq N$ for at least one~$j$;
in the algebra of symmetric polynomials, this ideal is generated by
the polynomials~$p_j$ with $j\geq N$ (see~\eqref{5.20}).  It must be
established that the action of~$e_{\ell}^*$ does not violate the
vanishing conditions on $(k+1)$-diagonals,
\begin{equation}
  \phi\Bigl(\,\underbrace{x,x,\dots,x}_{k+1},x_{k+2},\dots\Bigr)=0,
  \label{5.52}
\end{equation} 
or, more precisely, that the polynomial $e_{\ell}^*\phi$ satisfies
these conditions modulo the ideal~$\cJ(N)$.  To calculate the
right-hand side of~\eqref{5.50} for $x_0=x_1=\dots=x_{k}$, we split
the double sum in~\eqref{5.50} as
\begin{multline}
  \sum_{0\leq i<j\leq M}F_{\ell}(x_i,x_j)
  \bigl(\phi\bigl([\x]_i\bigr)+\phi\bigl([\x]_j\bigr)\bigr)=
  \\
  = \biggl(\,\sum_{0\leq i<j\leq k}+\sum_{0\leq i\leq k<j}+
  \sum_{k<i<j\leq M}\,\biggr)F_{\ell}(x_i,x_j)
  \bigl(\phi\bigl([\x]_i\bigr)+\phi\bigl([\x]_j\bigr)\bigr).
  \label{5.53}
\end{multline}
The third sum is readily seen to vanish on the diagonal in view
of~\eqref{5.52}.  Inserting~$x_0=x_1=\dots=x_k=x$ into the second sum,
we obtain terms of the form
\begin{equation}
  \frac{x^{\ell}-x_j^{\ell}}{x-x_j}
  \phi\Bigl(\,\underbrace{x,x,\dots,x}_{k},x_{k+1},\dots\Bigr).
  \label{5.54}
\end{equation} 
Again in view of~\eqref{5.52}, however, $x-x_j$ is a divisor of the
polynomial $\phi\Bigl(\,\underbrace{x,x,\dots,x}_{k},
x_{k+1},\dots\Bigr)$, and therefore each of these terms is in~$\cJ(N)$
(we recall that $\ell\geq N$).

It remains to consider the first sum in~\eqref{5.53}.  In this case,
$F_{\ell}(x,x)=\ell x^{\ell-1}$.  Each term in the sum lies in the
ideal for $\ell>N$, but for $\ell=N$, we have the terms
involving~$x^{\ell-1}$, which is not in the ideal.  However, these
terms can be summed up into the expression
\begin{equation}
  -2N\frac{k(k+1)}{2}x^{N-1}\phi
  \Bigl(\,\underbrace{x,x,\dots,x}_{k},x_{k+1},\dots\Bigr),
\end{equation} 
which is precisely canceled by the term
\begin{equation}
  +k(k+1)N\delta_{N,N}x^{N-1}\phi
  \Bigl(\,\underbrace{x,x,\dots,x}_{k},x_{k+1},\dots\Bigr),
\end{equation} 
which arises from the second term in the formula for~$e_N^*$,
Eq.~\eqref{5.50}.  This finishes the proof that
operators~\eqref{5.48}--\eqref{5.50} preserve relations~\eqref{5.52}
and therefore have a well-defined action on~$\wM^+_{r,k}[\iota]^*$.
Hence, the action is well defined on~$\wM^+_{r,k}[\iota]$ in that it
preserves the ideal generated by the left-hand sides of~\eqref{5.45}.

We note that the key role in the above cancellation is played by the
coefficient~$k$ in front of the second term in~\eqref{5.50}.  The same
coefficient becomes the level of the $\widehat{s\ell}(2)$
representation thereby constructed.  A combination of these facts
ensures the existence of the $\widehat{s\ell}(2)$ action.

The demonstration of the $\widehat{s\ell}(2)$ action on the
semi-infinite space is now completed following the same strategy as in
Sec.~\ref{sec:5.4}.  One first verifies that the action of positive
modes defined on different subspaces~$\wM^+_{r,k}[\iota]$ agrees with
the embeddings in~\eqref{5.44}.  The spectral flow then allows one to
define the action of negative modes (which again involves the argument
that along with~\eqref{5.44}, there exist similar filtrations with
$\wM^+_{r,k}[\iota]$ replaced by the spaces $\M^{(\mu)}_{r,k}[\iota]$
generated by the modes~$f_n$ with~$n\geq\mu$).  To complete the proof,
it only remains to verify the Serre relations.  For this, one first
establishes that the left-hand sides of the Serre relations commute
with all the modes~$f_n$ for $n\geq-\mu$, and because the modes~$f_n$
generate the entire semi-infinite space, it then follows that the
Serre relations are satisfied once they are satisfied on a particular
extremal state, which can be verified directly.  The Serre relations
guarantee that the algebra action constructed is the action of the
$\widehat{s\ell}(2)$ algebra.

\subsection{The isomorphism with the representation~$\mK_{r,p;\theta}$}
\label{sec:5.6}  
To continue with the $\N2$ algebra story, it remains to show that the
semi-infinite space, which we now know to be an $\N2$ module, is
isomorphic to a unitary $\N2$ representation.

There is a mapping
\begin{equation}
  \W_{r,p;\theta}\to\mK_{r,p;\theta},
  \label{5.55}
\end{equation} 
obtained by identifying each state~\eqref{3.1} with the corresponding
extremal state in~$\mK_{r,p;\theta}$ defined in Sec.~\ref{sec:2.4}
(and obviously, identifying each~$\cG_n$ with the corresponding $\N2$
algebra generator acting in the unitary representation). Obviously,
these mappings commute with the action of~$\cG_n$ and~$\mathsf{U}_{\pm
  p}$.

\begin{Thm}\label{thm:7} Mapping~\eqref{5.55} is an isomorphism of $\N2$
  modules.
\end{Thm}

\begin{prf}. We have the $\N2$ module $\W_{r,p;\theta}$, in
  which relations~\eqref{1.1} are satisfied. In this case, it is
  easiest to use the equivalence of the $\N2$ and $\widehat{s\ell}(2)$
  representation categories \cite{ref:54},~\cite{ref:55}. Applying the
  functor to $\W_{r,p;\theta}$, we obtain an
  $\widehat{s\ell}(2)$-module from the category~$\cO$ (more precisely,
  the spectral-flow orbit, whose length is equal to~$2$ on integrable
  representations), in which conditions~\eqref{4.39} are satisfied.
  Any such $\widehat{s\ell}(2)$-module is a direct sum of integrable
  (unitary) modules,\footnote{We recall that an $\widehat{s\ell}(2)$
    module is integrable if it decomposes into a sum of
    \textit{finite-dimensional} representations with respect to any of
    its $\widehat{s\ell}(2)$ subalgebras generated by $e_i$ and
    $f_{-i}$ (it actually suffices to establish this decomposition for
    \textit{two} such subalgebras); it is this property that follows
    from~\eqref{4.39}~\cite{ref:39}.}
  \begin{equation}
    \F(\W_{r,p;\cdot})=\bigoplus_{\alpha}\mAs_{(\alpha)}.
  \end{equation} 
  Applying the inverse functor, we obtain another sum of
  representations,
  \begin{equation}
    \W_{r,p;\theta}=\bigoplus_{\beta}\mK_{(\beta)},
  \end{equation} 
  where each module in the right-hand side is necessarily a unitary
  $\N2$ representation, i.e., some~$\mK_{r',p;\theta'}$, and the
  statement of the theorem follows by comparing the eigenvalues
  of~$\cL_0$ and~$\cH_0$.
\end{prf}

It follows from Theorem~\ref{thm:7} that the expression
in~\eqref{4.31} coincides with the corresponding unitary $\N2$
character.  Comparing this with the known expression for the same
character (see Sec.~\ref{sec:2.2}), we obtain the combinatorial
identity in the corollary of Theorem~\ref{thm:1} (see
Sec.~\ref{sec:1}).

As another corollary, the ``semi-infinite'' expression for the unitary
$\N2$ characters gives a formula for the string functions
$C^a_{r,p}(q)$ read off from representing the $\N2$ characters as
in~\eqref{2.9}.  To obtain this representation, we rewrite
Eq.~\eqref{4.33} (where we can set $\theta=0$) by splitting the
summation over~$n$ as
\begin{equation}
  \sum_{n\in\oZ}f(n)=\sum_{\ell\in\oZ}\sum_{a=0}^{p-3}
  f\bigl((p-2)\ell+ a\bigr).
\end{equation} 
We then shift each of the summation variables~$N_m$ as $N_m\mapsto
N_m+\ell$, which does not change the factors $(q)_{N_m-N_{m+1}}$.  The
summation over $\ell$ can therefore be performed, with the result that
the ``semi-infinite'' formula for the character takes the
form~\eqref{2.9} with the string function
\begin{equation}
  C^a_{r,p}(q)=
  \sum_{\substack{N_1\geq\dots\geq N_{p-2}\in\oZ \\ N_1+\dots+N_{p-2}=a}}
  \frac{q^{\sum_{m=1}^{p-2}N_m^2+\sum_{m=r}^{p-2}N_m}}
  {(q)_{N_1-N_2}(q)_{N_2-N_3}\dots(q)_{N_{p-3}-N_{p-2}}(q)_{\infty}}.
  \label{5.56}
\end{equation}

\section{Some related constructions}\label{sec:6}

\subsection{Positive bases: paths and the generalized Pascal
  triangles} \label{sec:6.1} Filtration~\eqref{5.1} allows us to
rewrite a representative of any state in the semi-infinite space
$\W_{r,p;\theta}$ through only nonnegative modes $\cG_{n\geq0}$.  The
next interesting problem is to construct a basis in each term of this
filtration.  We construct such bases in the finite-dimensional
subspaces $\M^+_{r,k}[\iota]$ (see~\eqref{5.44}) of the
$\widehat{s\ell}(2)$ semi-infinite space using some kind of a
``Demazure induction'' (cf.~\cite{ref:49}--\cite{ref:51}).  The
combination of bases in the corresponding~$\wM^+_{r,k}[\iota]$ spaces
gives a basis in the $\N2$ space~$\W^+_{r,p;\theta}[\iota]$.

\subsubsection{Positive bases in unitary $\widehat{s\ell}(2)$
  modules} \label{sec:6.1.1} We consider the filtration by
finite-dimensional subspaces in Eq.~\eqref{5.44}.  Basis vectors
in~$\wM^+_{r,k}[\iota]$ are in a $1:1$ correspondence with paths on a
rectangular lattice whose construction we now explain.  We label the
horizontal lines of the lattice by $\ell_0,\ell_1,\dots$ such
that~$\ell_0$ corresponds to the bottom.  On each line~$\ell_m$, we
label the sites as~$\ell_{m, n}$ with $n\geq0$.  We assign the
site~$\ell_{0,0}$ the symbol
$\Bigl[\,\underbrace{k+1,\dots,k+1}_{\iota+1},r\Bigr]$.  A path ending
at~$\ell_{m,n}$ is a connected sequence of links between~$\ell_{0,0}$
and~$\ell_{m,n}$ (see Fig.~\ref{fig:3}), with each site along the path
assigned an $[\dots]$ symbol as follows.  If the site~$\ell_{m,n}$ on
the path is already assigned a symbol $[a_1,\dots,a_{\nu}]$, where
$a_1\geq a_2\geq\dots\geq a_{\nu}$, the path can be continued from
this site via one of the following two steps whenever the
corresponding conditions are satisfied:
\begin{figure}[b]
  \begin{center}
    \unitlength=1pt
    \begin{picture}(140,100)
      \multiput(0,0)(20,0){8}{\line(0,1){100}}
      \multiput(0,0)(0,20){6}{\line(1,0){140}}
      {\linethickness{2pt}
        \put(0,0){\line(1,0){40.9}}
        \put(40,0){\line(0,1){20.8}}
        \put(40,20){\line(1,0){20.9}}
        \put(60,20){\line(0,1){40.8}}
        \put(60,60){\line(1,0){60.9}}
        \put(120,60){\line(0,1){20}}
        }
    \end{picture}
  \end{center}
  \caption[A path]{\label{fig:3} A path on the lattice. The starting
    point is assigned the symbol $[k+1,\dots,k+1,r]$, and the sites
    along the path are assigned $[\dots]$ symbols in accordance with
    the two rules.  These rules determine the possible ways of
    continuing the path.}
\end{figure}
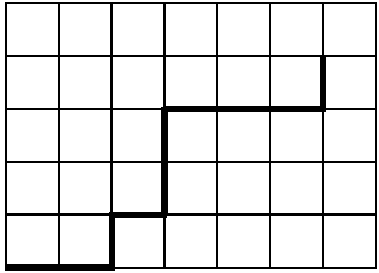
\begin{enumerate}
  
\item\label{step:down} Moving to the site $\ell_{m+1,n}$ and assigning
  it the symbol $[a_2,\dots,a_{\nu}]$ (provided $\nu\geq2$).
  
\item\label{step:right} Moving to the site $\ell_{m,n+1}$ and
  assigning it the symbol
  $[a_1,\dots,a_{\lambda-1},a_{\lambda}-1,a_{\lambda+1},\dots,
  a_{\nu}]$ if $a_{\lambda}\geq2$, where~$\lambda$ is defined by the
  conditions $a_1=\dots=a_{\lambda}>a_{\lambda+1}$ and $\lambda=\nu$
  if $a_1=\dots=a_{\nu}\geq2$.
\end{enumerate}

We let $\blacktriangle$ and $\blacktriangleright$ denote the link
created by taking the respective steps~\ref{step:down}
and~\ref{step:right}.  More precisely, we write
${\blacktriangleright}_m$ for each $\blacktriangleright$ link between
any two sites on~$\ell_m$.  We note that different paths assign
different $[\dots]$ symbols to the same site.

A path ending at $\ell_{m,n}$ is said to be \textit{admissible} if it
cannot be continued by step~~\ref{step:down} from~$\ell_{m,n}$ (it may
or may not be continued via step~\ref{step:right}).

Basis vectors in $\wM^+_{r,k}[\iota]$ are in a $1:1$ correspondence
with the admissible paths.  The path consisting of only the
$\blacktriangle$ links corresponds to the twisted highest-weight
vector $|r,k|\iota\rangle_{\widehat{s\ell}(2)}$ (see \eqref{4.40}
and~\eqref{4.41}).  For any other admissible path, let
${\blacktriangleright}_{j_0},{\blacktriangleright}_{j_1},
\dots,{\blacktriangleright}_{j_K}$ (where $0\leq j_0\leq\dots\leq
j_K$) be its $\blacktriangleright$ links.  The corresponding basis
vector is given by
\begin{equation}
  f_{j_0}f_{j_1}\dots f_{j_K}|r,k|\iota\rangle_{\widehat{s\ell}(2)}\in
  \wM^+_{r,k}[\iota].
  \label{6.1}
\end{equation}

This construction is based on the fact that~\cite{ref:42}
\begin{equation}
  \wM^+_{r,k}[\iota]\simeq
  \Gr\Bigl(\,
  \underbrace{\oC^{k+1}\otimes\dots\otimes\oC^{k+1}}_{\iota+1}
  \otimes\oC^{r}\Bigr)\equiv[k+1,\dots,k+1,r],
  \label{6.2}
\end{equation} 
where $\Gr$ means taking the graded object associated with a
filtration existing on the tensor product.  Traveling along the paths
then corresponds to ``traveling'' through the tensor product factors.
The last formula can also be viewed as an explanation of the
square-bracket notation.  For the representations with $r=1$, the last
tensor factor and, correspondingly, `$1$' in $[k+1,\dots,k+1,1]$ can
be dropped.  Indeed, it is easy to see that replacing the starting
symbol $\Bigl[\,\underbrace{k+1,\dots,k+1}_{\iota+1},1\Bigl]$ with
$\Bigl[\,\underbrace{k+1,\dots,k+1}_{\iota+1}\Bigr]$ does not change
the resulting vectors~\eqref{6.1}.

\begin{figure}[tb]
  \begin{center}
    \unitlength0.8pt
    \begin{picture}(400,250)(-40,0)
      \put(-30,0){
        \put(0,240){$[2]$}
        \put(2,228){$\blacktriangle$}
        \put(50,240){$\blacktriangleright[1]$}
        \put(0,5){
          \put(80,220){$[2]$}
          \put(83,205){$\blacktriangle$}
          \put(130,220){$\blacktriangleright[1]$}
          }
        \put(0,5){
          \put(100,205){$[2]$}
          \put(103,190){$\blacktriangle$}
          \put(140,205){$\blacktriangleright[1]$}
          }
        \put(150,195){$[1]$}
        \put(153,181){$\blacktriangle$}
        \put(170,185){$[1]$}
        \put(173,170){$\blacktriangle$}
        \put(188,175){$[2]$}
        \put(191,160){$\blacktriangle$}
        \put(213,175){$\blacktriangleright[1]$}
        \put(-5,-30){
          \put(240,240){$[1]$}
          \put(243,225){$\blacktriangle$}
          \put(252,230){$[1]$}
          \put(255,215){$\blacktriangle$}
          \put(264,220){$[1]$}
          \put(266,205){$\blacktriangle$}
          \put(276,210){$[1]$}
          \put(278,195){$\blacktriangle$}
          }
        \put(0,-20){
          \put(325,240){$[1]$}
          \put(328,230){$\blacktriangle$}
          \put(338,230){$[1]$}
          \put(341,220){$\blacktriangle$}
          \put(352,220){$[1]$}
          \put(354,210){$\blacktriangle$}
          }
        \put(0,-20){
          \put(420,220){$[1]$}
          \put(422,207){$\blacktriangle$}
          }
        \put(0,110){$[3,2]$}
        \put(5,98){$\blacktriangle$}
        \put(60,110){$\blacktriangleright[2,2]$}
        \put(0,5){
          \put(100,90){$[2,2]$}
          \put(109,75){$\blacktriangle$}
          \put(150,90){$\blacktriangleright[2,1]$}
          \put(230,90){$\blacktriangleright[1,1]$}
          }
        \put(140,110){$\blacktriangleright[2,1]$}
        \put(0,10){
          \put(176,70){$[2,2]$}
          \put(184,56){$\blacktriangle$}
          \put(240,70){$\blacktriangleright[2,1]$}
          \put(310,70){$\blacktriangleright[1,1]$}
          }
        \put(220,110){$\blacktriangleright[1,1]$}
        \put(0,5){
          \put(265,60){$[2,1]$}
          \put(275,45){$\blacktriangle$}
          \put(317,60){$\blacktriangleright[1,1]$}
          }
        \put(340,50){$[1,1]$}
        \put(350,35){$\blacktriangle$}
        \put(0,0){$[3,3,2]$}
        \put(80,0){$\blacktriangleright[3,2,2]$}
        \put(160,0){$\blacktriangleright[2,2,2]$}
        \put(240,0){$\blacktriangleright[2,2,1]$}
        \put(320,0){$\blacktriangleright[2,1,1]$}
        \put(400,0){$\blacktriangleright[1,1,1]$}
        \put(424,55){$[1,1]$}
        \put(434,40){$\blacktriangle$}
        }
    \end{picture}
  \end{center}
  \caption[Paths for $k=2$, $r=2$, and $\Nn=1$.]{\label{fig:4} The
    labels on paths for $k=2$, \ $r=2$, and $\iota=1$.  Each cluster
    of $[~]$-labels is attached to a lattice site and the respective
    $\blacktriangle$ or~$\blacktriangleright$ symbol shows by which
    step this label was obtained from a preceding one.  All the
    admissible paths can be drawn by connecting the labels following
    the $\blacktriangle$ and~$\blacktriangleright$ directions (all the
    paths start at the bottom-left corner).  }
\end{figure}
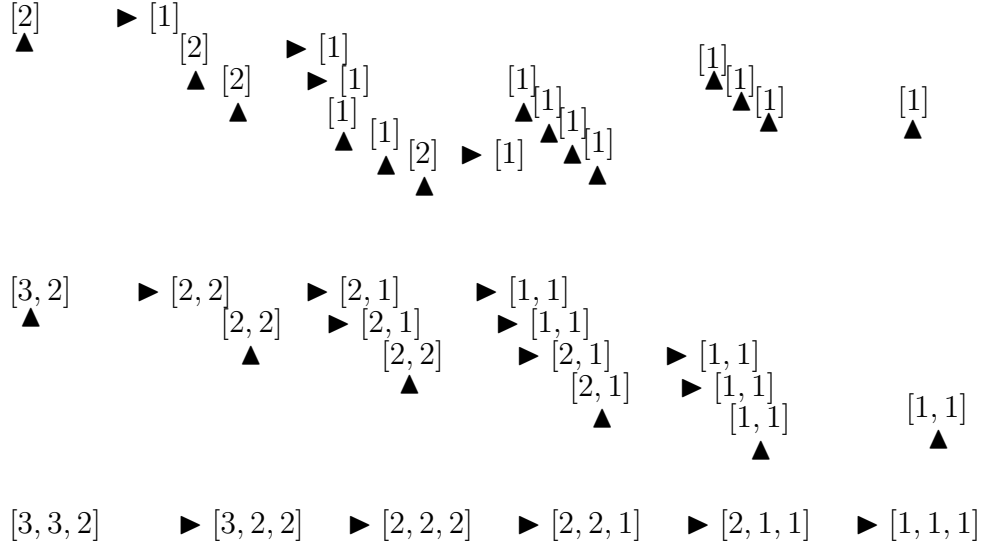

In Fig.~\ref{fig:4}, we consider the example where $k=2$, $r=2$, and
$\iota=1$, and the origin of paths~$\ell_{0,0}$ is therefore assigned
$[3,3,2]$.  The basis vectors read off from the admissible paths in
accordance with~\eqref{6.1} are given by
\begin{equation}
  \begin{matrix}
    f_2&f_1f_2&f_1^3&f_0^2f_1^2&f_0^5
    \\
    f_1&f_0f_2&f_0f_1^2&f_0^3f_1
    \\
    f_0&f_1^2&f_0^2f_1&f_0^4
    \\
    {}&f_0f_1&f_0^3
    \\
    {}&f_0^2&f_0^2f_2
  \end{matrix}
\end{equation} 
acting on the twisted highest-weight
state~$|2,2\,|\,1\rangle_{\widehat{s\ell}(2)}$.

\subsubsection{``Positive'' characters} \label{sec:6.1.2} The space
$\wM^+_{r,k}[\iota]$ is graded by the number of modes~$f_i$ applied to
the twisted highest-weight vector,
\begin{equation}
  \wM^+_{r,k}[\iota]=\bigoplus_{j=0}^{r+2k\iota}
  \wM^+_{r,k}[\iota;j],
  \label{6.3}
\end{equation} 
where each $\wM^+_{r,k}[\iota;j]$ is generated from
$|r,k|\iota\rangle_{\widehat{s\ell}(2)}$ by precisely~$j$
operators~$f_{\bullet}$.  The dimensions of $\M^+_{r,k}[\iota;j]$ are
arranged into a generalized Pascal triangle (cf.~\cite{ref:14}).

In the generalized Pascal triangle labeled by $k$ and $r$, the top row
consists of $r$ units, and each element in the $i$th row is the sum of
$k+1$ elements of the $(i-1)$th row: for even $k$, the sum includes
the element above the chosen one and $(k+2)/2 - 1$ of its neighbours
on each side; for odd $k$, it runs over $(k+1)/2$ elements
``north-west'' and $(k+1)/2$ elements ``north-east'' of the chosen
one.  For $k=1$, both cases $r=1$ and $r=2$ reduce to the standard
Pascal triangles (starting with~$1$ and $1,1$ in the top row
respectively); for $k=2$, the triangles are shown in Fig.~\ref{fig:5}.
\textit{The dimension of $\M^+_{r,k}[\iota;j]$ is read of from the
  $j$th entry of the $2\iota$th row in the Pascal triangle with the
  parameters~$k$ and~$r$}.

\begin{figure}[tb]
  \begin{equation*}
    \hskip-1mm
    \begin{matrix}
      1
      \\
      1\,1\,1
      \\
      1\,2\,3\,2\,1
      \\
      1\,3\,6\,7\,6\,3\,1
      \\
      1\,4\,10\,16\,19\,16\,10\,4\,1
      \\
      1\,5\,15\,30\,45\,51\,45\,30\,15\,5\,1
      \\
      \dots
      \\
      (r=1)
    \end{matrix}
    \quad
    \hskip-1mm
    \begin{matrix}
      1\,1
      \\
      1\,2\,2\,1
      \\
      1\,3\,5\,5\,3\,1
      \\
      1\,4\,9\,13\,13\,9\,4\,1
      \\
      1\,5\,14\,26\,35\,35\,26\,14\,5\,1
      \\
      1\,6\,20\,45\,75\,96\,96\,75\,45\,20\,6\,1
      \\
      \dots
      \\
      (r=2)
    \end{matrix}
    \quad
    \hskip-1mm
    \begin{matrix}
      1\,1\,1
      \\
      1\,2\,3\,2\,1
      \\
      1\,3\,6\,7\,6\,3\,1
      \\
      1\,4\,10\,16\,19\,16\,10\,4\,1
      \\
      1\,5\,15\,30\,45\,51\,45\,30\,15\,5\,1
      \\
      1\,6\,21\,50\,90\,126\,141\,126\,90\,50\,21\,6\,1
      \\
      \dots
      \\
      (r=3)
    \end{matrix}
  \end{equation*}
  \caption[Generalized Pascal triangles.]{\label{fig:5} Generalized
    Pascal triangles corresponding to unitary $\widehat{s\ell}(2)$
    representations of the level~$k=2$.  The $r=3$ case is a ``shift''
    of the $r=1$ one, in accordance with the spectral flow action on
    unitary representations, cf.~\cite{ref:55}.}
\end{figure}
        
Each space $\wM^+_{r,k}[\iota;j]$ is graded by the sum of modes
of~$f_i$.  The character of $\wM^+_{r,k}[\iota]$ can be written
through the $q$-supernomial coefficients
${\binom{\mathbf{L}_k}{a}}_{q}$, which are defined by the generating
functions~\cite{ref:52}
\begin{multline}
  T_{\mathbf{L}_k}(z,q)={}\sum_{a=0}^{\infty}z^a
  {\binom{\mathbf{L}_k}{a}}_{q}=
  \\
  ={} \sum_{L_1\geq N_1,L_2+N_1\geq N_2,\dots, L_{k}+N_{k-1}\geq
    N_{k}\geq0}z^{\sum_{i=1}^{k}N_i}
  q^{\sum_{i=1}^{k-1}N_{i+1}((\sum_{j=1}^{i}L_j)-N_i)}\times
  \\
  \times \bmatrix L_1 \\ N_1\endbmatrix_{q} \bmatrix L_2+N_1 \\ 
  N_2\endbmatrix_{q} \bmatrix L_3+N_2 \\ N_3\endbmatrix_{q}\dots
  \bmatrix L_{k}+N_{k-1} \\ {N_{k}}\endbmatrix_{q},
  \label{6.4}
\end{multline} 
where $\mathbf{L}_k=(L_1,L_2,\dots,L_{k})$ is a $k$-dimensional vector
with nonnegative integer entries and we use the standard notation
\begin{equation}
  \bmatrix n\\ m\endbmatrix_q=
  \begin{cases}
    \dfrac{(q)_n}{(q)_{n-m}(q)_m},&\quad n\geq m\geq0,
    \\
    0,&\quad\text{otherwise}.
  \end{cases}
  \label{6.5}
\end{equation} 
The supernomial coefficients are generating functions of partitions
admitting Durfee dissection with the defects
$(L_1,L_2,\dots,L_{k})$~\cite{ref:52}.

The character of $\wM^+_{r,k}[\iota]$ is given by~\cite{ref:42}
\begin{equation}
  \chr\wM^+_{r,k}[\iota](z,q)=
  z^{-k\iota-r+1}q^{k\iota^2+\iota(r-1)}T_{\mathbf{L}_k}(z,q^{-1}),
  \label{6.6}
\end{equation} 
where the $k$-dimensional vector $\mathbf{L}_k$ is chosen as
$\mathbf{L}_k=\Bigl(2\iota,0,0,\dots,0,
\underbrace{1,0,\dots,0}_{r-1}\,\Bigr)$.  After some algebra,
Eq.~\eqref{6.6} can be rewritten through $q$-binomial coefficients as
\begin{multline}
  \chr\wM^+_{r,k}[\iota](z,q)={}
  \sum_{\substack{N_1,N_2,\dots,N_{k} \\
      \iota\geq N_1\geq N_2\geq\dots\geq N_{k}\geq-\iota-1 \\
      N_{i}\geq-\iota\quad\text{for}\quad i\leq k+1-r}}
  z^{N_1+N_2+\dots+N_{k}}q^{\sum_{m=1}^{k}N_m^2+
    \sum_{m=k+2-r}^{k}N_m}\times
  \\
  \times \bmatrix 2\iota \\ \iota+N_1 \endbmatrix_{q} \bmatrix
  \iota+N_1 \\ \iota+N_2 \endbmatrix_{q}\dots \bmatrix \iota+N_{k-r}
  \\ \iota+N_{k+1-r}\endbmatrix_{q}\times
  \\
  \times \bmatrix \iota+N_{k+1-r}+1 \\ 
  \iota+N_{k+2-r}+1\endbmatrix_{q} \bmatrix \iota+N_{k+2-r}+1 \\ 
  \iota+N_{k+3-r}+1\endbmatrix_{q}\dots \bmatrix \iota+N_{k-1}+1 \\ 
  \iota+N_{k}+1\endbmatrix_{q}.
  \label{6.7}
\end{multline} 
The character in~\eqref{6.7} is normalized such that the
highest-weight vector $|r,k|0\rangle_{\widehat{s\ell}(2)}$ is in the
bigrade $(0,0)$.

\subsubsection{Positive bases and characters of
  $\W^+_{r,p;\theta}[\iota]$} \label{sec:6.1.3} 
For the $\N2$ spaces $\W^+_{r,p;\theta}[\iota]$ in Eq.~\eqref{5.1}, we
use the above construction to build a basis as follows.  The space
$\W^+_{r,p;\theta}[\iota]$ is graded by the number of
the~$\cG_{\bullet}$ operators,
\begin{equation}
  \W^+_{r,p;\theta}[\iota]=
  \bigoplus_{0\leq j\leq\frac{r-1+(p-2)(p\iota+\theta)}{p-1}}
  \W^+_{r,p;\theta}[\iota;j],
  \label{6.8}
\end{equation} 
where $\W^+_{r,p;\theta}[\iota;j]$ is the space obtained by
applying~$j$ operators~$\cG_{\bullet}$ to the extremal vector
$|r,p;\theta|\iota\rangle_{\infty/2}$. There exists an isomorphism
\textit{of vector spaces}
\begin{equation}
  \W^+_{r,p;\theta}[\iota;j]\simeq
  \wM^+_{r,p-2}[p\iota+\theta-j;j]
  \label{6.9}
\end{equation} 
induced by mapping the basis elements of
$\wM^+_{r,p-2}[p\iota+\theta-j;j]$ as
\begin{multline}
  (f_0)^{i_0}(f_1)^{i_1}\dots(f_j)^{i_j}\mapsto\cG_{0}\cG_{1}\dots
  \cG_{i_0-1}\cG_{i_0+1}\cG_{i_0+2}\dots
  \cG_{i_0+i_1}\cG_{i_0+i_1+2}\ldots\times
  \\
  \times\ldots \cG_{i_0+i_1+\dots+i_{j-1}+j}
  \dots\cG_{i_0+i_1+\dots+i_{j-1}+i_{j}+j-1}.
  \label{6.10}
\end{multline} 
This gives a basis in $\W^+_{r,p;\theta}[\iota]$.

\begin{figure}[tb]
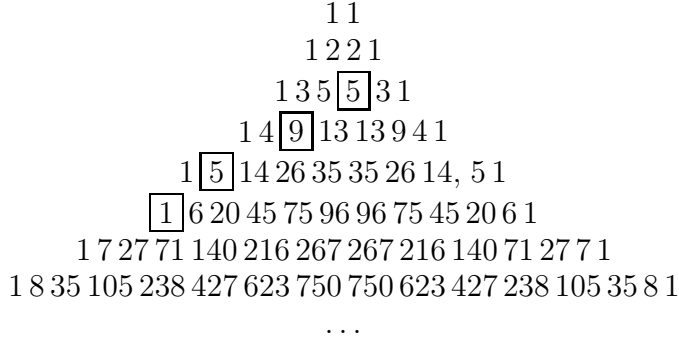

  \begin{equation*}
    \begin{matrix}
      1\,1
      \\
      1\,2\,2\,1
      \\
      1\,3\,5\,\boxed{5}\,3\,1
      \\
      1\,4\,\boxed{9}\,13\,13\,9\,4\,1
      \\
      1\,\boxed{5}\,14\,26\,35\,35\,26\,14,\,5\,1
      \\
      \boxed{1}\,6\,20\,45\,75\,96\,96\,75\,45\,20\,6\,1
      \\
      1\,7\,27\,71\,140\,216\,267\,267\,216\,140\,71\,27\,7\,1
      \\
      1\,8\,35\,105\,238\,427\,623\,750\,750\,623\,427\,238\,105\,35\,8\,1
      \\
      \dots
    \end{matrix}
  \end{equation*} 
  \caption[Knight moves]{\label{fig:6} Knight moves.  The dimensions of
    the spaces~$\W^+_{2,4;1}[1;j]$ are selected in the positions
    connected by knight moves in the generalized Pascal triangle.  }
\end{figure}

In accordance with~\eqref{6.9}, the dimensions
of~$\W^+_{r,p;\theta}[\iota;j]$ can also be read off from the
generalized Pascal triangles (see an example in Fig.~\ref{fig:6}).
These dimensions are given by the generalized Fibonacci numbers
$F^{r,p}_{p\iota+\theta}$ satisfying the defining relations
$F^{r,p}_i=F^{r,p}_{i-1}+F^{r,p}_{i-2}+\dots +F^{r,p}_{i-(p-1)}$.

We conjecture a character formula for~$\W^+_{r,p;\theta}[N]$.  We
normalize the analogue of~\eqref{6.7} such that the state
$|r,p;\theta|0\rangle_{\infty/2}$ is in the grade~$(0,0)$.

\begin{Conj}\label{Hypothesis} The characters of the subspaces
  involved in the positive filtration of unitary $\N2$ representations
  are given by
  \begin{equation}
    \chr\W^+_{r,p;\theta}[\iota](z,q)=z^{-(p-2)\iota-r+1}
    q^{\frac{p(p-2)\iota(\iota+1)}{2}-
      \iota(\frac{p(p-1)}{2}-r-\theta(p-2))}S_{\mathbf{L}_p}(z,q^{-1}),
    \label{6.12}
  \end{equation} 
  where
  \begin{equation}
    S_{\mathbf{L}_p}(z,q)=\sum_{a=0}^{\infty} z^aq^{\frac{a^2-a}{2}}
    \begin{pmatrix}
      {\mathbf{L}_p-(a,0,\dots,0)} \\ a
    \end{pmatrix}_q,
    \label{6.13}
  \end{equation} 
  with $\mathbf{L}_p=\Bigl(p\iota+\theta,0,0,\dots,0,
  \underbrace{1,0,\dots,0}_{r-1}\,\Bigr)$.
\end{Conj}

These characters can be rewritten as
\begin{multline}
  \chr\W^+_{r,p;\theta}[\iota](z,q)={}
  \sum_{\substack{N_1,N_2,\dots,N_{p-2}
      \\
      N_1+n\leq\iota+\theta-r+1,N_1\geq N_2\geq\dots\geq
      N_{p-2}\geq-\iota-1
      \\
      N_i\geq-\iota\quad\text{for}\quad i\leq p-r-1}}z^n
  q^{\frac{n^2-n}{2}-\theta
    n+\sum_{m=1}^{p-2}N_m^2+\sum_{m=r}^{p-2}N_m}\times
  \\
  \times \bmatrix 2\iota+\theta-r+1-n \\ \iota+N_1\endbmatrix_{q}
  \bmatrix \iota+N_1 \\ \iota+N_2\endbmatrix_{q}\dots \bmatrix
  \iota+N_{p-r-2} \\ \iota+N_{p-r-1}\endbmatrix_{q}\times
  \\
  \times \bmatrix \iota+N_{p-r-1}+1 \\ \iota+N_{p-r}+1\endbmatrix_{q}
  \bmatrix \iota+N_{p-r}+1 \\ \iota+N_{p-r+1}+1\endbmatrix_{q}\dots
  \bmatrix \iota+N_{p-3}+1 \\ \iota+N_{p-2}+1\endbmatrix_{q},
  \label{6.14}
\end{multline} 
where $n=\sum_{m=1}^{p-2}N_m$. It is easy to verify that this
expression has the correct limit as $\iota\rightarrow\infty$,
\begin{equation}
  \lim_{\iota\to\infty}\Bigl(z^{\frac{1-r+2\theta}{p}-\theta}
  q^{\frac12(1-\frac{2}{p})(\theta^2-\theta)+\theta\frac{r-1}{p}}
  \chr\W^+_{r,p;\theta}[\iota](z,q)\Bigr)=
  \chr\W_{r,p;\theta}(z,q),
  \label{6.15}
\end{equation} 
where the character $\chr\W_{r,p;\theta}(z,q)$ is given
by~\eqref{4.31}.

\subsection{The $\N2$ modular functor and functions on Riemann
  surfaces} \label{sec:6.2}
The representation of the unitary modules via semi-infinite forms
implies a relation between the~$\N2$ modular functor and the spaces of
skew-symmetric functions with prescribed singularities on Cartesian
powers of a genus-$g$ Riemann surface.

\subsubsection{$\N2$ correlation functions in the semi-infinite
  picture}\label{sec:6.2.1} Let $\cE_{n}^{g}$ be a genus-$g$ Riemann
surface with~$n$ marked points $P_1,\dots,P_n$ and let
$(\N2)^{\text{out}}$ denote the algebra generated by the part of the
$\N2$ currents $\cG(z)$, $\cQ(z)$,~$\cH(z)$, and~$\cL(z)$ that is
holomorphic outside the points $P_1,\dots,P_n$.  One then defines the
space of coinvariants
\begin{equation}
  \coinv=\frac{\mK_{r_1,p;\theta_1}\otimes
    \mK_{r_2,p;\theta_2}\otimes\dots\otimes\mK_{r_n,p;\theta_n}}
  {(\N2)^{\text{out}}\mK_{r_1,p;\theta_1}\otimes
    \mK_{r_2,p;\theta_2}\otimes\dots\otimes\mK_{r_n,p;\theta_n}},
  \label{6.16}
\end{equation} 
where $\mK_{r_i,p;\theta_i}$ are unitary representations and
$\mK_{r_1,p;\theta_1}\otimes\mK_{r_2,p;\theta_2}
\otimes\dots\otimes\mK_{r_n,p;\theta_n}$ is a representation of the
algebra $(\N2)_{P_1}\oplus\dots\oplus(\N2)_{P_n}$.  Similarly
to~\cite{ref:32}, this space of coinvariants can be shown to be
isomorphic to the space of coinvariants with respect to the algebra
generated by~$\cG(z)$ and~$\cH_0$,
\begin{equation}
  \coinv=\left(\frac{\mK_{r_1,p;\theta_1}\otimes
      \mK_{r_2,p;\theta_2}\otimes\dots\otimes\mK_{r_n,p;\theta_n}}
    {\g^{\text{out}}\mK_{r_1,p;\theta_1}\otimes
      \mK_{r_2,p;\theta_2}\otimes\dots\otimes
      \mK_{r_n,p;\theta_n}}\right)^0,
  \label{6.17}
\end{equation} 
where $\g^{\text{out}}$ is the algebra generated by the part of the
$\cG(z)$ current that is holomorphic outside $P_1,\dots,P_n$ and
$(\cdot)^0$ denotes the restriction to the zero-charge component (the
zero-charge restriction comes from taking the coinvariants with
respect to~$\cH_0$).

For a given set of representations $\mK_{r_1,p;\theta_1},\dots,
\mK_{r_n,p;\theta_n}$ placed at the points $P_1$,\dots, $P_n$ on the
Riemann surface and for fixed~$m$ and $\iota_1,\dots,\iota_n$, each
linear functional $\langle F|:\coinv\to\oC$ defines the function of of
$(x_1,\dots,x_m)\in\cE_{n}^{g}\times\dots\times\cE_{n}^{g}$ given by
\begin{equation}
  \langle F|\cG(x_1)\cG(x_2)\dots\cG(x_m)|
  \Phi^{\iota_1}_{r_1,\theta_1}(P_1)\otimes
  \Phi^{\iota_2}_{r_2,\theta_2}(P_2)\otimes\dots\otimes
  \Phi^{\iota_n}_{r_n,\theta_n}(P_n)\rangle_{p}^{g},
  \label{6.18}
\end{equation} 
where $\Phi^{\iota}_{r,\theta}$ are the operators corresponding to
extremal vectors~\eqref{3.1}.  In accordance with the chosen conformal
dimension of~$\cG$, expression~\eqref{6.18} is a 2-differential in
each variable.  We rewrite this expression in local coordinates by
separating the \textit{correlation function},
\begin{equation}
  \begin{split}
    &F^{p,g;\iota_1,\dots,\iota_n}_{r_1,r_2,\dots,r_n;
      \theta_1,\theta_2,\dots,\theta_n} (x_1,x_2,\dots,x_m\mid
    P_1,P_2,\dots,P_n)\,(dx_1)^2\, (dx_2)^2\dots(dx_m)^2=
    \\
    &\qquad\qquad =\langle F|\cG(x_1)\cG(x_2)\dots\cG(x_m)\mid
    \Phi^{\iota_1}_{r_1,\theta_1}(P_1)\otimes
    \Phi^{\iota_2}_{r_2,\theta_2}(P_2)\otimes\dots\otimes
    \Phi^{\iota_n}_{r_n,\theta_n}(P_n)\rangle_{p}^{g}.\qquad
  \end{split}
  \label{6.19}
\end{equation} 
The restriction to the zero charge component takes the form of the
constraint
\begin{equation}
  mp+n-\bigl(r_1+(p-2)(\theta_1+p\iota_1)\bigr)-
  \cdots-\bigl(r_n+(p-2)(\theta_n+p\iota_n)\bigr)=0.
  \label{6.20}
\end{equation}

It follows that the functions
$F^{p,g;\iota_1,\dots,\iota_n}_{r_1,r_2,\dots,r_n;
  \theta_1,\theta_2,\dots,\theta_n}$ defined in~\eqref{6.19} are
antisymmetric in $x_1,\dots,x_m$, are regular on
$\cE_{n}^{g}\times\dots\times\cE_{n}^{g}$ except at the points~$P_i$,
and possess the following properties.

\begin{enumerate}
\item \label{diag} On each $(p-1)$-diagonal
  $x_{i_1}=x_{i_2}=\dots=x_{i_{p-1}}$, one has
  \begin{equation}
    \frac{\partial^{p-2}}{\partial x_{i_1}^{p-2}}
    \frac{\partial^{p-3}}{\partial x_{i_2}^{p-3}}\dots
    \frac{\partial}{\partial x_{i_{p-2}}}
    F^{p,g;\iota_1,\dots,\iota_n}_{r_1,r_2,\dots,r_n;
      \theta_1,\theta_2,\dots,\theta_n}
    \biggr|_{x_{i_1}=x_{i_2}=\dots=x_{i_{p-1}}}=0.
    \label{6.21}
  \end{equation}
  
\item \label{zeros} For each $a$ such that $a<r_j$, the function
  \begin{equation}
    \frac{\partial^{a-1}}{\partial x_{i_a}^{a-1}}
    \frac{\partial^{a-2}}{\partial x_{i_{a-1}}^{a-2}}\dots
    \frac{\partial}{\partial x_{i_2}}
    F^{p,g;\iota_1,\dots,\iota_n}_{r_1,r_2,\dots,r_n;
      \theta_1,\theta_2,\dots,\theta_n}
    \label{6.22}
  \end{equation} 
  vanishes at $x_{i_1}=x_{i_2}=\dots=x_{i_{a}}=x\to P_j$ with an order
  not less than
  \begin{equation}
    -a(\theta_j+\iota_jp)+\frac{a(a-3)}{2}.
    \label{6.23}
  \end{equation} 
  For each~$a$ such that $r_j\leq a\leq p-2$, the function
  \begin{equation}
    \frac{\partial^{a}}{\partial x_{i_a}^{a}}\dots
    \frac{\partial^{r+1}}{\partial x_{i_{r+1}}^{r+1}}
    \frac{\partial^{r}}{\partial x_{i_{r}}^{r}}
    \frac{\partial^{r-2}}{\partial x_{i_{r-1}}^{r-2}}
    \frac{\partial^{r-1}}{\partial x_{i_{r-2}}^{r-1}}\dots
    \frac{\partial}{\partial x_{i_2}}
    F^{p,g;\iota_1,\dots,\iota_n}_{r_1,r_2,\dots,r_n;
      \theta_1,\theta_2,\dots,\theta_n}
    \label{6.24}
  \end{equation} 
  vanishes at $x_{i_1}=x_{i_2}=\dots=x_{i_{a}}=x\to P_j$ with the
  order not less than
  \begin{equation}
    -a(\theta_j+\iota_jp)+\frac{a(a-4r_j+3)}{2}-(r_j+1)(r_j-1)
    \label{6.25}
  \end{equation}
\end{enumerate}
(negative-order zeros are poles).  Condition~\eqref{6.21} follows from
the vanishing of~\eqref{1.1}, and~\eqref{6.23} and~\eqref{6.25} from
relations~\eqref{3.2}--\eqref{3.4}.

We let $\cF^{p,g;\iota_1,\dots,\iota_n}_{r_1,r_2,\dots,r_n;
  \theta_1,\theta_2,\dots,\theta_n}(m)$ denote the space of all
functions satisfying these properties; with Eq.~\eqref{6.20} assumed
to be satisfied, the notation is somewhat redundant; however, it is
useful to keep $m$ as a free parameter and determine some other label
from~\eqref{6.20}.

\begin{Thm}\label{thm:8}
  For sufficiently large~$m$, the assignment
  \begin{equation}
    \langle F|\mapsto
    F^{p,g;\iota_1,\dots,\iota_n}_{r_1,r_2,\dots,r_n;
      \theta_1,\theta_2,\dots,\theta_n}
    (\cdot,\dots,\cdot\mid P_1,P_2,\dots,P_n)
  \end{equation} 
  defined in~\eqref{6.19} establishes an isomorphism
  between~$\coinv^*$ and
  $\cF^{p,g;\iota_1,\dots,\iota_n}_{r_1,r_2,\dots,
    r_n;\theta_1,\theta_2,\dots,\theta_n}(m)$.  For these~$m$, the
  dimensions
  $d^{p,g}_{r_1,r_2,\dots,r_n;\theta_1,\theta_2,\dots,\theta_n}=
  \dim\cF^{p,g;\iota_1,\dots,\iota_n}_{r_1,r_2,\dots,r_n;
    \theta_1,\theta_2,\dots,\theta_n}(m)$ are independent of~$m$ and
  $\iota_1,\dots,\iota_n$.
\end{Thm}

The semi-infinite construction is essential in proving that
\textit{all} functions with the properties described above are the
$\N2$ correlation functions.  We do not give the proof here and
consider only two examples for illustration.  One example is a simple
enumeration of functions on a single-punctured elliptic curve with the
vacuum representation.  The dimension of the space of these functions
gives the number of primary fields in the corresponding minimal model.
For $p=3$ and $p=4$, we explicitly construct a basis in the functional
space.  The second example is with a three-punctured Riemann sphere,
where the dimensions of the corresponding functional spaces are
related to the $\N2$ fusion algebra; we digress in Sec~6.3 to derive
the unitary $\N2$ fusion algebra from the~$\widehat{s\ell}(2)$ unitary
fusion algebra.

\subsubsection{Tori with one marked point for small $p$}
\label{sec:6.2.2} For $p=3$, we consider the vacuum module associated
with a point on a torus (at the origin in the covering complex plane).
Setting $n=1$ and $r_1=1$ in condition~\eqref{6.20} then gives
$\iota=m$.  The independence from~$m$ already occurs starting with
$m=1$, and the space $\cF^{3,1}_{1;0}(1)$ (omitting the~$\iota$
superscript for brevity) consists of functions on the torus with a
pole at zero of an order~$\leq3$. We then have $d^{3,1}_{1;0}=3$, and
a basis in the space of such functions is
\begin{equation}
  1,\qquad\wp(x),\qquad\wp'(x),
  \label{6.26}
\end{equation} 
where
\begin{equation}
  \wp(x)=\frac{1}{x^2}+\sum'_{\omega\in\Lambda}
  \Bigl(\frac{1}{(x-\omega)^2}-\frac{1}{\omega^2}\Bigr)
  \label{6.27}
\end{equation} 
is the Weierstrass function, $\Lambda=\{m\omega_1+n\omega_2\mid
m,n\in\oZ\}$, \ $\omega_1,\omega_2\in\oC$, and
$\Im({\omega_1}/{\omega_2})>0$.

It is instructive to explicitly verify that the same dimension is also
obtained for $m=2$, i.e., to verify that
$\dim\cF^{3,1}_{1;0}(2)=d^{3,1}_{1;0}=3$, where the calculation is
entirely different because the basic condition~\ref{diag}
(see~\eqref{6.21}) applies in this case.  The corresponding space
$\cF^{3,1}_{1;0}(2)$ consists of antisymmetric functions of two
variables~$x_1$ and~$x_2$ with a pole of an order~$\leq6$ as
$x_1=x_2\to0$ and such that $\bigl({\partial} f(x_1,x_2)/{\partial
  x_{1}}\bigr)\bigr|_{x_1=x_2}=0$.  This space has a basis
\begin{align*}
  f_1(x_1,x_2)={}&\wp(x_1)^3-3\wp(x_1)^2\wp(x_2)+
  3\wp(x_1)\wp(x_2)^2-\wp(x_2)^3,
  \\
  f_2(x_1,x_2)={}&-g_2\wp'(x_1)-\frac{1}{2}g_1\wp(x_1)\wp'(x_1)+
  \wp(x_1)^3\wp'(x_1)-
  \\
  &-\frac{3}{2}g_1\wp(x_2)\wp'(x_1)+ 3\wp(x_2)^3\wp'(x_1)+
  g_2\wp'(x_2)+\frac{3}{2}g_1\wp(x_1)\wp'(x_2)-
  \\
  &-3\wp(x_1)^3\wp'(x_2)+
  \frac{1}{2}g_1\wp(x_2)\wp'(x_2)-\wp(x_2)^3\wp'(x_2),
  \\
  f_3(x_1,x_2)={}&g_2\wp(x_1)+g_1\wp(x_1)^2-g_2\wp(x_2)-2\wp(x_1)^3\wp(x_2)-
  g_1\wp(x_2)^2+
  \\
  &+2\wp(x_1)\wp(x_2)^3+
  \wp(x_1)\wp'(x_1)\wp'(x_2)-\wp(x_2)\wp'(x_1)\wp'(x_2),
\end{align*}
where $g_1=30\sum'_{\omega\in\Lambda}\tfrac{1}{\omega^4}$ and
$g_2=140\sum'_{\omega\in\Lambda}\tfrac{1}{\omega^6}$.  This recovers
the dimension $d^{3,1}_{1;0}=3$.

For $p=4$, condition~\eqref{6.20} becomes $2\iota=m$.  The first
``sufficiently large'' value of~$m$ is already $m=2$, with $\iota=1$.
The corresponding space~$\cF^{4,1}_{1;0}(2)$ consists of antisymmetric
functions of two variables with a pole of an order~$\leq4$ at zero and
such that the function $\bigl({\partial^{2}} f(x_1,x_2)/{\partial
  x_{1}^{2}}\bigr)\bigr|_{x_1=x_2=x}$ has a pole of the order~$\leq9$
as $x\to0$.  A basis in the space of such functions can be chosen as
\begin{multline}
  f_1=\wp(x_1)-\wp(x_2),\qquad f_2=\wp'(x_1)-\wp'(x_2),\qquad
  f_3=\wp(x_1)^2-\wp(x_2)^2,\\
  f_4=\wp(x_1)\wp'(x_2)-\wp(x_2)\wp'(x_1),\qquad
  f_5=\wp(x_1)^2\wp'(x_2)-\wp(x_2)^2\wp'(x_1),\\
  f_6=\wp(x_1)\wp(x_2)^2-\wp(x_2)\wp(x_1)^2,
\end{multline}
and we have~$d^{4,1}_{1;0}=6$.

These examples show that the dimensions of functional spaces coincide
with the number of primary fields in the respective minimal model
(i.e., with the dimensions of the modular functor for the torus).

\subsection{The unitary $\N2$ fusion algebra} \label{sec:6.3} We
consider the $\N2$ fusion algebra and then discuss its consequences
for the functional spaces.  The unitary $\N2$ fusion rules were
derived in~\cite{ref:53} from the Verlinde hypothesis.  We give an
alternative derivation, which starts with the $\widehat{s\ell}(2)$
fusion algebra; this derivation does not rely on the Verlinde theorem
statement for the $\N2$ algebra, and the coincidence with the result
in~\cite{ref:53} may have an independent interest.

The fusion algebra $\FA{\widehat{s\ell}(2)}(k)$ of the unitary
level-$k$ $\widehat{s\ell}(2)$ representations $\mAs_{r,k}$ is given
by~\cite{ref:62}
\begin{equation}
  \mAs_{r_1,k}\sfusion\mAs_{r_2,k}=
  \bigoplus_{\substack{r_3=|r_1-r_2|+1 \\ \text{step}=2 }}^{k+1-|r_1+r_2-k-2|}
  \mAs_{r_3,k},\qquad1\leq r_i\leq k+1.
  \label{6.29}
\end{equation} 
It is easy to apply the method of~\cite{ref:54} to~\eqref{6.29}:
tensoring the $\widehat{s\ell}(2)$ modules with the module~$\Omega$
over free fermions and applying~\eqref{2.11}, we use
$\widehat{s\ell}(2)$ fusion rules and then collect the terms on the
right-hand side so as to identify some $\mK_{r',p;\theta'}$
representations.  This gives the $\N2$ fusion algebra
\begin{equation}
  \mK_{r_1,p;\theta_1}\nfusion
  \mK_{r_2,p;\theta_2}=
  \bigoplus_{\substack{r_3=|r_1-r_2|+1\\
      \text{step}=2 }}^{p-1-|r_1+r_2-p|}
  \mK_{r_3,p;\theta_1+\theta_2+\frac{1}{2}(1-r_1-r_2+r_3)},
  \quad
  1\leq r_1,r_2\leq p-1,~\theta_1,\theta_2\in\oZ_p.
  \label{6.30}
\end{equation} 
\begin{Rem}[the spectral flow]\label{rem:4} In the fusion algebra,
  $\mK_{1;0}$ is the identity and $\Theta=\mK_{p-1;0}$ is the spectral
  flow operator acting on representations in accordance
  with~\eqref{6.30} as
  \begin{equation}
    \mK_{p-1,p;0}\nfusion\mK_{r_2,p;\theta}=
    \mK_{p-r_2,p;\theta+1-r_2}\approx\mK_{r_2,p;\theta+1}.
    \label{6.31}
  \end{equation} 
  We also have $\underbrace{\mK_{p-1,p;0} \nfusion\dots
    \nfusion\mK_{p-1,p;0}}_p=\mK_{1,p;0}\equiv1$, as must be the case
  with the spectral flow on unitary representations.
\end{Rem}

The above fusion algebra is only for the Neveu--Schwarz sector.  To
extend the fusion to the Ramond sector, \textit{it suffices to add a
  single element $\mK_{1,p;\frac{1}{2}}$ such that}
\begin{equation}
  \mK_{1,p;\frac12}\nfusion\mK_{r,p;\theta}=
  \mK_{r,p;\theta+\frac12}.
  \label{6.32}
\end{equation} 
This relation completely determines the fusion involving Ramond sector
representations.  Moreover, the fusion can be formally extended to any
other ``sector'' with the fractional twists $\theta={\beta}/{\alpha}$
by adding a single field $\mK_{1,p;\frac{1}{\alpha}}$ such that
\begin{equation}
  \mK_{1,p;\frac{1}{\alpha}}\nfusion
  \mK_{r,p;\theta}=\mK_{r,p;\theta+\frac{1}{\alpha}},
  \label{6.33}
\end{equation} 
which then determines all the fusion rules for
$\mK_{r,p;\frac{2}{\alpha}+\oZ},\dots,
\mK_{r,p;\frac{\alpha-1}{\alpha}+\oZ}$.

\begin{Rem}\label{rem:5} Fusion rules~\eqref{6.30} mean that the
  three-point function is nonvanishing if and only if
  \begin{multline}
      r+r'+r''-2\theta-2\theta'-2\theta''-3=0, \\
      |r'-r''|<r<r'+r'',\qquad r+r'+r''<2p,\qquad
      r+r'+r''\equiv1\bmod2.
    \label{6.34}
  \end{multline} 
  This agrees with the result in ~\cite{ref:53} under the
  Neveu--Schwarz sector correspondence
  \begin{equation}
    k=r-\theta-\frac12,\qquad j=
    \theta+\frac12,\qquad1\leq r\leq p-1,\qquad
    0\leq\theta\leq r-1
    \label{6.35}
  \end{equation} 
  between our parameterization of the unitary $\N2$ modules and the
  parameterization used in~\cite{ref:60},~\cite{ref:53}.
\end{Rem}

We now show how the correspondence between the $\widehat{s\ell}(2)$
and $\N2$ fusion algebras derived above fits into the general scheme
between the modular functors expressed by~\eqref{2.13}.
In~\eqref{2.13}, we take~$\cE$ to be a torus and recall that there is
a basis in the modular functor on the torus whose elements are given
by unitary representation characters (such bases, more precisely,
depend on a cycle on the torus).  The fusion algebra is then an
operation on the modular functor for the torus.

It turns out that the correspondence in Eq.~\eqref{2.13} agrees with
natural structures on the modular functors, in the present case, with
the fusion algebra.  With the modular functors for the torus
identified with the respective fusion algebras $\FA{\aA}(\kappa)$ for
$\aA=\widehat{s\ell}(2)$ and $\N2$, Eq.~\eqref{2.13} becomes
\begin{equation}
  \FA{\N2}(p)=\Coinv_{\langle R_1\otimes P_1\rangle}
  \bigl(\Inv_{\langle R_2\otimes P_2\rangle}
  \bigl(\FA{\widehat{s\ell}(2)}(p-2)\otimes
  \FA{\text{free}}(p)\bigr)\bigr),
  \label{6.36}
\end{equation} 
where the invariants and coinvariants are taken with respect to the
subalgebras generated by the elements $R_1$, $R_2$, $P_1$, and $P_2$
explicitly constructed in what follows.  The algebra
$\FA{\text{free}}(p)$, which is the fusion algebra for the algebra of
vertex operators associated with the lattice $\sqrt{2p}\,\oZ$
(see~\eqref{2.11}), is isomorphic to the group algebra of the cyclic
group~$\oZ_{2p}$ (as a linear space, it is represented by
theta-functions of the level $2p=2(k+2)$, as can be seen from
Eq.~\eqref{2.10} for the characters).  It carries an action of the
Heisenberg group associated with half-periods, and we let~$P_1$
and~$P_2$ denote the corresponding generators (such that
$P_1P_2=(-1)^pP_2P_1$): if~$\nu$, with $\nu^{2p}=1$, is a generator
of~$\oZ_{2p}$, we have
\begin{align}
  P_1(\nu^i)&=\nu^{i+p},\label{6.37}
  \\
  P_2(\nu^i)&=(-1)^i\nu^i.
  \label{6.38}
\end{align}

\begin{figure}[tb]
  \begin{equation*}
    \begin{array}{lcccccccc}
      \mL_{1,2}\tensor\nu^*\quad
      &\bullet&\circ&\bullet&\circ&\bullet&\circ&\bullet&\circ\\
      \mL_{2,2}\tensor\nu^*
      &\circ&\bullet&\circ&\bullet&\circ&\bullet&\circ&\bullet\\
      \mL_{3,2}\tensor\nu^*
      &\bullet&\circ&\bullet&\circ&\bullet&\circ&\bullet&\circ\\
    \end{array}
  \end{equation*}
  \caption{\label{fig:7} The tensor product 
    $\FA{\widehat{s\ell}(2)}(p-2)\otimes \FA{\text{free}}(p)$ for
    $p=4$ (the $\widehat{s\ell}(2)$ level $k=2$).  The dots represent
    $\mL_{r,2}\otimes\nu^i$ for $r=1,2,3$ and $i=0,\dots,7$ (with~$i$
    labeling columns). The solid dots are those for which $r+i-1$ is
    even.}
\end{figure}

Using the chosen basis, the Heisenberg group action on the
$\widehat{s\ell}(2)$ modular functor can be explicitly described as
\begin{align}
  &R_1(\mAs_{r,k})=\mAs_{k-r+2,k},
  \label{6.39}
  \\
  &R_2(\mAs_{r,k})=(-1)^{r-1}\mAs_{r,k}
  \label{6.40}
\end{align}
(we recall that basis elements are identified with unitary
$\widehat{s\ell}(2)$ representations).  We note that~$R_2$ is an
automorphism of the fusion algebra and~$R_1$ is the spectral flow
transform realized via fusion, $R_1(\mAs_{r,k})= \mL_{k+1,k}\sfusion
\mL_{r,k}$.  It has an important property that (omitting the fusion
operation sign) $(\mAs_{k+1,k}a)(\mAs_{k+1,k}b)=ab$ for any
$a,b\in\FA{\widehat{s\ell}(2)}(k)$.  Taking the diagonal action of the
Heisenberg group, we now see that $R_1\otimes P_1$ indeed commutes
with $R_2\otimes P_2$ (i.e., the central element acts identically).
In the tensor product
$\FA{\widehat{s\ell}(2)}(p-2)\otimes\FA{\text{free}}(p)$, we then take
the invariants with respect to $R_2\otimes P_2$, i.e., restrict to the
elements of the form $\mAs_{r,k}\otimes\nu^i$ with even $r+i-1$ (see
Fig.~\ref{fig:7}).  The construction is then completed by taking
\textit{co}invariants with respect to the action of~$R_1\otimes P_1$.

We now label the elements of the space constructed on the right-hand
side of~\eqref{6.36} as
\begin{equation}
  \mK_{r,p;\theta}=\mAs_{r,p-2}\otimes\nu^{2\theta-r+1}
  \bmod R_1\otimes P_1
  \label{6.41}
\end{equation} 
and identify~$\mK_{r,p;\theta}$ with the corresponding generator of
the $\N2$ fusion algebra.  The identification under the action of
$R_1\otimes P_1$ means that~\eqref{2.7} is satisfied by construction;
the periodicity under the spectral flow transform by~$p$ is also
obvious.  Moreover, calculating
\begin{equation}
  (\mAs_{r_1,k}\otimes\nu^{\theta_1})
  \nfusion
  (\mAs_{r_2,k}\otimes\nu^{\theta_2})=
  \biggl(\mAs_{r_1,p-2}\sfusion
  \mAs_{r_2,p-2}\biggr)\otimes\nu^{\theta_1+\theta_2}
\end{equation} 
and using~\eqref{6.29}, we obtain the unitary $\N2$ fusion
algebra~\eqref{6.30}. Therefore, the relation between the
$\widehat{s\ell}(2)$ and $\N2$ fusion algebras can be considered as a
particular case of the general relation~\eqref{2.13} that extends the
equivalence of categories~\cite{ref:55},~\cite{ref:54} to modular
functors.

Returning to the functional spaces introduced in Sec.~\ref{sec:6.2},
we reformulate the $\N2$ fusion algebra as a statement on the
dimensions of functional spaces.  The structure constants of the
fusion algebra coincide with the dimensions of the modular functor on
the three-punctured $\oC\oP^1$.  Recalling that this modular functor
is related to the functional spaces via Theorem~\ref{thm:8}, we obtain
(omitting the~$\iota$ superscripts for brevity)

\begin{Prop}\label{Proposition} For $m\geq p-1$, the dimensions of the
  spaces $\cF^{p,0}_{r,r',r'';\theta,\theta',\theta''}(m)$ and
  $\cF^{p,0}_{r,r';\theta,\theta'}(m)$ are given by
  \begin{equation}
    d^{p,0}_{r,r',r'';\theta,\theta',\theta''}=
    \dim\cF^{p,0}_{r,r',r'';\theta,\theta',\theta''}(m)=
    \begin{cases}
      1,&\quad\text{conditions~\eqref{6.34} are satisfied},
      \\
      0,&\quad\text{otherwise},
    \end{cases}
    \label{6.42}
  \end{equation}
  and
  \begin{equation}
    d^{p,0}_{r,r';\theta,\theta'}=\dim\cF^{p,0}_{r,r';\theta,\theta'}(m)=
    \begin{cases}
      1,&\quad r'=r\quad\text{and}\quad\theta'=r-\theta-1,
      \\
      0,&\quad\text{otherwise},
    \end{cases}
    \label{6.43}
  \end{equation} 
  where $1\leq r,r',r''\leq p-1$ and $0\leq\theta,\theta',\theta''\leq
  r-1$.
\end{Prop}

\section{Concluding remarks} \label{sec:7} Possibly the most
conspicuous feature of semi-infinite constructions is the asymmetry
between the upper-triangular and lower-triangular generators, i.e.,
for the $\N2$ algebra, between the fermions~$\cG$ and~$\cQ$: starting
with a module generated by the modes~$\cG_n$ subject to the conditions
$\partial^{p-2}\cG(z)\dots\partial\cG(z)\cG(z)=0$, we could then (with
much effort) reconstruct the action of the other algebra generators,
including~$\cQ_m$.  However, the relation
$\partial^{p-2}\cQ(z)\dots\partial\cQ(z)\cQ(z)=0$ is certainly
satisfied for the current constructed from the~$\cQ_m$ generators; it
is interesting to investigate consequences of this relation for the
corresponding functional spaces.

We expect that the methods developed above for the $\N2$
superconformal algebra can also be useful in other semi-infinite
constructions. In similar constructs, depending on the chosen
constraints (the analogue of the relations $S^{p}_a=0$) a module
structure can be found on the semi-infinite space.  Even in the case
with only one (bosonic or fermionic) field satisfying the constraints,
interesting representations of W algebras can thus be obtained.
Nontrivial relations between $\widehat{s\ell}(2\,|\,1)$ and
\hbox{$\N2$}~\cite{ref:63} and $\widehat{s\ell}(2)$~\cite{ref:64}
representation theories allow us to expect interesting semi-infinite
constructions for a class of $\widehat{s\ell}(2\,|\,1)$
representations.

\bigskip

\noindent
\textbf{Acknowledgments.}  The authors are grateful to
J.~Figueroa-O'Farrill, A.~K.~Pogrebkov, I.~Yu.~Se\-ga\-lov,
V.~A.~Sirota, A.~Taormina, and R.~Weston for discussions and useful
remarks.  The work is partially supported by the RFBR
Grants~98-01-01155 and 99-01-01169, by the Russian Federation
President Grant~99-15-96037, and by INTAS-OPEN-97-1312.

\end{document}